\documentclass[10pt,a4paper]{article}
\usepackage[textwidth=153mm,top=25mm,bottom=25mm]{geometry}
\usepackage[T1]{fontenc}
\usepackage{lmodern}
\usepackage{amsmath,amssymb,bm,multirow}
\usepackage{graphicx,xcolor}
\usepackage[numbers,sort&compress]{natbib}
\usepackage[hidelinks]{hyperref}
\newcommand*\dd{\mathop{}\!\mathrm{d}}

\title{Quantum-vortex excitons beyond band topology}
\author{\parbox{\dimexpr\textwidth-2\tabcolsep\relax}{\centering\normalsize
Jin-Hyung Choi$^{1,2}$, Sang-Hoon Han$^2$, Young-Kwon Han$^3$,\\
Jun-Won Rhim$^{2,*}$, Sun-Woo Kim$^{1,4,*}$ and Joshua J. P. Thompson$^{5,*}$\\[1ex]
{\small
$^1$Department of Physics, Hanyang University, Seoul 04763, Republic of Korea\\[0.3ex]
$^2$Department of Physics, Ajou University, Suwon 16499, Republic of Korea\\[0.3ex]
$^3$Asia Pacific Center for Theoretical Physics, Pohang 37673, Republic of Korea\\[0.3ex]
$^4$Hanyang Institute for Quantum Science and Quantum Technology, Hanyang University, Seoul 04763, Republic of Korea\\[0.3ex]
$^5$Department of Materials Science and Metallurgy, University of Cambridge,
27 Charles Babbage Road, Cambridge CB3 0FS, United Kingdom\\[0.3ex]
$^*$Authors to whom correspondence should be addressed.\\[0.6ex]
E-mail: jwrhim@ajou.ac.kr, sunwookim@hanyang.ac.kr, jjt56@cam.ac.uk
}}}
\date{}
\begin{document}
\maketitle

\begin{abstract}
Optically exciting a typical low-dimensional semiconductor produces a nodeless \(s\)-type exciton as its lowest-energy bound state. Here, we explain how Coulomb phase matching stabilises a quantum-vortex exciton as the lowest bound state in flat Chern bands. Using a prototypical Yin--Yang kagome lattice, we demonstrate that increasing the spin--orbit coupling drives a transition of the lowest exciton from a quantum-vortex state to a zero-winding state, while leaving the band Chern numbers unchanged. We trace this behaviour to a gauge-invariant combination of exciton phase differences and Bloch-overlap phases that can reduce the Coulomb energy. By comparing states with identical wavefunction amplitudes, we isolate this phase contribution and show that it is sufficient to drive the reversal in exciton ordering. We classify exciton topology by the momentum-space wavefunction
winding relative to the conduction--valence Chern-number difference,
connecting the competing states to circular-polarisation selection rules. We then show that the resulting reordering can be tracked directly in the optical absorption spectrum, through the transition from a dark vortex exciton to a bright zero-winding exciton. These findings establish a microscopic link between exciton topology, Coulomb binding, and optical response that extends beyond the information contained in the underlying electronic band Chern numbers.
\end{abstract}

\section{Introduction}
Excitons are bound electron--hole pairs central to semiconductor
optics~\cite{kira2006many,wang2018colloquium,perea2022exciton}.
In topologically trivial systems, excitons exist in a non-hydrogenic
Rydberg series~\cite{berkelbach2013theory,chernikov2014binding}, with the
lowest and thermally dominant exciton being a nodeless, $s$-like
state~\cite{elliott1957intensity}.
Various computational and theoretical approaches have explored the
optical response of these
excitons~\cite{onida2002electronic,rohlfing2000electron,albrecht1998abinitio,benedict1998optical,berghauser2014analytical,qiu2013optical},
as well as their transport~\cite{chang2026ultrafast,rosati2020negative},
dynamics~\cite{selig2018dark}, and response to external
fields~\cite{rosati2026microscopic,scharf2016excitonic}.
Although excitonic properties are typically dictated by the interplay
of the electron--hole kinetic energies and the screening environment
in and around the material, the underlying Bloch-state properties of
the electron and hole have increasingly been identified as key factors
in determining excitonic properties.

Naturally, this has opened the door for the study of the impact of quantum geometry and band topology on the excitonic landscape~\cite{srivastava2015geometry,zhou2015berry,trushin2018selfrotating}. Research at the intersection of excitons and topology~\cite{wu2017topological,davenport2026graphene,chen2017chiral} ranges from generating topological excitons from trivial electronic bands~\cite{davenport2024interaction,hwang2026alchemy} to predictions of exciton transport enhanced by quantum geometry in topological organic nanoribbons~\cite{jankowski2025excitonic,thompson2025transport}.
These phenomena become particularly relevant in flat Chern
bands~\cite{tang2011high,sun2011nearly}, where exciton wavefunctions
can extend across the Brillouin zone and carry nonzero phase
winding~\cite{xie2024theory}.
Here, exciton topology refers to this winding interpreted relative
to the underlying nontrivial Bloch-state background, with the
conduction--valence Chern-number difference providing the reference.

Crucially, exciton topology has direct optical consequences,
giving rise to distinctive polarization-dependent selection
rules~\cite{cao2018unifying,zhang2018optical,lozano2025optical}.
These rules reflect the interplay between the phase winding of
the exciton wavefunction and that of the optical transition
matrix element.
This interplay can make states with different windings optically
bright or dark and allow them to couple to different circular
polarizations.
The sensitivity of optical response to exciton wavefunction
structure has been explored in bilayer graphene~\cite{ju2017tunable}
and topological
insulators~\cite{garate2011excitons,efimkin2013resonant,hou2023exchange,maisel2023single,kung2019chiral}.
These connections make polarization-resolved spectroscopy a
means of probing exciton wavefunction topology.

However, the topology of the constituent bands alone does not determine
which of these exciton states is energetically favoured. Establishing this
ordering is essential for identifying the thermally dominant states
and understanding which topological and optical properties become
relevant at low energies. Such an analysis must account for the full
screened Coulomb interaction, which is nonlocal in momentum space, with
its contributions depending jointly on the exciton wavefunction and
the underlying Bloch states. Here we investigate how this interaction selects exciton states in flat Chern
bands and connects their topology to their energies and optical
responses.

Specifically, we analyse excitons with zero total momentum in the
nearly flat bands of the Yin--Yang kagome lattice, a proposed setting
for flat-band excitonic
physics~\cite{sethi2021flat,zhou2022excited,sethi2023condensate}.
We classify exciton states by their winding relative to the
conduction--valence Chern-number difference. With concentrated Berry
curvature at weak but finite spin--orbit coupling, the lowest exciton
is a Chern-locked quantum vortex, with winding equal to this difference.
Increasing the coupling magnitude at fixed sign spreads the curvature
and favours a zero-winding state, although the band Chern numbers remain
unchanged.

We explain this reordering through phase matching in the full projected
Coulomb interaction. The gauge-invariant combination of exciton and
Bloch-overlap phases determines the interaction energy of each
momentum pair, which can be lowered by phase matching at fixed weight. We find that a quantum vortex exciton forms in order to compensate the phase stemming from the Bloch-overlap of the electron and hole. Crucially, we show that the phase pattern  can change with spin--orbit coupling to favour the zero-winding exciton.
Calculated circularly polarized absorption directly connects
this energetic selection to optical selection rules, leading to the lowest-energy exciton that can be tuned from dark to bright. The analysis thus
links quantum geometry to exciton topology and binding through the
full Coulomb interaction.

\begin{figure} [!t]
    \centering
    \includegraphics[width=0.70\linewidth]{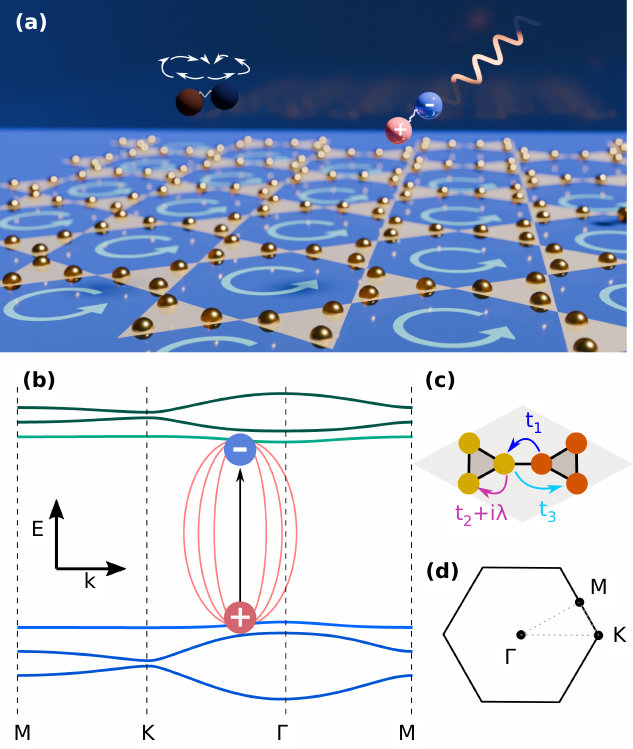} 
    \caption{
{\bf Quantum-vortex excitons in flat Chern bands: a Yin--Yang kagome realization.}
(a) Schematic illustration of bright and dark excitons above the lattice (gold).
The light red and blue pair represents a bright exciton coupled to light (orange wave), and the dark red and blue pair represents a dark exciton.
The arrows illustrate the vortex phase structure.
Optical activity depends on the relative phase structure of the exciton wavefunction and the interband optical matrix element.
(b) Schematic electronic band structure along a high-symmetry path, with the electron and hole in the nearly flat conduction and valence bands.
(c) Tight-binding model and the hopping and spin--orbit couplings considered here. The spin--orbit coupling is denoted by $\lambda_{\mathrm{soc}}$ in the text.
(d) Hexagonal Brillouin zone of the Yin--Yang kagome lattice.
}
    \label{fig:fig1}
\end{figure}

\begin{figure}[!htbp]
\centering
\includegraphics[width=\linewidth]{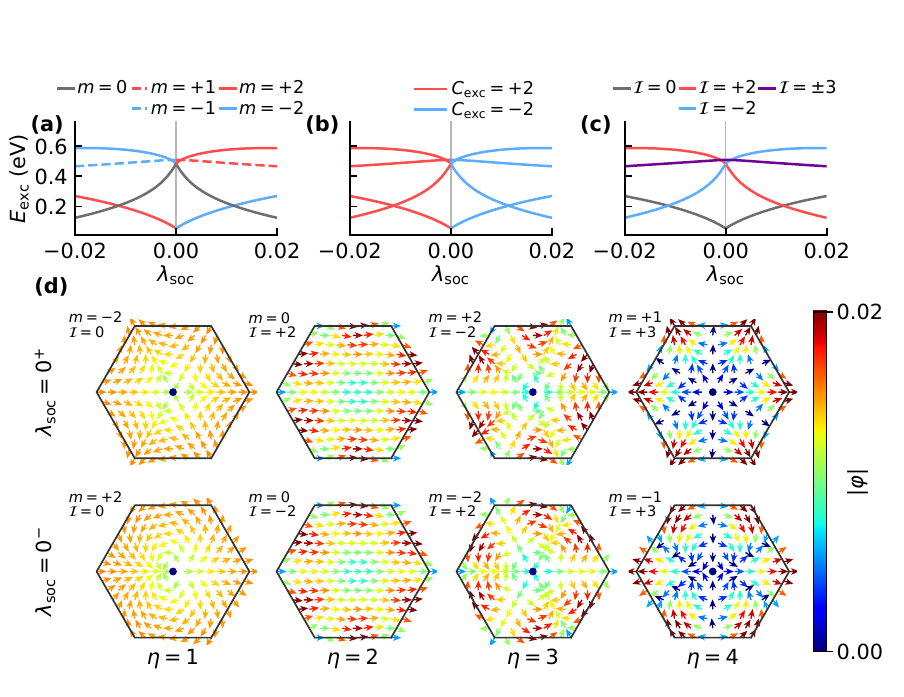}
\caption{
\textbf{Exciton spectrum and relative winding sectors.}
(a)--(c) The same energy spectrum $E^{(\eta)}$ versus spin--orbit
coupling $\lambda_{\mathrm{soc}}$, labelled by (a) coefficient winding
$m$, (b) the common band background $C_{\mathrm{exc}}=C_c-C_v$, and
(c) the relative index $\mathcal I=[m-C_{\mathrm{exc}}]_6$.
Here $C_{\mathrm{exc}}=+2$ at negative SOC and $-2$ at positive SOC.
The SOC axes are in eV. Displayed winding labels are reduced modulo
six, whereas table~\ref{tab:qv_topology} gives the integer windings.
(d) The first four wavefunctions, in increasing energy order, at
$\lambda_{\mathrm{soc}}=+10^{-5}\,\mathrm{eV}$ (top) and
$-10^{-5}\,\mathrm{eV}$ (bottom). Arrow direction and colour encode
$\arg\varphi^{(\eta)}$ and $|\varphi^{(\eta)}|$, respectively.
The exact $\Gamma$ point is absent from the mesh. Core dots indicate
the zero required for a continuous vortex wavefunction in a regular
local Bloch frame, not an additional numerical sample.
}
\label{fig:qv}
\end{figure}

\section{Quantum-vortex excitons in nearly flat Chern bands}
\label{sec:qv}

We study zero-centre-of-mass excitons between the nearly flat valence
and conduction bands of the diatomic Yin--Yang kagome lattice
\cite{sethi2021flat,zhou2022excited}. The six-orbital model has $C_6$
rotation symmetry. Figures~\ref{fig:fig1}(b) and (c) show the selected
band pair and lattice couplings. Spin--orbit coupling (SOC),
$\lambda_{\mathrm{soc}}$, changes the band energies and Bloch states
through the complex hoppings. Varying its magnitude at fixed sign
allows us to follow exciton reordering within a single band Chern
phase. We consider the $t_2=0$ limit and its near-zero numerical
realization (see sections~S1.1 and S1.3 of the Supplementary
Information (SI) for the Hamiltonian and parameters).

Within the projected electron--hole formulation
\cite{albrecht1998abinitio,rohlfing2000electron,onida2002electronic},
we retain one valence band, one conduction band and a statically
screened direct attraction. The momentum-space Wannier equation is
\begin{equation}
\sum_{\mathbf k'\in\mathrm{FBZ}}
W_{\mathbf k\mathbf k'}\varphi^{(\eta)}(\mathbf k')
=E^{(\eta)}\varphi^{(\eta)}(\mathbf k),
\label{eq:wannier}
\end{equation}
where $\mathrm{FBZ}$ denotes the first Brillouin zone,
$W_{\mathbf k\mathbf k'}=\delta_{\mathbf k\mathbf k'}\Delta_{\mathbf k}
+K_{\mathbf k\mathbf k'}$, and
$\Delta_{\mathbf k}=\epsilon_{c\mathbf k}-\epsilon_{v\mathbf k}$.
The band energy and cell-periodic Bloch state are denoted by
$\epsilon_{n\mathbf k}$ and $|u_{n\mathbf k}\rangle$, with $n=c,v$.
The coefficient $\varphi^{(\eta)}(\mathbf k)$ weights the pair
formed by promoting a valence electron to the conduction band at
the same momentum $\mathbf k$. The index $\eta$ ranks exciton
eigenstates by increasing energy at each SOC. Their coefficients obey
$\sum_{\mathbf k\in\mathrm{FBZ}}|\varphi^{(\eta)}(\mathbf k)|^2=1$
on the uniform numerical mesh. The eigenvalue $E^{(\eta)}$ is an
excitation energy. A bound state lies below the continuum threshold
$\min_{\mathbf k}\Delta_{\mathbf k}$.

The projected interaction kernel sums reciprocal images,
\begin{equation}
K_{\mathbf k\mathbf k'}=\sum_{\mathbf G}
V_{\mathbf k,\mathbf k'+\mathbf G},
\label{eq:interaction_kernel}
\end{equation}
with momentum-resolved links
\begin{equation}
V_{\mathbf k\mathbf k'}=
\frac{\Delta A}{(2\pi)^2}U(|\mathbf k-\mathbf k'|)
\Lambda_{\mathbf k,\mathbf k'}.
\label{eq:interaction_link}
\end{equation}
Here $\Delta A$ is the common momentum-space integration area,
repeated in each retained reciprocal cell. The Bloch-overlap form
factor is \cite{srivastava2015geometry,choi2026radius}
\begin{equation}
\Lambda_{\mathbf k,\mathbf k'}=
\langle u_{c\mathbf k}|u_{c\mathbf k'}\rangle
\langle u_{v\mathbf k'}|u_{v\mathbf k}\rangle.
\label{eq:form_factor}
\end{equation}
Finite overlaps retain the Bloch-wavefunction geometry, including
its singular behaviour near band touching
\cite{rhim2019classification,rhim2020distance}.
We use the attractive Rytova--Keldysh potential
$U(q)=-e^2/[2\epsilon_0\epsilon_s q(1+r_0q)]$ for $q>0$
\cite{rytova1967screened,keldysh}, where $\epsilon_s$ is the effective relative permittivity of the
surrounding media and $r_0$ the screening length.
The sum in equation~\eqref{eq:interaction_kernel} retains the central
image and its six nearest neighbours. The scalar coefficients obey
$\varphi(\mathbf k'+\mathbf G)=\varphi(\mathbf k')$, while the Bloch
vectors retain their reciprocal sewing matrices. The screening
parameters, finite-$q$ cutoff and continuation are specified in
sections~S1.2 and S1.3 of the SI.

For any state with the same discrete normalization, the total energy is
\begin{equation}
E[\varphi]=\sum_{\mathbf k\in\mathrm{FBZ}}
\Delta_{\mathbf k}|\varphi(\mathbf k)|^2+K[\varphi],
\label{eq:energy_functional}
\end{equation}
where the interaction expectation is
\begin{equation}
K[\varphi]=\sum_{\mathbf k,\mathbf k'\in\mathrm{FBZ}}\sum_{\mathbf G}
V_{\mathbf k,\mathbf k'+\mathbf G}[\varphi],
\label{eq:interaction_functional}
\end{equation}
and $V_{\mathbf k\mathbf k'}[\varphi]
\equiv\varphi^*(\mathbf k)V_{\mathbf k\mathbf k'}\varphi(\mathbf k')$.
Individual links can be complex, whereas the complete expectations
of the Hermitian operators $W$ and $K$ are real. The sums include
all ordered pairs, and the integration weight is already in $V$.
For an eigenstate, $E[\varphi^{(\eta)}]=E^{(\eta)}$.
Weak dispersion of $\Delta_{\mathbf k}$ allows the exciton to sample
a broad momentum region, making the Bloch-overlap geometry relevant
to binding. Setting $\Lambda=1$ would discard this geometric
contribution to the projected interaction.

Figure~\ref{fig:qv} connects the energy spectrum to the internal
phase textures. At small but finite SOC, the lowest exciton is
separated from the excited manifold (see Figure~\ref{fig:qv}(a)) and
has a phase vortex centred at $\Gamma$ (see Figure~\ref{fig:qv}(d)).
Its circulation reverses with the SOC sign. For a wavefunction
continuous in a regular local Bloch frame, nonzero winding on
arbitrarily small contours around the core requires
$|\varphi^{(1)}(\Gamma)|=0$. We call such a state a quantum-vortex
(QV) exciton. The core marker in Figure~\ref{fig:qv}(d) represents
this continuous-field interpretation.

For a complex function $f$ that is nonzero on a closed,
counterclockwise contour $c$, its integer winding is
\begin{equation}
w_c(f)=\frac{1}{2\pi\mathrm i}\oint_c f^{-1}\dd f
\in\mathbb Z.
\label{eq:winding}
\end{equation}
This integer is unchanged under deformations that cross no zero or
singularity. We use a $C_6$-invariant contour $c_\Gamma$ around
$\Gamma$ and write $m^{(\eta)}=w_{c_\Gamma}(\varphi^{(\eta)})$,
holding the contour and Bloch frame fixed across the compared
states (see section~S2.2 of the SI). On the small positive-SOC side,
the first four states have $m^{(\eta)}=(-2,0,-4,1)$,
with the signs reversed on the negative side. Thus the lowest state is a
vortex, while the next state has zero coefficient winding.
The integers are retained when describing continuous phase textures,
whereas the labels in Figure~\ref{fig:qv} are reduced modulo six.

To compare these textures with their common band background, define
$A_n=\mathrm i\langle u_{n\mathbf k}|\mathrm d u_{n\mathbf k}\rangle$,
$F_n=\mathrm d A_n$ and
$C_n=(2\pi)^{-1}\int_{\mathrm{FBZ}}F_n$ \cite{xiao2010berry}.
The electron--hole connection and curvature are
$A_{\mathrm{exc}}=A_c-A_v$ and $F_{\mathrm{exc}}=F_c-F_v$, giving
$
C_{\mathrm{exc}}=C_c-C_v.
$
This invariant describes the internal electron--hole Bloch frame,
not an exciton centre-of-mass dispersion. A nonzero Chern number
obstructs a globally smooth periodic frame
\cite{brouder2007exponential}. We use the fixed calculation frame
with its reciprocal sewing convention. Here $C_{\mathrm{exc}}=+2$
for negative SOC and $-2$ for positive SOC
(see Figure~\ref{fig:qv}(b)). The lowest state satisfies
$m^{(1)}=C_{\mathrm{exc}}$ at small nonzero SOC, a relation we call
Chern locking (see section~S2.3 of the SI for the corresponding
geometry and wavefunctions).

The band Chern-number difference is common to all excitons of
the selected band pair, whereas their coefficient windings can differ.
To distinguish their internal textures relative to that background,
we define the topological index
\begin{equation}
\mathcal I^{(\eta)}=[m^{(\eta)}-C_{\mathrm{exc}}]_6,
\label{eq:relative_index}
\end{equation}
in the fixed rotation-sewn Bloch frame, where $[j]_6$ denotes
the class of the integer $j$ modulo six.
For a continuously followed exciton state, this index remains
unchanged under smooth deformations that preserve $C_6$ symmetry
and the rotation-sewing convention, provided each constituent
band remains isolated and the coefficient stays continuous and
nonzero on $c_\Gamma$.
Continuous, single-valued Bloch-frame rephasings on $c_\Gamma$
that preserve this convention change $m$ only by multiples of
six and hence leave $\mathcal I$ unchanged
(see sections~S2.2 and S2.4 of the SI).
The labels apply to the nonzero-SOC sides. The isolated-band
Chern numbers and relative index are not assigned at the band
touching at $\lambda_{\mathrm{soc}}=0$.
The physical significance of this classification for the
interaction energy is explained in the next section through
Coulomb phase matching.

\begin{table}[!t]
\caption{
Integer windings $m^{(\eta)}$ and relative topological
classes $\mathcal{I}^{(\eta)}=[m^{(\eta)}-C_{\mathrm{exc}}]_6$
in the calculation frame
for the first ten exciton states on the two small, nonzero-SOC sides.
The labels $+3$ and $-3$ denote the same class.
Here, $C_{\mathrm{exc}}=+2$ for $\lambda_{\mathrm{soc}}\rightarrow0^-$
and $C_{\mathrm{exc}}=-2$ for $\lambda_{\mathrm{soc}}\rightarrow0^+$.
}
\label{tab:qv_topology}
\centering
\setlength{\tabcolsep}{4pt}
\renewcommand{\arraystretch}{1.15}

\begin{tabular}{ccrrrrrrrrrr}
\hline

&
\multirow{2}{*}{$\lambda_{\mathrm{soc}}$}
&
\multicolumn{10}{c}{Exciton state $\eta$}
\\

&
&
$1$ & $2$ & $3$ & $4$ & $5$
& $6$ & $7$ & $8$ & $9$ & $10$
\\
\hline

\multirow{2}{*}{$m^{(\eta)}$}
&
$0^-$
&
$2$ & $0$ & $4$ & $-1$ & $1$
& $3$ & $2$ & $0$ & $4$ & $-1$
\\

&
$0^+$
&
$-2$ & $0$ & $-4$ & $1$ & $-1$
& $-3$ & $-2$ & $0$ & $-4$ & $1$
\\
\hline

\multirow{2}{*}{$\mathcal{I}^{(\eta)}$}
&
$0^-$
&
$0$ & $-2$ & $2$ & $-3$ & $-1$
& $1$ & $0$ & $-2$ & $2$ & $-3$
\\

&
$0^+$
&
$0$ & $2$ & $-2$ & $3$ & $1$
& $-1$ & $0$ & $2$ & $-2$ & $3$\\
\hline
\end{tabular}
\end{table}

Table~\ref{tab:qv_topology} makes the distinction between the two topological indices explicit.
Here, the Chern-locked state has $\mathcal I=0$ on both SOC sides, despite
its opposite circulation, while the zero-winding state has
$\mathcal I=+2$ at positive SOC and $-2$ at negative SOC. Several
eigenstates can share the same topological index. We therefore write
$\varphi_{\mathcal I}$ for the selected eigenstate with a given topological index
and retain $\eta$ when its energy rank is needed. In the present convention, the topological class also fixes the sixfold
rotation character of a state with definite symmetry
(see section~S3.4 of the SI), providing the connection to optical
selection developed in section~\ref{sec:optical}.

As $|\lambda_{\mathrm{soc}}|$ increases at fixed sign, the
zero-winding state crosses below the Chern-locked QV while the
constituent band Chern numbers remain unchanged
(see Figure~\ref{fig:qv}(c)). The topological index thus provides a
consistent way to follow the competing states through a change
in energy order. Section~\ref{sec:phase-matching} examines how
phase matching in the projected Coulomb interaction contributes
to this competition.

\section{Phase matching and the energetic ordering of topological excitons}
\label{sec:phase-matching}
\label{sec:qg_origin}

The energetic competition between excitons with different values of
the topological index introduced in section~\ref{sec:qv} depends on how
their phases align with the underlying Bloch geometry.
In the electron--hole eigenbasis, the diagonal
contribution $\sum_{\mathbf k}\Delta_{\mathbf k}|\varphi(\mathbf k)|^2$
is independent of the relative phases at fixed amplitudes.
The projected Coulomb interaction instead couples different momenta,
making those phases energetically significant via the Bloch form factor. The nature of the exciton is thus specified by the competition between
these form-factors and the electron--hole Bloch frame. Therefore their phase structures must be considered jointly.

For a selected state $\varphi_{\mathcal I}$, we define the relative
coefficient phase as
$\Delta\theta_{\mathcal I}(\mathbf k,\mathbf k')=
\operatorname{Arg}[\varphi_{\mathcal I}^*(\mathbf k)
\varphi_{\mathcal I}(\mathbf k')]$.
The complex contribution to the interaction energy then has the
combined phase
\begin{equation}
\begin{split}
\chi_{\mathcal I}(\mathbf k,\mathbf k')
&\equiv\operatorname{Arg}[-V_{\mathbf k\mathbf k'}[\varphi_{\mathcal I}]]\\
&=\Delta\theta_{\mathcal I}(\mathbf k,\mathbf k')
+\operatorname{Arg}\Lambda_{\mathbf k\mathbf k'}\pmod{2\pi}.
\end{split}
\label{eq:s3-phase-addition}
\end{equation}
Here $\operatorname{Arg}\in[-\pi,\pi)$ is defined for a nonzero
argument, and the minus sign removes the phase of the attractive
interaction $U<0$. Writing
$a_{\mathcal I}(\mathbf k,\mathbf k')
\equiv|V_{\mathbf k\mathbf k'}[\varphi_{\mathcal I}]|$,
we define the real interaction-energy contribution associated
with the ordered momentum pair $(\mathbf k,\mathbf k')$ as
\begin{equation}
\begin{split}
V^{\mathrm{pair}}_{\mathbf k\mathbf k'}[\varphi_{\mathcal I}]
&\equiv
\operatorname{Re}V_{\mathbf k\mathbf k'}[\varphi_{\mathcal I}]\\
&=-a_{\mathcal I}(\mathbf k,\mathbf k')
\cos\chi_{\mathcal I}(\mathbf k,\mathbf k').
\end{split}
\label{eq:s3-phase-loss}
\end{equation}
The nonnegative weight $a_{\mathcal I}$ contains the quadrature
measure, interaction strength, form-factor magnitude and both
coefficient amplitudes. At fixed weight, phase matching at
$\chi_{\mathcal I}=0$ gives the most attractive contribution,
$V^{\mathrm{pair}}=-a_{\mathcal I}$.
We define the increase relative to this aligned value,
$a_{\mathcal I}(1-\cos\chi_{\mathcal I})$, as the
phase-deviation cost, hereafter the deviation cost.
A complex contribution and its conjugate sum to
$2V^{\mathrm{pair}}=-2a_{\mathcal I}\cos\chi_{\mathcal I}$.
The full interaction expectation in
Equation~\eqref{eq:interaction_functional} already includes all
ordered pairs and retained reciprocal images, without an additional
factor of two (see section~S3.1 of the SI).

This interaction phase is independent of the Bloch gauge.
Under $|u_{n\mathbf k}\rangle\mapsto
e^{\mathrm i\beta_n(\mathbf k)}|u_{n\mathbf k}\rangle$, let
$\zeta=\beta_v-\beta_c$. Representing the same physical state
requires $\varphi'(\mathbf k)=e^{\mathrm i\zeta(\mathbf k)}
\varphi(\mathbf k)$, while
$\Lambda'_{\mathbf k\mathbf k'}=
e^{\mathrm i[\zeta(\mathbf k)-\zeta(\mathbf k')]}
\Lambda_{\mathbf k\mathbf k'}$.
The changes in $\Delta\theta_{\mathcal I}$ and
$\operatorname{Arg}\Lambda$ cancel, leaving the entire complex
contribution $V'[\varphi']=V[\varphi]$ unchanged.
The Wannier equation preserves this covariance and determines
eigenstates with a gauge-independent spectrum (see section~S2.1 of the SI).
Changing coefficient phases without changing the Bloch frame, by contrast,
generally changes the physical state and its interaction energy.

The local meaning of phase matching can be understood by comparing
nearby momenta. Consider a smooth local Bloch frame where
$\varphi_{\mathcal I}\ne0$, and let $\theta_{\mathcal I}$ be a smooth
phase of the exciton coefficient. For
$\delta\mathbf k=\mathbf k'-\mathbf k$, the Bloch-overlap phase is
$\operatorname{Arg}\Lambda_{\mathbf k,\mathbf k+\delta\mathbf k}
=-A_{\mathrm{exc}}(\delta\mathbf k)+O(|\delta\mathbf k|^2)$.
Combining this with the change in the exciton phase gives
$\chi_{\mathcal I}\simeq
(\mathrm d\theta_{\mathcal I}-A_{\mathrm{exc}})(\delta\mathbf k)$
to first order. Thus phase matching requires the change in the
exciton phase to compensate the phase contributed by the Bloch overlaps.
The Berry connection specifies this compensation for neighbouring
momenta.

The covariant derivative $D=\mathrm d-\mathrm iA_{\mathrm{exc}}$
accounts for this Bloch-frame contribution when describing changes
in the wavefunction. Writing
$\varphi_{\mathcal I}=r_{\mathcal I}e^{\mathrm i\theta_{\mathcal I}}$
gives
$|D\varphi_{\mathcal I}|^2=|\mathrm d r_{\mathcal I}|^2
+r_{\mathcal I}^2|\mathrm d\theta_{\mathcal I}-A_{\mathrm{exc}}|^2$.
The first term describes amplitude variations, while the second
measures the local phase mismatch after accounting for the Bloch frame.
Nontrivial Berry phases around closed momentum-space loops constrain
whether this mismatch can vanish everywhere
(see section~S2.1 of the SI).
This geometric interpretation connects with earlier studies of
Berry-phase effects on exciton spectra \cite{zhou2015berry}.

When an expansion about neighbouring momenta is justified, a local
energy term proportional to $|D\varphi_{\mathcal I}|^2$ can describe
the leading isotropic gradient contribution, including both amplitude
variations and phase mismatch. The projected Coulomb interaction,
however, also couples momenta separated by finite distances in
reciprocal space. Their contributions depend on the form factor
between those momenta, which is not generally captured by the local
expansion. We therefore evaluate the interaction using the full
form factor and its pair weights
$a_{\mathcal I}(\mathbf k,\mathbf k')$ and combined phases
$\chi_{\mathcal I}(\mathbf k,\mathbf k')$.

\begin{figure}[!htbp]
\centering
\includegraphics[width=0.70\linewidth]{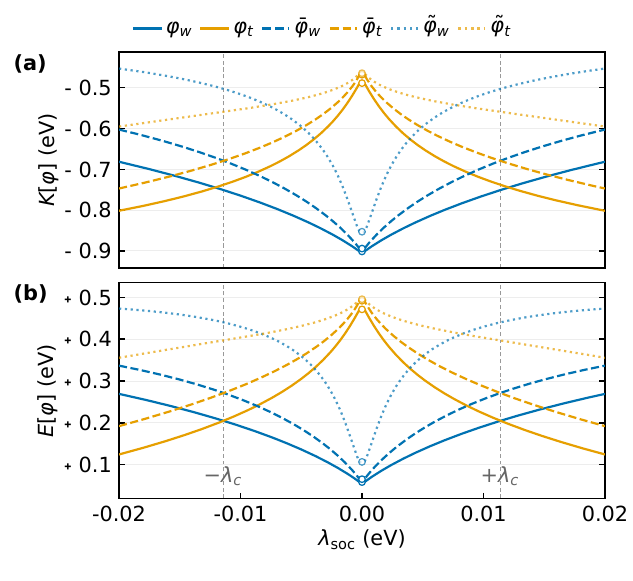}
\caption{
\textbf{Phase-preserving amplitude controls.}
(a) Interaction energy $K[\varphi]$ and (b) total energy
$E[\varphi]$ versus SOC.
Blue ($w$) denotes the Chern-locked sector ($\mathcal I=0$),
whereas orange ($t$) denotes the zero-winding sector
($\mathcal I=+2$ at positive SOC and $-2$ at negative SOC).
Solid curves show native eigenstates $\varphi$.
Dashed ($\bar{\varphi}$) and dotted ($\tilde{\varphi}$) curves
show normalized trials with constant amplitudes and amplitudes
taken from the selected state in the other sector, respectively, while retaining
the native phases.
Trial energies are expectation values, not additional eigenvalues.
Gray vertical lines mark the native crossings at $\pm\lambda_c$.
Open circles at zero SOC average the nearest nonzero-SOC samples.
}
\label{fig:phase-energy-controls}
\end{figure}

To examine how the energy trends differ between topological sectors,
Figure~\ref{fig:phase-energy-controls} compares native eigenstates
with phase-preserving amplitude controls, quantifying the energetic
role of the combined phase.
The interaction and total energies are shown in
Figures~\ref{fig:phase-energy-controls}(a) and (b), respectively.
For normalized states
$\varphi_{\mathcal I}=r_{\mathcal I}e^{\mathrm i\theta_{\mathcal I}}$,
we define
$\bar\varphi_{\mathcal I}=N_k^{-1/2}e^{\mathrm i\theta_{\mathcal I}}$
and
$\tilde\varphi_{\mathcal I}=r_{\mathcal J}e^{\mathrm i\theta_{\mathcal I}}$,
where $N_k$ is the number of sampled FBZ momenta and
$\mathcal J$ denotes the competing sector at the same SOC.
These normalized trials replace the native amplitude by a
constant or by the competing amplitude, respectively, while
preserving the parent phase wherever defined
(see section~S3.2 of the SI).

For the uniform-amplitude trials,
equation~\eqref{eq:s3-phase-loss} gives the common link weight
$\bar a_{\mathbf k,\mathbf k'+\mathbf G}
=|V_{\mathbf k,\mathbf k'+\mathbf G}|/N_k$.
Their diagonal band-energy expectations are also identical,
so at fixed SOC
\begin{equation}
\begin{aligned}
&E[\bar\varphi_{\mathcal I}]-E[\bar\varphi_{\mathcal J}]
=K[\bar\varphi_{\mathcal I}]-K[\bar\varphi_{\mathcal J}]\\
&\qquad=-\sum_{\mathbf k,\mathbf k',\mathbf G}
\bar a_{\mathbf k,\mathbf k'+\mathbf G}
\bigl(\cos\chi_{\mathcal I}-\cos\chi_{\mathcal J}\bigr).
\end{aligned}
\label{eq:uniform-amplitude-phase-comparison}
\end{equation}
Here $\chi_{\mathcal I}$ abbreviates
$\chi_{\mathcal I}(\mathbf k,\mathbf k'+\mathbf G)$,
and the sums follow equation~\eqref{eq:interaction_functional}.
This identity isolates the energetic effect of phase matching
at equal weights. Comparing $\varphi_{\mathcal I}$,
$\bar\varphi_{\mathcal I}$ and $\tilde\varphi_{\mathcal I}$
instead tests amplitude sensitivity at fixed phase.
Comparing $K$ and $E$ identifies the additional contribution
from the diagonal band energy.

The solid, dashed and dotted curves in
Figures~\ref{fig:phase-energy-controls}(a) and (b) retain the
contrasting SOC trends under both amplitude replacements.
Here the Chern-locked exciton is favoured near zero SOC,
whereas the zero-winding exciton reduces in energy (more negative) as
$|\lambda_{\mathrm{soc}}|$ increases.
The uniform-amplitude trials also exhibit an energy-order reversal
(dashed curves in Figure~\ref{fig:phase-energy-controls}(b)),
demonstrating that the competition does not require distinct
native amplitude profiles.
Amplitude redistribution nevertheless shifts the crossing,
and the common interaction weights evolve with SOC
(see sections~S3.2 and S3.3 of the SI).
We next examine the momentum-resolved origin of these trends.

\begin{figure}[!htbp]
\centering
\includegraphics[width=\columnwidth,height=178.5mm,keepaspectratio]{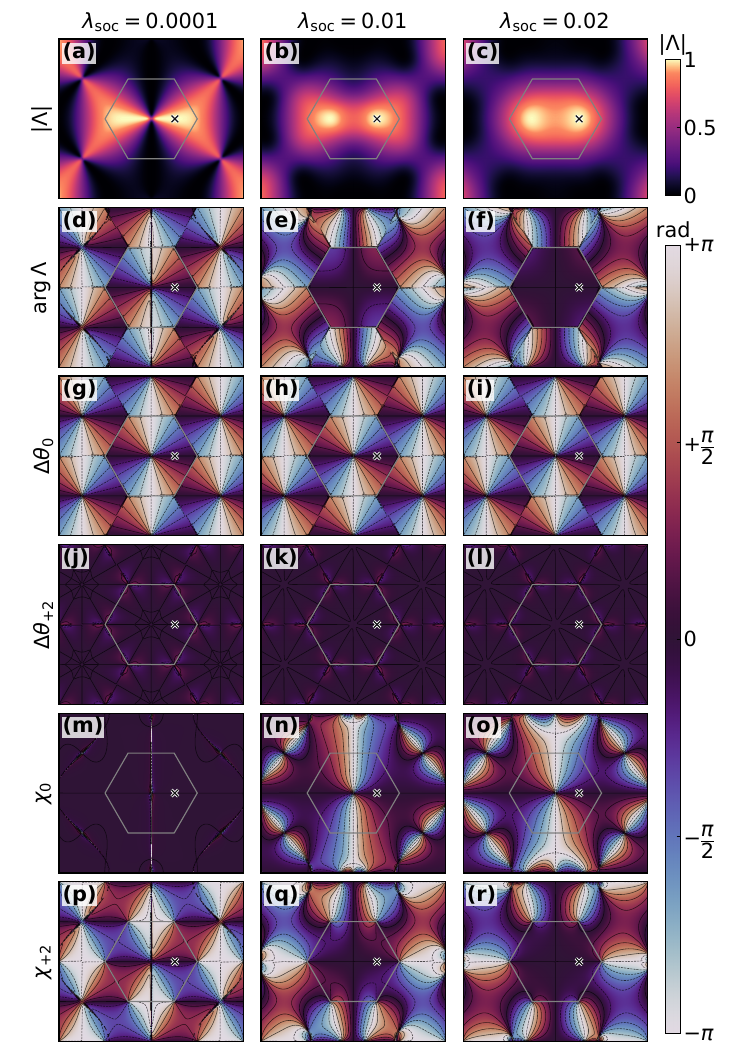}
\caption{
\textbf{Phase matching at fixed band topology.}
Columns show $\lambda_{\mathrm{soc}}=+0.0001$, $+0.01$ and
$+0.02\,\mathrm{eV}$, with $C_{\mathrm{exc}}=-2$ throughout.
The target momentum $\mathbf k'$ varies at fixed source
$a_0\mathbf k=(1.0771,0)$ (cross). Grey hexagons mark the FBZ.
Panels (a)--(c) show the form-factor magnitude $|\Lambda|$.
Panels (d)--(f) show the form-factor phase $\operatorname{Arg}\Lambda$.
Panels (g)--(i) and (j)--(l) show the relative coefficient phases
$\Delta\theta_0$ and $\Delta\theta_{+2}$, respectively.
Panels (m)--(o) and (p)--(r) show the combined interaction phases
$\chi_0$ and $\chi_{+2}$, respectively.
The selected state in the $\mathcal I=0$ sector is Chern locked, whereas
$\mathcal I=+2$ has zero coefficient winding at positive SOC.
Separate phases depend on the Bloch frame, whereas their combination
in equation~\eqref{eq:s3-phase-addition} is gauge invariant
where the interaction contribution is nonzero.
Phases are shown in $[-\pi,\pi)$ with $\pi/6$ contours.
}
\label{fig:exciton-form-factor-phases}
\end{figure}

Figure~\ref{fig:exciton-form-factor-phases} compares the
Chern-locked $\mathcal I=0$ exciton with the zero-winding
$\mathcal I=+2$ exciton at positive SOC.
Each map fixes the source momentum at the cross and varies the
target momentum, with the reciprocal continuation of
equation~\eqref{eq:interaction_kernel} outside the FBZ.
At $\lambda_{\mathrm{soc}}=+0.0001\,\mathrm{eV}$, the form-factor
phase has a pronounced angular texture
(Figure~\ref{fig:exciton-form-factor-phases}(d)).
In the displayed frame, its leading behaviour over much of the FBZ
interior is $\operatorname{Arg}\Lambda\simeq
-C_{\mathrm{exc}}\delta\phi$, where
$\delta\phi=\phi_{\mathbf k'}-\phi_{\mathbf k}$ and
$\phi_{\mathbf k}=\operatorname{Arg}(k_x+\mathrm i k_y)$.
Together with $\Delta\theta_{\mathcal I}\simeq
m_{\mathcal I}\delta\phi$, this gives
$\chi_{\mathcal I}\simeq(m_{\mathcal I}-C_{\mathrm{exc}})\delta\phi$
modulo $2\pi$, using the integer difference without reduction
modulo six (see section~S2.5 of the SI).

In the regime $0<|\lambda_{\mathrm{soc}}|\ll\lambda_c$, the
Chern-locked state $m_0=C_{\mathrm{exc}}$ therefore approaches
the phase-matched minimum of equation~\eqref{eq:s3-phase-loss}
at fixed pair weight.
For neighbouring points on a loop encircling the vortex core,
$\chi_0\simeq0$ gives
$(\mathrm d\theta_0-A_{\mathrm{exc}})_\parallel\simeq0$,
where $\parallel$ denotes the tangential component.
If the amplitude is also approximately constant along the loop,
this becomes $D_\parallel\varphi_0\simeq0$, expressing approximate
parallel transport around the vortex core.
The exact condition $D_\parallel\varphi_0=0$ is recovered in the
ideal angular and constant-amplitude limit.

For $\lambda_{\rm soc}>0$, the winding $m_0=-2=C_{\mathrm{exc}}$
compensates the form-factor phase, leaving $\chi_0$ close to zero
over most of the interior
(Figure~\ref{fig:exciton-form-factor-phases}(m)).
For $\varphi_{+2}$, the nearly constant coefficient phase
(Figure~\ref{fig:exciton-form-factor-phases}(j)) leaves the
residual mismatch $\chi_{+2}\simeq2\delta\phi$ shown in
Figure~\ref{fig:exciton-form-factor-phases}(p).
At equal weight, this mismatch weakens the attraction and
accounts for the phase-matching advantage of the Chern-locked
state at small nonzero SOC.

At $\lambda_{\mathrm{soc}}=+0.01$ and $+0.02\,\mathrm{eV}$,
the form-factor phase is small over a broad interior region
(Figures~\ref{fig:exciton-form-factor-phases}(e) and (f)),
while the coefficient-phase patterns remain nearly unchanged.
The resulting loss of compensation for $\varphi_0$ produces
the mismatch in Figures~\ref{fig:exciton-form-factor-phases}(n)
and (o). The same evolution improves alignment for
$\varphi_{+2}$ in Figures~\ref{fig:exciton-form-factor-phases}(q)
and (r), where the overlap magnitude is substantial
(figures~\ref{fig:exciton-form-factor-phases}(b) and (c)).

\begingroup
\clubpenalty=10000
The redistribution of Berry curvature provides a geometric
explanation of this evolution. At small SOC, the curvature is
concentrated near $\Gamma$. Increasing SOC broadens this
distribution while leaving its full-BZ integral unchanged
(see section~S2.3 of the SI).
On a core-surrounding loop $c_R$ carrying the same coefficient
winding, the normalized covariant phase circulation is
$m_{\mathcal I}-(2\pi)^{-1}\oint_{c_R}A_{\mathrm{exc}}$.
In a frame smooth over the enclosed region, the connection
integral equals the enclosed Berry flux.
As the curvature spreads beyond a fixed interior loop, this
flux becomes less negative at positive SOC.
The circulation consequently moves away from zero for
$m_0=-2$, but toward zero for $m_{+2}=0$:
the geometric phase compensates the vortex winding less
effectively while reducing the mismatch of the zero-winding state.\par
\endgroup

The curvature spreads beyond a smaller loop at smaller SOC.
This provides a geometric interpretation of the earlier departure
from $\operatorname{Arg}\Lambda\simeq-C_{\mathrm{exc}}\delta\phi$
in the near-core phase maps (see sections~S3.4 and S3.5 of the SI).
The loop picture describes the local geometric trend,
while the displayed finite-momentum phases and energy calculations
retain the full form factor.

Through the link energy $-a_{\mathcal I}\cos\chi_{\mathcal I}$,
loss of phase matching raises the real interaction contribution
at fixed weight, whereas improved matching lowers it.
The full energy sums and common-amplitude controls in
Figure~\ref{fig:phase-energy-controls} quantify the resulting
competition and ordering reversal, giving the microscopic basis
for the crossing in Figure~\ref{fig:qv}(c) at unchanged
$C_{\mathrm{exc}}$.
The native crossing also reflects the evolving link weights
and diagonal band-energy contribution.

\begin{figure}[!htbp]
\centering
\includegraphics[width=\columnwidth]{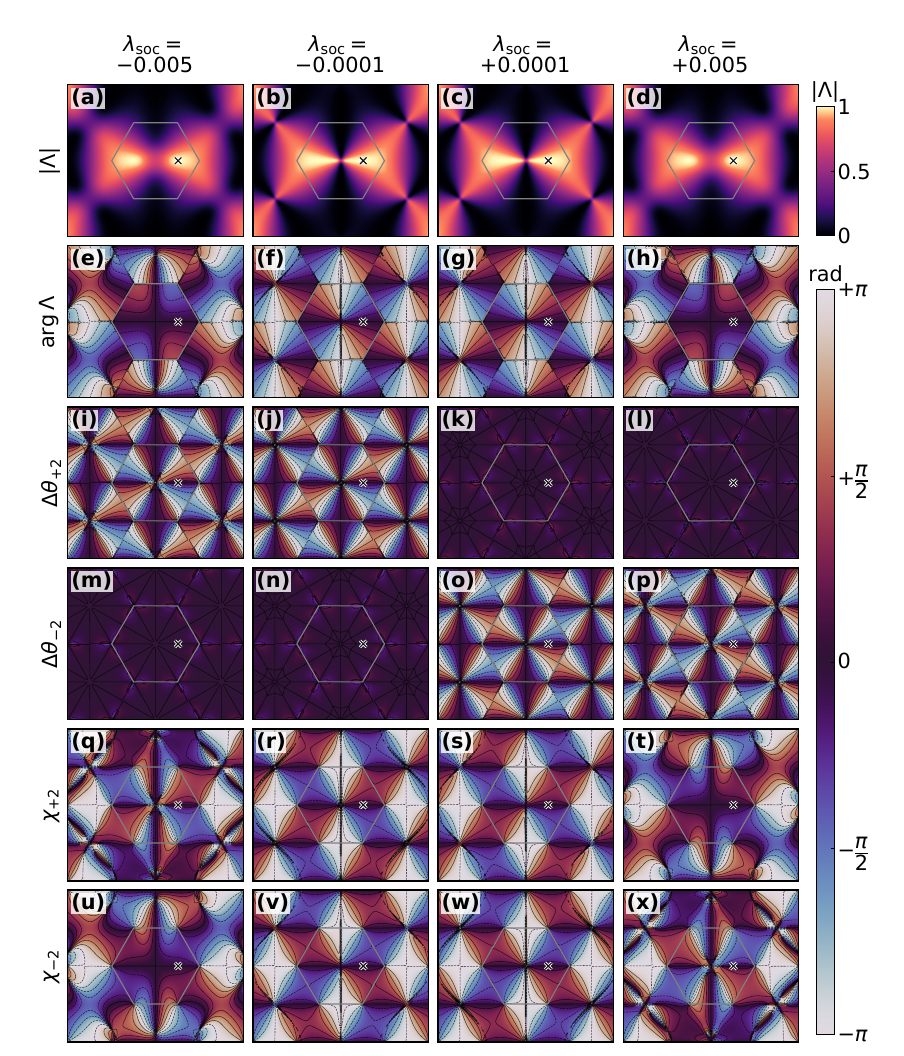}
\caption{
\textbf{Relative interaction phases across SOC reversal.}
Columns show $\lambda_{\mathrm{soc}}=-0.005$, $-0.0001$,
$+0.0001$ and $+0.005\,\mathrm{eV}$.
The fixed source is $a_0\mathbf k=(1.0771,0)$ (cross),
and grey hexagons mark the FBZ.
(a)--(d) show the form-factor magnitude $|\Lambda|$.
(e)--(h) show the form-factor phase $\operatorname{Arg}\Lambda$.
(i)--(l) and (m)--(p) show the relative coefficient phases
$\Delta\theta_{+2}$ and $\Delta\theta_{-2}$, respectively.
(q)--(t) and (u)--(x) show the combined interaction phases
$\chi_{+2}$ and $\chi_{-2}$, respectively.
The latter retain similar patterns within the same relative
sector across small SOC reversal: compare panels (r) and (s) for $+2$,
and (v) and (w) for $-2$.
Subscripts denote the relative classes of section~\ref{sec:qv}
on the nonzero-SOC sides. Phase and target-momentum conventions
are those of Figure~\ref{fig:exciton-form-factor-phases}.
}
\label{fig:relative-sector-phases}
\end{figure}

The preceding phase maps follow the
competition from small positive SOC to beyond the level crossing
within a fixed band topology.
We now turn to the transition at $\lambda_{\mathrm{soc}}=0$,
where the band background changes from $C_{\mathrm{exc}}=+2$
to $-2$. The question is how the relative exciton sectors connect
across this change. The separate form-factor and coefficient
phases can change abruptly in the chosen frames, so their
individual textures do not identify that correspondence.
Figure~\ref{fig:relative-sector-phases} instead compares their
gauge-invariant combination $\chi_{\mathcal I}$.

In the inner columns, the $\mathcal I=+2$ coefficient winding
changes from $m=4$ to $0$, while $\mathcal I=-2$ changes from
$m=0$ to $-4$. Together with the change in $C_{\mathrm{exc}}$,
these shifts preserve the integer differences
$m_{+2}-C_{\mathrm{exc}}=2$ and $m_{-2}-C_{\mathrm{exc}}=-2$.
The leading combined textures consequently remain
$\chi_{\pm2}\simeq\pm2\delta\phi$ modulo $2\pi$.
This same-sector correspondence across SOC reversal is visible in
figures~\ref{fig:relative-sector-phases}(r) and (s) for $\mathcal I=+2$,
and in Figures~\ref{fig:relative-sector-phases}(v) and (w) for $\mathcal I=-2$.
By contrast, the zero-winding states at negative and positive SOC
have opposite relative indices and residual phase chiralities,
as seen by comparing Figures~\ref{fig:relative-sector-phases}(v) and (s).
The relative index therefore organizes the
interaction-phase correspondence across the two band backgrounds
more directly than the coefficient winding alone.
The similar patterns support a common limiting texture where
the interaction contribution is nonzero. Gauge invariance
itself does not establish SOC continuity. The class convention
remains that of section~\ref{sec:qv}, and the isolated-band
indices are not assigned at the band touching itself
(see sections~S2.4 and S3.4 of the SI).

At fixed small SOC, the two relative sectors display approximately
opposite residual phases of similar magnitude. This is apparent when comparing
Figures~\ref{fig:relative-sector-phases}(r) and (v) at negative SOC,
and Figures~\ref{fig:relative-sector-phases}(s) and (w) at positive SOC. Since
$\cos\chi=\cos(-\chi)$, exactly opposite phases give equal real interaction contributions at equal weight. The approximate opposition displayed here explains how the two opposite index excitons can approach the same energy near zero SOC. Degeneracy requires the corresponding full-energy relation and follows when the selected states share a symmetry-preserving zero-SOC limit (see section~S3.4 of the SI).
The displayed finite-SOC states retain a splitting.

At $|\lambda_{\mathrm{soc}}|=0.005\,\mathrm{eV}$,
the two sectors differ in phase magnitude and angular profile
as well as sign, as shown by Figures~\ref{fig:relative-sector-phases}(q)
and (u) at negative SOC and Figures~\ref{fig:relative-sector-phases}(t)
and (x) at positive SOC. The zero-winding member,
$\mathcal I=-2$ at negative SOC and $\mathcal I=+2$ at positive
SOC, becomes more closely phase matched in the central region,
whereas the opposite sector retains a structured mismatch.
The resulting difference in their cosine factors contributes
to the growing energy separation, together with the evolving
interaction weights and diagonal term.
Thus a relative sector can retain its identity while its
interaction-phase texture and energy rank change.

Equal energies nevertheless do not make the opposite relative
sectors physically equivalent: their coherent optical transition
amplitudes distinguish circular-polarization channels, as shown
in section~\ref{sec:optical}.

\begin{figure}[!htbp]
\centering
\centering
    \includegraphics[width=0.70\linewidth]{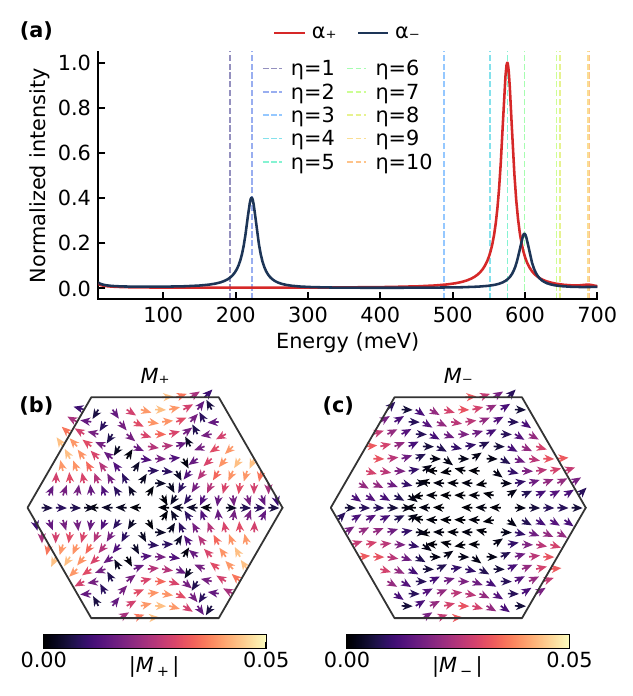} 
    \caption{
    {\bf Polarization-resolved absorption and optical matrix-element textures.}
    (a) Normalized excitonic absorption spectra $\alpha_+$ and $\alpha_-$ for right- and left-circularly polarized light, respectively, at $\lambda_{\mathrm{soc}}=+0.010\,\mathrm{eV}$, normalized by a common maximum over both polarizations. Vertical dashed lines mark the energies $E^{(\eta)}$ of the first ten exciton states, $\eta=1,\ldots,10$. Only a subset of these levels carries appreciable oscillator strength for each circular polarization.
    (b) and (c) Complex-arrow plots of the circular components of the interband optical matrix element, $M_+$ and $M_-$, respectively. The arrow orientation represents the phase $\arg M_\pm(\mathbf{k})$, while the colour represents the magnitude $|M_\pm(\mathbf{k})|$. Their distinct momentum-space phase textures enter the polarization-dependent optical selection rule together with the exciton-wavefunction winding.
    }
    \label{fig:aM}
\end{figure}

\begin{figure}[!htbp]
\centering
    \includegraphics[width=\linewidth,height=178.5mm,keepaspectratio]{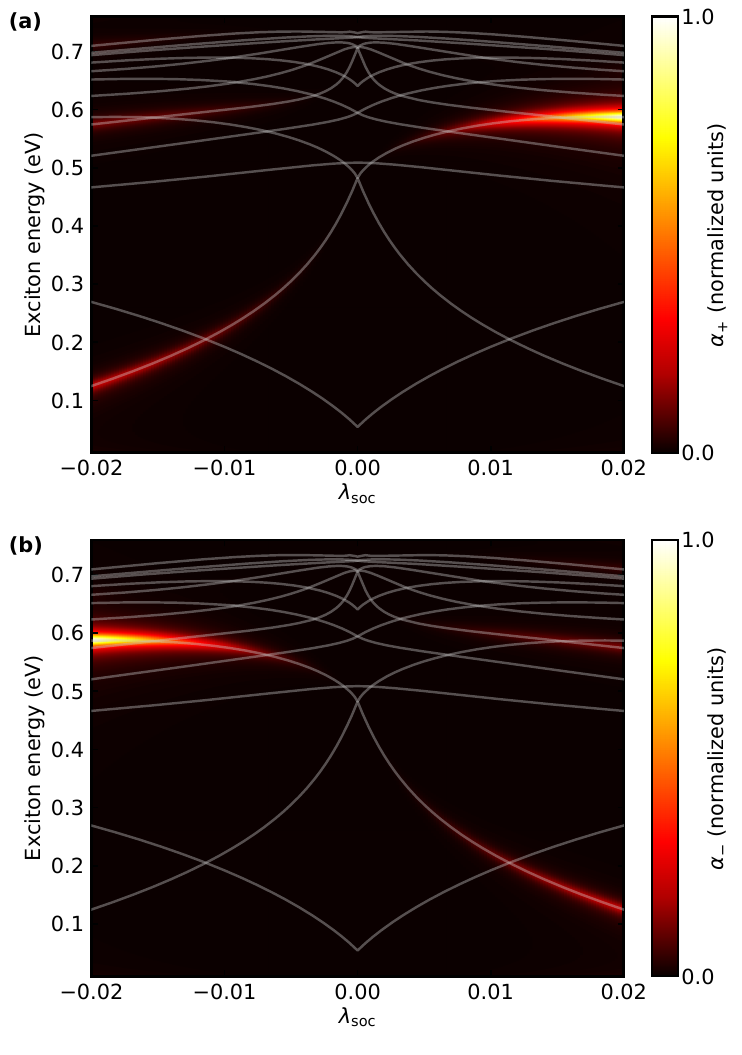}
    \caption{
    {\bf Polarization-resolved optical signatures of quantum-vortex excitons.}
    Exciton absorption spectra as a function of the spin--orbit coupling $\lambda_{\mathrm{soc}}$ (in eV) for (a) $\sigma_{+}$- and (b) $\sigma_{-}$-polarized light. The colour scale represents $\alpha_{\pm}$ normalized separately by the maximum over each complete energy--SOC panel, while the grey curves indicate the exciton energies $E^{(\eta)}$ obtained from the Wannier equation. Bright spectral weight appears only for exciton states in sectors allowed by the polarization-dependent optical selection rule. Reversing the sign of $\lambda_{\mathrm{soc}}$ reverses the chirality of the quantum-vortex exciton and correspondingly interchanges the dominant
    $\sigma_{+}$ and $\sigma_{-}$ absorption channels.
    }
    \label{fig:absorption_map}
\end{figure}

\section{Optical signatures of quantum-vortex excitons}
\label{sec:optical}

Polarization-resolved absorption distinguishes the relative sectors
discussed in section~\ref{sec:phase-matching} through their transition
amplitudes, providing information beyond their energies.
Optical access to states that are dark in one channel is well
established. For example, two-photon spectroscopy has resolved dark
excitonic states in monolayer WS$_2$ \cite{ye2014dark}.
For the competing relative sectors, we calculate the
Elliott-type optical response \cite{elliott1957intensity,wu2017topological},
\begin{equation}
\alpha_{\sigma}(\omega)
=
\frac{\mathcal N}{\hbar\omega}
\operatorname{Im}
\left[
\sum_{\eta}
\frac{\Theta_{\sigma}^{(\eta)}}
{E^{(\eta)}-\hbar\omega-\mathrm{i}\gamma}
\right],
\label{eq:absorption}
\end{equation}
where $\sigma=\pm$ labels the circular polarizations,
$\hbar\omega$ is the photon energy, $\mathcal N$ is an overall
frequency-independent scale, and $\gamma=10\,\mathrm{meV}$ is the
half width of the Lorentzian denominator. The oscillator weight is
$\Theta_{\sigma}^{(\eta)}=|\mathcal T_{\sigma,\eta}|^2$, with
$\mathcal T_{\sigma,\eta}=\sum_{\mathbf k}
M_\sigma(\mathbf k)\varphi^{(\eta)*}(\mathbf k)$.
Here $M_\mu(\mathbf k)=\langle u_{c\mathbf k}|\hat O_\mu(\mathbf k)
|u_{v\mathbf k}\rangle$ for $\mu=x,y$, where $\hat O_\mu$ is the
bond-based optical operator,
and $M_\pm=M_x\pm\mathrm{i}M_y$.
The absorption spectra in Figure~\ref{fig:aM}(a) share a common
normalization maximum for both polarizations, whereas
figures~\ref{fig:absorption_map}(a) and (b) are normalized separately
over each complete energy--SOC panel (see sections~S5.1 and S5.2
of the SI for the optical operator, line shape and normalization).

The transition function is
$T_{\sigma,\eta}(\mathbf k)=M_\sigma(\mathbf k)
\varphi^{(\eta)*}(\mathbf k)$.
For the isolated conduction--valence pair at $t_2=0$, the projected
optical matrix elements have definite sixfold rotation characters
(see section~S5.3 of the SI for the derivation).
Combining the phase of the interband optical matrix element with that
of the envelope gives the winding-based optical selection rule
\cite{cao2018unifying,zhang2018optical}.
On a common contour $c$ where both factors are nonzero,
$w_c(T_{\sigma,\eta})=w_c(M_\sigma)-w_c(\varphi^{(\eta)})$.
For a transition function with a definite $C_6$ rotation character,
contributions from six symmetry-related momentum regions cancel unless
that character is trivial. On a $C_6$-invariant contour, its winding identifies the character
modulo six, so the necessary condition is
$w_c(T_{\sigma,\eta})\equiv0\pmod6$
(see section~S5.4 of the SI for the symmetry derivation).
Using the relative index defined in section~\ref{sec:qv}, this becomes
\begin{equation}
\mathcal I^{(\eta)}
\equiv w_c(M_\sigma)-C_{\mathrm{exc}}\pmod6.
\label{eq:optical_relative_index}
\end{equation}
Here the common contour carries the same exciton winding class as
$c_\Gamma$. This form connects the sectors used to describe the energy
ordering to the optical channels that can access them. A permitted
class can still have a small oscillator weight because the full
coherent sum also depends on amplitudes and further phase cancellation.

For small negative SOC, the optical winding classes are
$w_c(M_+)\equiv0$ and $w_c(M_-)\equiv4$ modulo six, while for small
positive SOC they are $w_c(M_+)\equiv-4$ and $w_c(M_-)\equiv0$.
Together with the corresponding $C_{\mathrm{exc}}=+2$ and $-2$,
equation~\eqref{eq:optical_relative_index} selects
$\mathcal I=-2$ for $\sigma_+$ and $\mathcal I=+2$ for $\sigma_-$
on both sides of zero SOC. The Chern-locked $\mathcal I=0$ state is
therefore dark in either circular polarization in this regime.
For example, table~\ref{tab:qv_topology} shows that near
$\lambda_{\mathrm{soc}}=0^+$ the $\eta=2$ state belongs to the
$\sigma_-$-allowed class, whereas $\eta=3$ belongs to the
$\sigma_+$-allowed class.

The $\mathcal I=+2$ states whose combined phases appear in
figures~\ref{fig:relative-sector-phases}(r) and (s) belong to the
$\sigma_-$-allowed sector, whereas the $\mathcal I=-2$ states represented
in Figures~\ref{fig:relative-sector-phases}(v) and (w) belong to the
$\sigma_+$-allowed sector.
The sectors remain optically distinct even when their energies
approach one another and their opposite interaction phases give
similar cosine factors. Absorption probes the coherent function
$M_\sigma\varphi_{\mathcal I}^*$ rather than the interaction phase
$\chi_{\mathcal I}$ itself. The persistence of the allowed
helicity within each sector across SOC reversal therefore provides
an observable consequence of the relative classification.

Figure~\ref{fig:aM}(a) shows the finite-SOC absorption response at
$\lambda_{\mathrm{soc}}=+0.010\,\mathrm{eV}$.
The low-energy $\sigma_-$ resonance is associated with $\eta=2$,
the dominant $\sigma_+$ resonance with $\eta=5$, and a higher
$\sigma_-$ resonance with $\eta=6$. Here $\eta$ denotes the energy rank
at the displayed coupling and changes when states reorder.
The distinct phase textures of $M_+$ and $M_-$ are displayed in
figures~\ref{fig:aM}(b) and (c), respectively.
The many levels without appreciable resonances in Figure~\ref{fig:aM}(a)
illustrate the restrictions imposed by
equation~\eqref{eq:optical_relative_index}.

Figure~\ref{fig:absorption_map} places these resonances within the
full SOC-dependent spectrum. At fixed positive SOC sign, the bright
zero-winding state crosses the dark Chern-locked state as
$|\lambda_{\mathrm{soc}}|$ increases (see Figure~\ref{fig:absorption_map}(b)), becoming
the lowest exciton beyond the crossing discussed in
section~\ref{sec:phase-matching}. The tracked states retain their
sector identities while their energetic order changes. The lowest
exciton consequently becomes accessible in the $\sigma_-$ channel.
The corresponding crossing at negative SOC is visible in the
$\sigma_+$ channel in Figure~\ref{fig:absorption_map}(a).
Reversing the SOC sign also reverses
$C_{\mathrm{exc}}$ and the relative index of the zero-winding state,
interchanging its allowed circular polarization. Thus the optical
spectra connect the interaction-driven reordering of the relative
sectors to a change in the optical accessibility of the lowest state.

\section{Discussion}
\label{sec:discussion}

Our analysis explains why the Coulomb attraction can favour a vortex exciton
over a state without winding. The vortex phase can compensate the
Bloch-overlap phase, lowering the interaction energy through phase matching
between different momentum components. In the Yin--Yang kagome (YYK) model, this compensation
favours a Chern-locked vortex at weak finite spin--orbit coupling. Increasing
the SOC magnitude at fixed sign changes the Bloch geometry and makes the
zero-winding state energetically preferred without changing the band Chern
numbers. The reversal survives when the competing phase patterns are
assigned identical normalized amplitudes, providing a quantitative test of
the phase mechanism. Calculated circularly polarized absorption connects
this reordering to the distinct optical accessibility of the relative
winding sectors.

Chemically accessible framework structures offer routes towards a YYK
realization. In nearly flat bands, the small dispersion allows binding to
sample Bloch-state overlaps across a broad momentum region.
Nickel bis(dithiolene) nanosheets have been synthesized~\cite{kambe2013pi},
and YYK bands are predicted in a Li-intercalated
bilayer~\cite{zhou2020giant}. Anilato-based aluminium frameworks offer
another proposed route to $\pi$-orbital YYK bands~\cite{ni2020pi}, while
calculations for a covalent organic framework suggest that iodine
incorporation can open SOC-induced topological gaps~\cite{gao2024enantiomorphic}.
These candidates motivate calculations using material-specific Bloch states
and screened interactions, followed by spectroscopic tests of the predicted
exciton energies and polarization dependence.

Previous studies connect exciton topology to Wannier localization and
transport~\cite{davenport2026berryology,jankowski2025excitonic,thompson2025transport}.
Explaining which relative-winding sector is energetically favoured is not trivial:
neither a topological diagnosis nor a computed spectrum alone identifies
the interaction contributions responsible for the ordering.
Local geometric expansions explain spectral shifts and
splittings~\cite{zhou2015berry,srivastava2015geometry}.
Such energy descriptions, formulated through Berry connections, covariant
derivatives or Berry curvature, approximate the interaction using geometry
between neighbouring momenta. Finite-momentum channel analyses provide
further energetic mechanisms in continuum models or idealized curvature
settings~\cite{garate2011excitons,allocca2018fingerprints,panigrahi2026valley}.
Flat-Chern-band theories also yield limiting Coulomb vortex states and
contact-interaction binding energies, including ordering
changes~\cite{xie2024theory,lozano2025optical}.
For YYK, the tested local-gradient reconstruction has a
cutoff-dependent stiffness. A prescribed rescaling retains a
qualitative crossing but does not reproduce the small-SOC structure
of the full spectrum. A separate fixed-state energy decomposition
shows that omitting finite-transfer contributions can reverse the
ordering even when the diagonal energy and exact interaction
baseline are retained (see sections~S4.2, S4.3 and S4.5 of the SI).
These comparisons establish the need
to analyse the screened Coulomb interaction at finite momentum separation,
with both the combined phases and their weights retained.

The same energetic picture also helps explain the shape of the exciton
wavefunction. Normalized variations of uniform-amplitude trials favour
depletion near $\Gamma$ for the Chern-locked phase pattern and enhancement
for the zero-winding pattern at larger positive SOC
(see section~S3.5 of the SI). Geometry therefore enters both the preferred
winding and its momentum-space amplitude profile through the interaction.
Together with the optical selection rules, this gives the relative
classification a physical meaning beyond identifying distinct textures.
More broadly, these results motivate extending the finite-momentum phase
analysis to other composite excitations whose binding samples a broad
region of momentum space. In YYK, this analysis connects exciton topology
to binding and optical response beyond what the constituent band Chern
numbers alone specify.

\section*{Acknowledgments}
The authors thank Wojciech Jankowski for useful discussions. This work was supported by the Nano \& Material Technology Development Program through the National Research Foundation of Korea (NRF), funded by Ministry of Science and ICT (No. RS-2026-25593133), and was also supported by Korea Institute for Advancement of Technology (KIAT) grant funded by the Korea Government (MOTIE) (RS-2024-00410066, HRD Program for Industrial Innovation). J.J.P.T. acknowledges support from an EPSRC Programme Grant EP/W017091/1.  J.W.R. was supported by the National Research Foundation of Korea
(NRF) Grant funded by the Korean government (MSIT) (grant
no. RS-2023-NR068116) and the Ministry of Education (grant
no. RS-2023-00285390). J.W.R. and J.H.C. were supported by the National Research Foundation of Korea
(NRF) Grant funded by the Korean government (MSIT) (grant
no. RS-2026-25597121).
We acknowledge the assistance of ChatGPT in editing the English and improving the clarity of this manuscript. The authors reviewed and revised the AI-assisted text and take full responsibility for the manuscript.

\section*{Author contributions}
All authors discussed the results and contributed to the writing of the manuscript. The final form of the manuscript, including the Supplementary Information, benefitted from input from all authors.

\section*{Competing interests}
The authors declare no competing interests.

\makeatletter
\immediate\write\@auxout{\string\citation{apsrevControl}}
\makeatother
\bibliographystyle{apsrev4-2}
\bibliography{bib}
\end{document}

\maketitle

\noindent This Supplementary Information specifies the projected model,
numerical conventions, relative-winding classification, phase-matching
analysis and optical response used in the main text. All excitons
considered here have zero centre-of-mass crystal momentum.
A general envelope is written $\varphi$; a superscript
$(\eta)$ selects an energy-ranked eigenstate and a subscript
$\mathcal I$ selects a branch of the stated relative class.
\tableofcontents
\clearpage

\section{Microscopic model and projected exciton equation}
\label{sec:si-model}

\subsection{Six-orbital Yin--Yang kagome Hamiltonian}

We use the six-orbital diatomic Yin--Yang kagome tight-binding model
\cite{sethi2021flat,zhou2022excited}, with the orbital convention and
parameters of the main-text calculations specified below.
To distinguish the numerical momentum coordinates from
physical wavevectors, let $\widetilde{\mathbf k}=a_0\mathbf k$, where
$a_0$ is the length scale specified below. The dimensionless vectors
$\mathbf a_1=(\sqrt{3}/2,-1/2)$,
$\mathbf a_2=(\sqrt{3}/2,1/2)$, and $\mathbf a_3=(0,1)$ define
$\kappa_j=\widetilde{\mathbf k}\cdot\mathbf a_j$.
In the orbital convention of the calculation, the six embedding
vectors in units of $a_0$ are
$(\mathbf 0,-\mathbf a_1,-\mathbf a_2,
\mathbf 0,\mathbf a_1,\mathbf a_2)$.
Writing $\mathbf 1_d$ for the identity in $d$ dimensions,
the Bloch Hamiltonian for this embedding in the ordered six-orbital basis is
\begin{equation}
H(\widetilde{\mathbf k})
=\epsilon_{\rm orb}\mathbf 1_6+
\begin{pmatrix}
D_+(\widetilde{\mathbf k})&B(\widetilde{\mathbf k})\\
B(\widetilde{\mathbf k})&D_-(\widetilde{\mathbf k})
\end{pmatrix}.
\label{eq:si-bloch-hamiltonian}
\end{equation}
The real symmetric interblock matrix is
\begin{equation}
B(\widetilde{\mathbf k})=
\begin{pmatrix}
t_1&t_3\cos\kappa_1&t_3\cos\kappa_2\\
t_3\cos\kappa_1&t_1&t_3\cos\kappa_3\\
t_3\cos\kappa_2&t_3\cos\kappa_3&t_1
\end{pmatrix}.
\label{eq:si-interblock}
\end{equation}
Both $D_s$, with $s=\pm1$, have zero diagonal. Their independent
off-diagonal entries are
\begin{equation}
(D_s)_{12}=t_2\cos\kappa_1+\mathrm{i}\lambda_{\mathrm{soc}}
\mathrm e^{\mathrm{i}s\kappa_1},
\label{eq:si-d12}
\end{equation}
\begin{equation}
(D_s)_{13}=t_2\cos\kappa_2-\mathrm{i}\lambda_{\mathrm{soc}}
\mathrm e^{\mathrm{i}s\kappa_2},
\label{eq:si-d13}
\end{equation}
\begin{equation}
(D_s)_{23}=t_2\cos\kappa_3+\mathrm{i}\lambda_{\mathrm{soc}}
\mathrm e^{\mathrm{i}s\kappa_3}.
\label{eq:si-d23}
\end{equation}
Hermiticity fixes $(D_s)_{ji}=(D_s)_{ij}^*$.
The labels $+$ and $-$ distinguish the two orbital blocks in this
representation. They are not labels of the optical polarizations.
The common on-site shift $\epsilon_{\rm orb}$ cancels from the
conduction--valence energy difference.

Diagonalizing $H$ gives the band energies $\epsilon_{n\mathbf k}$
and normalized cell-periodic Bloch vectors $|u_{n\mathbf k}\rangle$.
The projected exciton calculation retains the third and fourth bands
in ascending energy order as the valence band $v$ and conduction band
$c$, respectively. The remaining filled bands enter the reference
state but do not provide additional electron--hole channels in this
two-band projection.

\subsection{Zero-centre-of-mass pair and screened direct interaction}

We use the projected equation and energy functionals of main-text
section II; the conventions needed to reproduce the calculation are
specified here. Let $\hat a_{n\mathbf k}^{\dagger}$ create an electron in a Bloch
state and let $|\Omega\rangle$ be the filled-valence-band reference
state. We write
$\hat e_{\mathbf k}^{\dagger}=\hat a_{c\mathbf k}^{\dagger}$ and
$\hat h_{\mathbf k}^{\dagger}=\hat a_{v\mathbf k}$.
The hole index denotes the momentum of the removed valence electron;
the crystal momentum of a pair is therefore
$\mathbf Q=\mathbf k_e-\mathbf k_h$.
The optically relevant sector studied here has exactly $\mathbf Q=0$,
and its exciton state is
\begin{equation}
|\Phi^{(\eta)}\rangle
=\sum_{\mathbf k\in\mathrm{FBZ}}
\varphi^{(\eta)}(\mathbf k)
\hat a_{c\mathbf k}^{\dagger}\hat a_{v\mathbf k}|\Omega\rangle.
\label{eq:si-pair-state}
\end{equation}
The discrete coefficients obey
\begin{equation}
\sum_{\mathbf k\in\mathrm{FBZ}}|\varphi^{(\eta)}(\mathbf k)|^2=1.
\label{eq:si-pair-norm}
\end{equation}
No small-relative-momentum approximation is imposed on this state.

Within the projected electron--hole eigenvalue formulation
\cite{rohlfing2000electron,onida2002electronic}, the model retains the
statically screened direct attraction. Projection of the density interaction onto the selected
bands gives the form factor
\begin{equation}
\Lambda_{\mathbf k,\mathbf p}
=\langle u_{c\mathbf k}|u_{c\mathbf p}\rangle
\langle u_{v\mathbf p}|u_{v\mathbf k}\rangle.
\label{eq:si-overlap}
\end{equation}
The second factor has reversed bra and ket order because it acts on
the missing valence electron. For a target momentum
$\mathbf p=\mathbf k'+\mathbf G$, the physical momentum transfer
is $q=|\mathbf k-\mathbf k'-\mathbf G|$.
We use the attractive Rytova--Keldysh interaction \cite{rytova1967screened,keldysh}
\begin{equation}
U(q)=-\frac{e^2}{2\epsilon_0\epsilon_s q(1+r_0q)}.
\label{eq:si-rk}
\end{equation}
Here $\epsilon_s=(\epsilon_1+\epsilon_2)/2$ is the average relative
permittivity of the surrounding media and
$r_0=\epsilon_wL_w/(2\epsilon_s)$, with layer relative permittivity
$\epsilon_w$ and effective thickness $L_w$.
The function $U$ contains no momentum integration weight.

On the uniform production mesh, the weighted momentum-resolved link is
\begin{equation}
V_{\mathbf k,\mathbf k'+\mathbf G}
=\frac{\Delta A_{\mathbf k'}}{(2\pi)^2}
U(|\mathbf k-\mathbf k'-\mathbf G|)
\Lambda_{\mathbf k,\mathbf k'+\mathbf G}.
\label{eq:si-weighted-link}
\end{equation}
The integration area $\Delta A_{\mathbf k'}$ is expressed in physical
momentum units. For a trial coefficient $\varphi$, its complex link contribution is
$V_{\mathbf k,\mathbf p}[\varphi]\equiv
\varphi^*(\mathbf k)V_{\mathbf k,\mathbf p}\varphi(\mathbf p)$.
Summing over the retained reciprocal images gives
\begin{equation}
K_{\mathbf k\mathbf k'}
=\sum_{\mathbf G}V_{\mathbf k,\mathbf k'+\mathbf G}.
\label{eq:si-kernel}
\end{equation}
Thus the projected Wannier equation reads
\begin{equation}
\sum_{\mathbf k'}
\left[\Delta_{\mathbf k}\delta_{\mathbf k\mathbf k'}
+K_{\mathbf k\mathbf k'}\right]
\varphi^{(\eta)}(\mathbf k')
=E^{(\eta)}\varphi^{(\eta)}(\mathbf k),
\label{eq:si-wannier}
\end{equation}
where $\Delta_{\mathbf k}=\epsilon_{c\mathbf k}-\epsilon_{v\mathbf k}$.
An eigenstate below the continuum has binding energy
$E_b^{(\eta)}=\min_{\mathbf k}\Delta_{\mathbf k}-E^{(\eta)}$.

For any normalized trial state, the interaction expectation is
\begin{equation}
K[\varphi]=\sum_{\mathbf k,\mathbf k'\in\mathrm{FBZ}}
\varphi^*(\mathbf k)K_{\mathbf k\mathbf k'}\varphi(\mathbf k'),
\label{eq:si-k-functional}
\end{equation}
and the total energy is
\begin{equation}
E[\varphi]=\sum_{\mathbf k}\Delta_{\mathbf k}|\varphi(\mathbf k)|^2
+K[\varphi].
\label{eq:si-energy-functional}
\end{equation}
These sums contain all ordered momentum pairs. There is no additional
factor of two, factor of one half, or mesh-area factor multiplying the
expectation value. Individual links
$\varphi^*(\mathbf k)V_{\mathbf k,\mathbf k'+\mathbf G}
\varphi(\mathbf k')$ can be complex, whereas the expectation of the
Hermitian kernel is real.

\subsection{Reciprocal continuation and numerical parameters}

The numerical momentum coordinates are dimensionless. A transfer
$\widetilde q=|\widetilde{\mathbf k}-\widetilde{\mathbf k}'
-\widetilde{\mathbf G}|$ corresponds to the physical transfer
$q=\widetilde q/a_0$ used in Eq.~\eqref{eq:si-rk}.
The hexagonal FBZ has circumradius $2\pi/3$ in these coordinates.
Sixfold rotation of a triangular mesh with division parameter $N=35$
gives $N_k=3672$ sampled momenta and excludes the exact $\Gamma$ point.
These are the native-mesh settings; the additional refined-grid
calculation in section~\ref{sec:si-near-core} uses the weighted
normalization and mesh specified there.
The common integration area is
\begin{equation}
\Delta A=\sin(\pi/3)
\left(\frac{2\pi}{3Na_0}\right)^2.
\label{eq:si-mesh-area}
\end{equation}
This area enters Eq.~\eqref{eq:si-weighted-link} once, while the
wavefunction uses the unweighted discrete normalization in
Eq.~\eqref{eq:si-pair-norm}.

The reciprocal-image sum includes $\widetilde{\mathbf G}=0$ and
the six shortest reciprocal vectors
\begin{equation}
\widetilde{\mathbf G}_j
=\frac{2\pi}{\sqrt{3}}
\bigl(\cos(j\pi/3),\sin(j\pi/3)\bigr),
\qquad j\in\{0,\ldots,5\}.
\label{eq:si-reciprocal-vectors}
\end{equation}
The Rytova--Keldysh crossover momentum $q_*=1/r_0$ is approximately
one quarter of the FBZ width between opposite edges,
$2\pi/(\sqrt{3}a_0)$.
Motivated by the $|U(q)|\propto q^{-2}$ decay for $q\gg q_*$,
we retain the central reciprocal image and its six nearest neighbours.
The cell-periodic Bloch states are continued using
$|u_{n,\mathbf k'+\mathbf G}\rangle=T_{\mathbf G}
|u_{n\mathbf k'}\rangle$.
For the orbital embedding of Eq.~\eqref{eq:si-bloch-hamiltonian},
$T_0=\mathbf 1_6$. The diagonal entries of $T_{\mathbf G_j}$ are
$(1,-1,-1,1,-1,-1)$ for $j=0,3$,
$(1,1,-1,1,1,-1)$ for $j=1,4$, and
$(1,-1,1,1,-1,1)$ for $j=2,5$.
Consequently,
\begin{equation}
\Lambda_{\mathbf k,\mathbf k'+\mathbf G}
=\langle u_{c\mathbf k}|T_{\mathbf G}|u_{c\mathbf k'}\rangle
\langle u_{v\mathbf k}|T_{\mathbf G}|u_{v\mathbf k'}\rangle^*.
\label{eq:si-continued-overlap}
\end{equation}
The coefficient and integration weight are repeated under this
continuation: $\varphi(\mathbf k'+\mathbf G)=\varphi(\mathbf k')$.
There is no averaging over the seven reciprocal images.

The production parameters are listed in Table~\ref{tab:si-parameters}.
The production calculation uses $t_2=-10^{-7}\,\mathrm{eV}$ as a
near-zero realization of the limit discussed in the main text.
The exact projected-character argument below is established at
$t_2=0$.
In assembling the interaction, transfers with
$\widetilde q<10^{-3}$ are omitted, including the diagonal
zero-transfer contribution. The assembled Wannier matrix is replaced
by its Hermitian part before diagonalization; the same Hermitian
interaction matrix is used in the energy expectations and amplitude
relations below. The production solver
uses complex single precision for this matrix and retains the lowest
50 exciton eigenstates. Energies are reported in electronvolts in the
manuscript.

\begin{table}[!htbp]
\centering
\caption{Parameters of the production model and screened interaction.
The spin--orbit coupling $\lambda_{\mathrm{soc}}$ is varied as
specified in the main-text figures.}
\label{tab:si-parameters}
\begin{tabular}{ll}
\hline
Parameter & Value\\
\hline
$t_1$ & $0.54\,\mathrm{eV}$\\
$t_2$ & $-10^{-7}\,\mathrm{eV}$\\
$t_3$ & $0.06\,\mathrm{eV}$\\
$\epsilon_{\rm orb}$ & $0.45\,\mathrm{eV}$\\
$a_0$ & $0.318\,\mathrm{nm}$\\
$L_w$ & $0.400\,\mathrm{nm}$\\
$\epsilon_s$ & $4.5$\\
$\epsilon_w$ & $8.0$\\
Mesh division parameter $N$ & $35$\\
Number of sampled momenta $N_k$ & $3672$\\
Number of reciprocal images & $7$\\
Minimum retained $\widetilde q$ & $10^{-3}$\\
\hline
\end{tabular}
\end{table}

\section{Bloch-frame covariance and the sixfold relative winding class}
\label{sec:si-gauge-index}

\subsection{Frame covariance of the projected exciton problem}

We use the single conduction--valence channel of the main text
at nonzero spin--orbit coupling. Its electron--hole Bloch frame
is $P_{\mathbf k}=|u_{c\mathbf k}\rangle\langle u_{v\mathbf k}|$.
Under $|u_{n\mathbf k}\rangle\mapsto
e^{\mathrm i\beta_n(\mathbf k)}|u_{n\mathbf k}\rangle$, define
$\zeta(\mathbf k)=\beta_v(\mathbf k)-\beta_c(\mathbf k)$.
The frame acquires the phase $e^{-\mathrm i\zeta}$, and the
exciton coefficient transforms as
\begin{equation}
\varphi^{(\eta)\prime}(\mathbf k)
=e^{\mathrm i\zeta(\mathbf k)}\varphi^{(\eta)}(\mathbf k).
\label{eq:si-envelope-gauge}
\end{equation}
This compensation leaves the physical electron--hole state
unchanged. The band Berry connections are
$A_n=\mathrm i\langle u_n|\mathrm d u_n\rangle$.
Their difference therefore obeys
\begin{equation}
A_{\rm exc}'=A_{\rm exc}+\mathrm d\zeta,
\label{eq:si-connection-gauge}
\end{equation}
where $A_{\rm exc}=A_c-A_v$. Writing the band curvature as
$F_n=\mathrm dA_n$, the electron--hole curvature is
$F_{\rm exc}=F_c-F_v$, equivalently $\mathrm dA_{\rm exc}$.
With $D=\mathrm d-\mathrm i A_{\rm exc}$, the corresponding
covariant derivative satisfies
\begin{equation}
D'\varphi'=e^{\mathrm i\zeta}D\varphi.
\label{eq:si-covariant-derivative}
\end{equation}

The finite-momentum form factor transforms according to
\begin{equation}
\Lambda'_{\mathbf k,\mathbf k'}
=e^{\mathrm i[\zeta(\mathbf k)-\zeta(\mathbf k')]}
 \Lambda_{\mathbf k,\mathbf k'}.
\label{eq:si-form-factor-gauge}
\end{equation}
The interaction kernel has the same transformation law when
the reciprocal continuation is transformed consistently.
Consequently, the product
$\varphi^*(\mathbf k)\Lambda_{\mathbf k,\mathbf k'}
\varphi(\mathbf k')$ and the complete interaction expectation
are gauge invariant. In particular, the combined phase
$\chi(\mathbf k,\mathbf k')=
\operatorname{Arg}[-V_{\mathbf k\mathbf k'}[\varphi]]$
is invariant wherever its argument is nonzero, although the
form-factor and envelope phases depend separately on the frame.

The same covariance has a local differential limit. In a
smooth region, the expansion
$\Lambda_{\mathbf k,\mathbf k+\boldsymbol\delta}
=1-\mathrm i\delta^\mu A_{{\rm exc},\mu}(\mathbf k)
+O(|\boldsymbol\delta|^2)$ gives
\begin{equation}
\Lambda_{\mathbf k,\mathbf k+\boldsymbol\delta}
\varphi(\mathbf k+\boldsymbol\delta)-\varphi(\mathbf k)
=\delta^\mu D_\mu\varphi(\mathbf k)
+O(|\boldsymbol\delta|^2).
\label{eq:si-finite-link-local-limit}
\end{equation}
This relation identifies the connection contained in the
finite-momentum form factor. The energies reported in this
work are evaluated from the full projected interaction sums.

\subsection{Core winding in the calculation frame}

We retain the numerical $C_6$-adapted Bloch gauge used to
construct the stored Wannier eigenstates.

The numerical frame is constructed in the seed sector
$|k_y|\leq k_x/\sqrt{3}$. Each band eigenvector is aligned with its
$\mathrm{M}_1$ reference by removing the principal phase of their overlap.
The phase mismatch between the upper edge and the orbital rotation of the
lower edge is corrected across the sector in proportion to
$\sqrt{3}k_y/(2k_x)$, with fixed offsets $r_v\pi/3$ and $-r_c\pi/3$ for
the valence and conduction bands. The other sectors are generated as
\begin{equation}
 |u_v(R_6^s\mathbf{k})\rangle
 = e^{isr_v\pi/3}\widehat R_6^{\,s}|u_v(\mathbf{k})\rangle,
 \qquad s\in\{0,\ldots,5\},
\label{eq:si-valence-sector}
\end{equation}
\begin{equation}
 |u_c(R_6^s\mathbf{k})\rangle
 = e^{-isr_c\pi/3}\widehat R_6^{\,s}|u_c(\mathbf{k})\rangle.
\label{eq:si-conduction-sector}
\end{equation}
Here $R_6$ rotates momentum by $\pi/3$, and $\widehat R_6$ is its fixed
orbital representation in the Hamiltonian basis. The integer pairs are
$(r_v,r_c)=(2,-1)$ for
$\lambda_{\mathrm{soc}}>0$ and $(-2,1)$ for
$\lambda_{\mathrm{soc}}<0$.

Let $c_\Gamma$ be
the counterclockwise, $C_6$-invariant contour surrounding
$\Gamma$ used in the winding analysis. For a coefficient
nonzero on this contour, define
\begin{equation}
m^{(\eta)}_\Gamma
=\frac{1}{2\pi\mathrm i}\oint_{c_\Gamma}
[\varphi^{(\eta)}]^{-1}\mathrm d\varphi^{(\eta)}.
\label{eq:si-core-winding}
\end{equation}
For the core-winding diagnostic, 720 counterclockwise target angles on a
circle of radius $0.20|\Gamma\mathrm{K}|$ are mapped to the nearest native
momentum points, removing repeated indices in their first-occurrence
order. The number of distinct contour points therefore depends on the
mesh. We sum the principal phase increments
$\operatorname{Arg}[\varphi_{j+1}\varphi_j^*]$, including the closing
link, and divide by $2\pi$. The diagnostic is accepted only for finite
amplitudes with $\min|\varphi|>10^{-8}\max|\varphi|$, phase steps below
$0.95\pi$ in magnitude, and a winding within $10^{-5}$ of an integer.
This calculation uses native complex amplitudes before plotting
interpolation.

The contour winding is an integer. The main text writes $m^{(\eta)}$ for
$m^{(\eta)}_\Gamma$. The contour specifies the internal
core structure being classified; the band Chern-number
difference is determined separately by the full-BZ curvature,
\begin{equation}
C_{\rm exc}
=\frac{1}{2\pi}\int_{\rm FBZ}(F_c-F_v).
\label{eq:si-band-chern-difference}
\end{equation}
We classify the states by the relative winding class
\begin{equation}
\mathcal I^{(\eta)}
=\bigl[m^{(\eta)}_\Gamma-C_{\rm exc}\bigr]_6.
\label{eq:si-relative-class}
\end{equation}
Here $[n]_6$ denotes the residue class of an integer modulo
six. We use representatives in
$\{-2,-1,0,1,2,3\}$; the labels $+3$ and $-3$ denote the
same class and may be displayed interchangeably. Integer windings and class labels are retained
separately. Thus windings that differ by $6\ell$, with
$\ell\in\mathbb Z$, give the same relative class.

The spectral and table labels use this core contour and calculation
frame. Interior contours whose integer windings differ by a multiple
of six assign the same relative class.

\subsection{Berry geometry and relative-class wavefunctions}
\label{sec:si-geometry}

\begin{figure}[p]
\centering
\includegraphics[width=150mm]{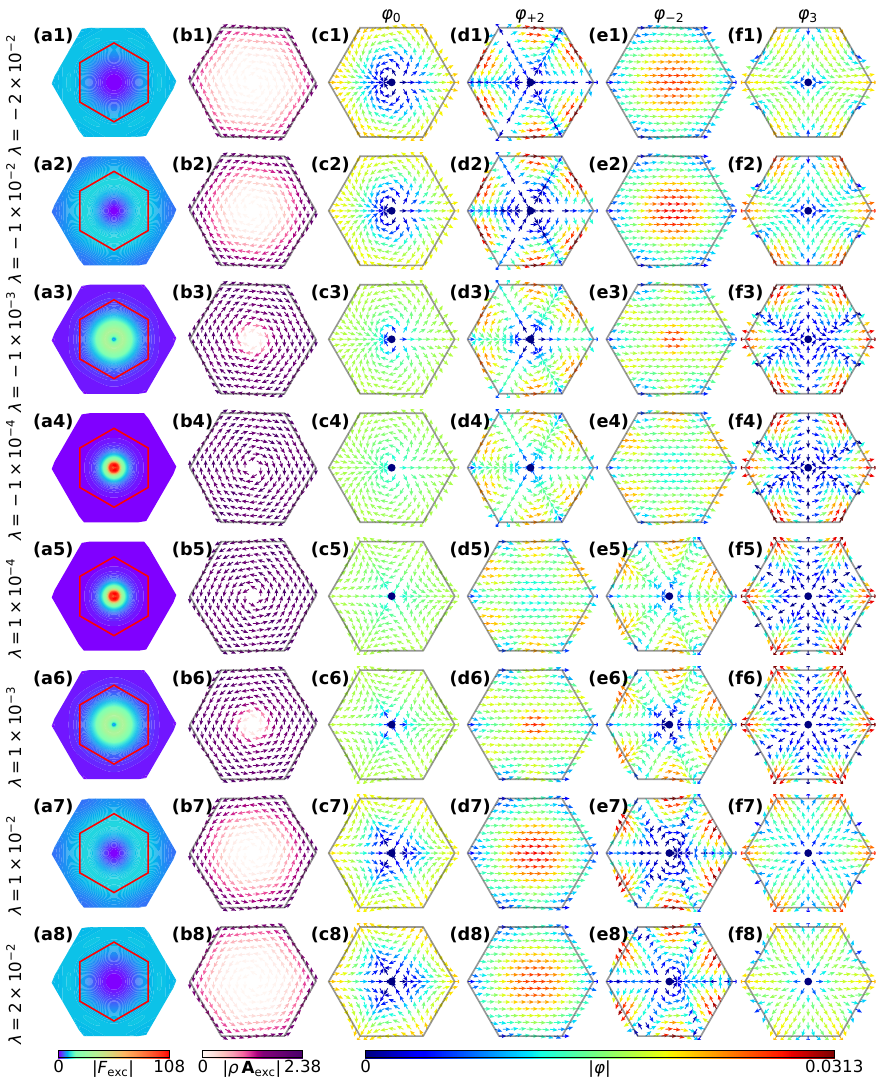}
\caption{Electron--hole Berry geometry and exciton wavefunctions
for the relative classes $\mathcal I=0,+2,-2,3$.
The SOC numbering is aligned with
figures~\ref{fig:si-all-sector-phases} and
\ref{fig:si-all-sector-phases-source2}.
Numeric suffixes 1--8 label rows from top to bottom at
$\lambda_{\mathrm{soc}}=-0.02,-0.01,-0.001,-0.0001,
+0.0001,+0.001,+0.01,+0.02\,\mathrm{eV}$, respectively;
the same suffix identifies the corresponding SOC column in
those phase maps. Letters (a)--(f) identify columns showing
$|F_{\mathrm{exc}}|$, $\rho\mathbf A_{\mathrm{exc}}$,
$\varphi_0$, $\varphi_{+2}$, $\varphi_{-2}$ and $\varphi_3$,
respectively. The dimensionless radial factor $\rho$ is defined
in the text. Arrows in (b1)--(b8) represent real connection
vectors; those in columns (c)--(f) encode complex phases with
fixed display lengths. Colors give the magnitudes indicated by
the shared bars. The wavefunctions use the common $K_1$
phase-reference convention, interpolation and displayed
$\Gamma$ values described in the text.
The abbreviated SOC headings denote $\lambda_{\mathrm{soc}}$.}
\label{fig:si-geometry}
\end{figure}

Figure~\ref{fig:si-geometry} compares the electron--hole Berry geometry
and the selected exciton branches in the relative classes
$\mathcal I=0,+2,-2,3$ across positive and negative SOC. Here $\varphi_{\mathcal I}$
denotes a selected low-energy branch rather than a fixed energy rank.
The band background has $C_{\mathrm{exc}}=+2$ for negative SOC
and $-2$ for positive SOC. These signs follow from the signed
curvature integral in equation~\eqref{eq:si-band-chern-difference};
the magnitude plotted in column (a) does not distinguish them.
Let $\phi_{\mathbf k}=\operatorname{Arg}
(\widetilde k_x+\mathrm i\widetilde k_y)$ denote the azimuthal
angle about $\Gamma$ in the displayed coordinates, with
$\operatorname{Arg}\in[-\pi,\pi)$ for nonzero arguments.
For this figure, let $\widetilde{\mathbf k}=\rho\mathbf e_\rho$
with $\rho=|\widetilde{\mathbf k}|$ and define the polar unit vectors
$\mathbf e_\rho=(\cos\phi_{\mathbf k},\sin\phi_{\mathbf k})$ and
$\mathbf e_\phi=(-\sin\phi_{\mathbf k},\cos\phi_{\mathbf k})$.
All plotted derivatives and radial factors refer to the dimensionless
momentum coordinates $\widetilde{\mathbf k}$.
Writing the connection one-form in polar coordinates gives
\begin{equation}
A_{\mathrm{exc}}=A_\rho\,\dd\rho+A_\phi\,\dd\phi_{\mathbf k}.
\label{eq:si-polar-connection}
\end{equation}
The coefficients are
$A_\rho=\mathbf A_{\mathrm{exc}}\cdot\mathbf e_\rho$ and
$A_\phi=\rho\mathbf A_{\mathrm{exc}}\cdot\mathbf e_\phi$.
In particular, the angular coefficient contains a factor of $\rho$.
Column (b) shows the radius-weighted Cartesian connection
$\rho\mathbf A_{\mathrm{exc}}$. In column (a), the plotted
$|F_{\mathrm{exc}}|$ denotes the magnitude of the scalar
coefficient of the curvature two-form in the dimensionless
Cartesian momentum coordinates.

In the stated calculation frame, the maps show the concentration of Berry curvature near $\Gamma$
at small nonzero SOC and the accompanying changes of the selected
coefficient phases. The Chern-locked branch $\varphi_0$ has
a depleted core and nonzero winding, whereas the zero-winding
branch, $\varphi_{-2}$ for negative SOC and $\varphi_{+2}$
for positive SOC, has enhanced weight near $\Gamma$. The
phase-matching interpretation of these profiles is developed
in section~\ref{sec:si-near-core}. 
The fields are interpolated for display. Exact $\Gamma$ is excluded
from the production mesh. The displayed wavefunction at this point
is extrapolated for zero winding and set to zero for a nonzero
vortex winding; it is not an additional computed eigenvector sample.
For plotting, each wavefunction is given a constant overall phase
so that its value at the mesh representative of $K_1$ is positive
and real. The momentum coordinates and real connection vectors
are rotated together by $30^\circ$. Complex-arrow directions encode
the wavefunction phase rather than a spatial vector and receive no
additional phase rotation.

The radial phase reversals visible in panels (d2,e7) provide
the spatial context for the interaction-phase changes analysed
in section~\ref{sec:si-soc-reversal}.

\clearpage

\subsection{Gauge scope of the class labels}

The numerical class convention and the covariance of the
physical exciton problem have distinct roles. For a rephasing
$g(\mathbf k)=e^{\mathrm i\zeta(\mathbf k)}$ that is
single-valued and nonzero on $c_\Gamma$, its winding is
$n_g=(2\pi\mathrm i)^{-1}\oint_{c_\Gamma}g^{-1}\mathrm d g$.
Equation~\eqref{eq:si-envelope-gauge} gives
\begin{equation}
m_\Gamma'=m_\Gamma+n_g.
\label{eq:si-core-winding-gauge}
\end{equation}
The band Chern difference is unchanged. A rephasing that
extends smoothly and without a zero over the disk enclosed
by $c_\Gamma$ has $n_g=0$ and preserves the integer winding.

The relative class survives rephasings that preserve the calculation
frame's rotation-sewing convention. To state this for general $C_n$
symmetry, let $R_n$ denote a rotation by $2\pi/n$ and define the
pair sewing phase by
\begin{equation}
\widehat R_nP_{\mathbf k}\widehat R_n^{-1}
=B_{\rm exc}(\mathbf k)P_{R_n\mathbf k}.
\label{eq:si-rotation-sewing}
\end{equation}
Under the frame rephasing defined above, this phase transforms as
\begin{equation}
B_{\rm exc}'(\mathbf k)
=g(R_n\mathbf k)g^*(\mathbf k)B_{\rm exc}(\mathbf k).
\label{eq:si-rotation-sewing-gauge}
\end{equation}
Preserving the sewing convention requires $g(R_n\mathbf k)=g(\mathbf k)$.
On a rotation-invariant contour on which $g$ is continuous and nonzero,
its phase increment repeats over $n$ rotation-related segments. Its integer winding
therefore has the form $n_g=nh$, with $h\in\mathbb Z$, giving
\begin{equation}
\bigl[m_\Gamma'-C_{\rm exc}\bigr]_n
=\bigl[m_\Gamma-C_{\rm exc}\bigr]_n.
\label{eq:si-compatible-gauge-class}
\end{equation}
For the YYK model, $n=6$. Thus the reported class is preserved by
rephasings compatible with the stated rotation sewing, while the
physical state and kernel obey the general covariance relations above.

\subsection{Integer lifts and continuous phase matching}

For the phase maps we retain the actual integer difference
in the specified frame and contour,
\begin{equation}
\nu_s=m_{\Gamma,s}-C_{\rm exc}.
\label{eq:si-integer-lift}
\end{equation}
Its class is $\mathcal I_s=[\nu_s]_6$.
For a particular state, define the coefficient coherence phase by
$\Delta\theta_s(\mathbf k,\mathbf k')=
\operatorname{Arg}[\varphi_s^*(\mathbf k)\varphi_s(\mathbf k')]$
where both coefficients are nonzero. With the azimuths defined in
section~\ref{sec:si-geometry}, the leading small-SOC angular
textures identified in main-text section III give
\begin{equation}
\chi_s(\mathbf k,\mathbf k')
\simeq\nu_s(\phi_{\mathbf k'}-\phi_{\mathbf k})
\pmod{2\pi}.
\label{eq:si-continuous-phase-lift}
\end{equation}
This is an approximation to the observed finite-link phase, not
a replacement for its exact definition.
At rotation-related separations $\Delta\phi=j\pi/3$, adding
$6\ell$ to $\nu_s$ leaves $\exp(\mathrm i\nu_s\Delta\phi)$
unchanged. This need not hold at general angular separations.
The modulo-six class organizes symmetry sectors, while the
integer lift retains information needed for continuous angular
textures. All reported energies use the computed phases and
amplitudes in the full momentum sums.

\section{Finite-momentum phase matching and amplitude controls}
\label{sec:si-phase}

\subsection{Exact phase dependence of a retained interaction link}

Let $s$ label a specified normalized state $\varphi_s$.
The label selects the envelope used in the interaction functional;
it is not itself a winding number. Below, $s=w,t$ identifies the
Chern-locked and zero-winding branches. When a branch is selected
by its relative class, we instead write $\varphi_{\mathcal I}$,
$a_{\mathcal I}$ and $\chi_{\mathcal I}$. This notation refers to
the selected branch within that class, not to an energy rank.
For each retained interaction link, define its nonnegative weight by
\begin{equation}
a_s(\mathbf k,\mathbf k')
=\bigl|V_{\mathbf k\mathbf k'}[\varphi_s]\bigr|.
\label{eq:si-link-weight}
\end{equation}
Where $a_s>0$, the combined phase defined in
section~\ref{sec:si-gauge-index} is
$\chi_s=\operatorname{Arg}[-V_{\mathbf k\mathbf k'}[\varphi_s]]$.
The real contribution of the attractive link is
\begin{equation}
\operatorname{Re}V_{\mathbf k\mathbf k'}[\varphi_s]
=-a_s(\mathbf k,\mathbf k')\cos\chi_s(\mathbf k,\mathbf k').
\label{eq:si-link-phase}
\end{equation}
At fixed $a_s$, the phase-aligned value is $-a_s$. The loss of
attractive energy relative to this value is
\begin{equation}
\operatorname{Re}V_{\mathbf k\mathbf k'}[\varphi_s]+a_s
=a_s(1-\cos\chi_s).
\label{eq:si-link-loss}
\end{equation}
This expression is nonnegative and holds at finite momentum
separation. A zero-weight link contributes zero, without requiring
a phase to be assigned to it. Each weight contains the quadrature
measure, the Coulomb interaction, the form-factor magnitude and both
wavefunction amplitudes. The interaction expectation sums the real
contributions over all ordered FBZ pairs and retained reciprocal
images, so this sum carries no additional factor of two.

Simultaneous alignment of all links is constrained by the Bloch
geometry and the common coefficient phases. The amplitude controls
below compare these phases at fixed link weights, while the full
SOC-dependent energy retains both $a_s$ and $\chi_s$.

\subsection{Phase-preserving trial states and energy comparisons}
\label{sec:si-amplitude-controls}

\begin{figure}[!htbp]
\centering
\resizebox{\textwidth}{!}{%
\setlength{\unitlength}{1bp}%
\begin{picture}(425.196850,221.102362)
\put(0,0){\includegraphics[width=425.196850bp]{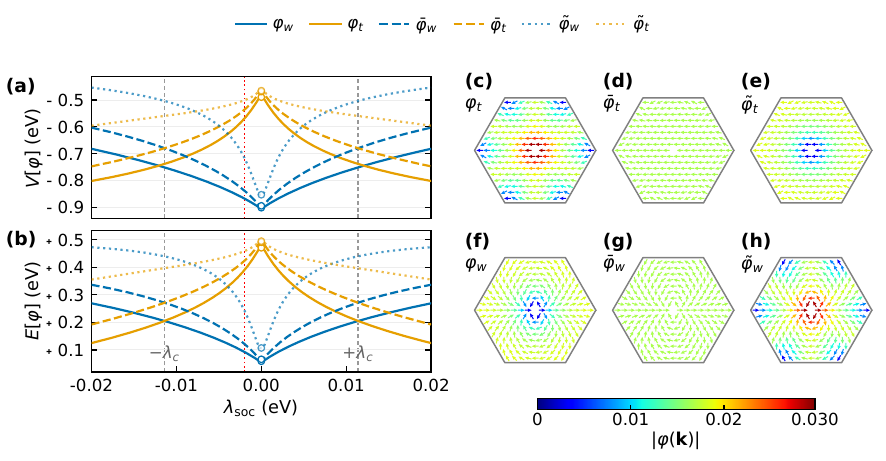}}
\put(10,128){\color{white}\rule{11bp}{43bp}}
\put(15.5,149.5){\makebox(0,0){\rotatebox{90}{\fontsize{9}{11}\selectfont $K[\varphi]$ (eV)}}}
\put(10,55){\color{white}\rule{11bp}{43bp}}
\put(15.5,76.5){\makebox(0,0){\rotatebox{90}{\fontsize{9}{11}\selectfont $E[\varphi]$ (eV)}}}
\end{picture}}
\caption{Relative-winding-resolved interaction and total energies, with
phase-preserving amplitude controls. Panels (a,b) show the
interaction energy $K[\varphi]$ and total energy $E[\varphi]$, respectively.
Solid, dashed and dotted
curves show native, constant-amplitude and exchanged-amplitude states.
Blue and orange curves denote the Chern-locked ($w$) and zero-winding
($t$) phase families, respectively.
grey dashed vertical lines mark the native-eigenstate crossings at
$\lambda_{\mathrm{soc}}=\pm\lambda_c$, with
$\lambda_c\simeq0.0114\,\mathrm{eV}$.
The red dotted vertical line in each energy panel marks
$\lambda_{\mathrm{soc}}=-0.002\,\mathrm{eV}$, used for all
wavefunction maps on the right.
Panels (c)--(e) show $\varphi_t$, $\bar\varphi_t$ and
$\tilde\varphi_t$; (f)--(h) show $\varphi_w$, $\bar\varphi_w$
and $\tilde\varphi_w$, respectively.
Fixed-length arrows encode phase and colors encode magnitude in the
calculation frame. The hexagons mark the FBZ; the maps use a common
216-point subset of the full energy-summation mesh.
Open circles at zero SOC
use the average of the two nearest nonzero samples,
$\lambda_{\mathrm{soc}}=\pm0.0002\,\mathrm{eV}$, and are not separately
computed zero-SOC states.}
\label{fig:si-six-states}
\end{figure}

To distinguish the energetic roles of phase and amplitude, we compare
the selected Chern-locked ($w$) and zero-winding ($t$) branches with
trial states that retain their phases while replacing their amplitudes.
These branch labels are used in figure~\ref{fig:si-six-states}.
In the main-text index notation, $\varphi_w$ is $\varphi_0$ on both
SOC sides, while $\varphi_t$ is $\varphi_{+2}$ at positive SOC and
$\varphi_{-2}$ at negative SOC. The same correspondence applies to
the bar and tilde states.
The constant-amplitude counterparts are
\begin{equation}
\bar\varphi_s(\mathbf k)
=\frac{1}{\sqrt{N_k}}\frac{\varphi_s(\mathbf k)}{|\varphi_s(\mathbf k)|},
\qquad s\in\{w,t\}.
\label{eq:si-bar-state}
\end{equation}
Here $N_k$ is the number of sampled FBZ momenta, with discrete
normalization $\sum_{\mathbf k}|\varphi_s(\mathbf k)|^2=1$.
The exchanged-amplitude states are
\begin{equation}
\tilde\varphi_w(\mathbf k)
=|\varphi_t(\mathbf k)|\frac{\varphi_w(\mathbf k)}{|\varphi_w(\mathbf k)|},
\label{eq:si-tilde-w}
\end{equation}
\begin{equation}
\tilde\varphi_t(\mathbf k)
=|\varphi_w(\mathbf k)|\frac{\varphi_t(\mathbf k)}{|\varphi_t(\mathbf k)|}.
\label{eq:si-tilde-t}
\end{equation}
The bar and tilde states retain the parent phase with a constant
or exchanged amplitude, respectively. They are normalized when the
parent states are normalized and are defined at nonzero parent
amplitudes on the sampled mesh. At fixed SOC they retain $\chi_s$,
with the corresponding link weights evaluated from
equation~\eqref{eq:si-link-weight}.

The pairs $(\varphi_w,\tilde\varphi_t)$,
$(\tilde\varphi_w,\varphi_t)$ and
$(\bar\varphi_w,\bar\varphi_t)$ have identical amplitudes at every
sampled momentum. At fixed SOC, Bloch frame and kernel, they therefore
have the same magnitude of each retained interaction link, including
the reciprocal images. If
$\Delta[\varphi]=\sum_{\mathbf k}(\epsilon_{c\mathbf k}-\epsilon_{v\mathbf k})|\varphi(\mathbf k)|^2$
denotes the diagonal band-energy expectation, then $\Delta$ is identical
within each pair. Each pair's total-energy difference therefore equals
its interaction-energy difference and isolates the combined phases
in equation~\eqref{eq:si-link-phase} at equal link weights.

Figure~\ref{fig:si-six-states}(a,b) shows that the three amplitude
profiles within each phase family preserve the characteristic SOC
dependence of the interaction and total energies. The $w$-phase
states have lower energies near zero SOC, and their energies rise
with $|\lambda_{\mathrm{soc}}|$. The $t$-phase states instead gain
energy lowering as the SOC magnitude grows. The native eigenstate
branches cross near $\lambda_{\mathrm{soc}}=\pm0.0114\,\mathrm{eV}$.
Amplitude replacement changes the energy values and the locations
of crossings between the trial states, while preserving the
contrasting trends of these two phase families. The red dotted
line at $\lambda_{\mathrm{soc}}=-0.002\,\mathrm{eV}$ selects
the wavefunction maps: panels (c)--(e) compare the three $t$-phase
states, and (f)--(h) compare the three $w$-phase states.
Within each row, the phase texture is retained while the magnitude
is native, constant or taken from the other branch.
Together with the equal-amplitude comparisons, these results identify
the phase structure relative to the Bloch background as a principal
factor in the energetic ordering and SOC trends of the selected
branches. Across SOC, the full energy also retains the changing
link weights and diagonal band-energy contributions.

Table~\ref{tab:si-control-energies} quantifies the comparison at
$\lambda_{\mathrm{soc}}=-0.0002\,\mathrm{eV}$.
The energy separation between the two phase sectors exceeds the
changes caused by amplitude replacement, while their diagonal
contributions are nearly equal. The separation is therefore
dominated by the projected interaction.

\begin{table}[!htbp]
\centering
\caption{Amplitude-control energies at
$\lambda_{\mathrm{soc}}=-0.0002\,\mathrm{eV}$, in eV.
$E-K$ denotes the diagonal band-energy contribution.}
\label{tab:si-control-energies}
\begin{tabular}{lrrr}
\hline
State & $K$ & $E$ & $E-K$\\
\hline
$\varphi_w$ & $-0.901233$ & $0.058761$ & $0.959994$\\
$\bar\varphi_w$ & $-0.894280$ & $0.065709$ & $0.959989$\\
$\tilde\varphi_w$ & $-0.853058$ & $0.106931$ & $0.959989$\\
$\varphi_t$ & $-0.489169$ & $0.470821$ & $0.959989$\\
$\bar\varphi_t$ & $-0.468627$ & $0.491362$ & $0.959989$\\
$\tilde\varphi_t$ & $-0.464927$ & $0.495067$ & $0.959994$\\
\hline
\end{tabular}
\end{table}

For the constant-amplitude pair, the common link weight is
$a_{\mathrm{flat}}(\mathbf k,\mathbf k')=|V_{\mathbf k\mathbf k'}|/N_k$.
The controls retain the phases $\chi_w$ and $\chi_t$ of their
parent states. Their exact difference on a retained link is
\begin{equation}
\operatorname{Re}V_{\mathbf k\mathbf k'}[\bar\varphi_t]
-\operatorname{Re}V_{\mathbf k\mathbf k'}[\bar\varphi_w]
=a_{\mathrm{flat}}(\cos\chi_w-\cos\chi_t).
\label{eq:si-equal-amplitude}
\end{equation}
In the small negative-SOC regime, the leading angular textures in
figure~\ref{fig:si-all-sector-phases}(g4,i4) are $\chi_w\simeq0$ and
$\chi_t\simeq-2(\phi_{\mathbf k'}-\phi_{\mathbf k})$ modulo $2\pi$.
The corresponding approximation is
\begin{equation}
\operatorname{Re}V_{\mathbf k\mathbf k'}[\bar\varphi_t]
-\operatorname{Re}V_{\mathbf k\mathbf k'}[\bar\varphi_w]
\simeq a_{\mathrm{flat}}\{1-\cos[2(\phi_{\mathbf k'}-\phi_{\mathbf k})]\}.
\label{eq:si-angular-control}
\end{equation}
This leading contribution is nonnegative and favors the Chern-locked
phase. The approximation describes the observed angular texture;
the tabulated energies retain the complete weighted momentum sums.

\subsection{Additional relative-winding classes}

\begin{figure}[p]
\centering
\resizebox{\textwidth}{!}{%
\setlength{\unitlength}{1bp}%
\begin{picture}(425.196850,391.181102)
\put(0,0){\includegraphics[width=425.196850bp]{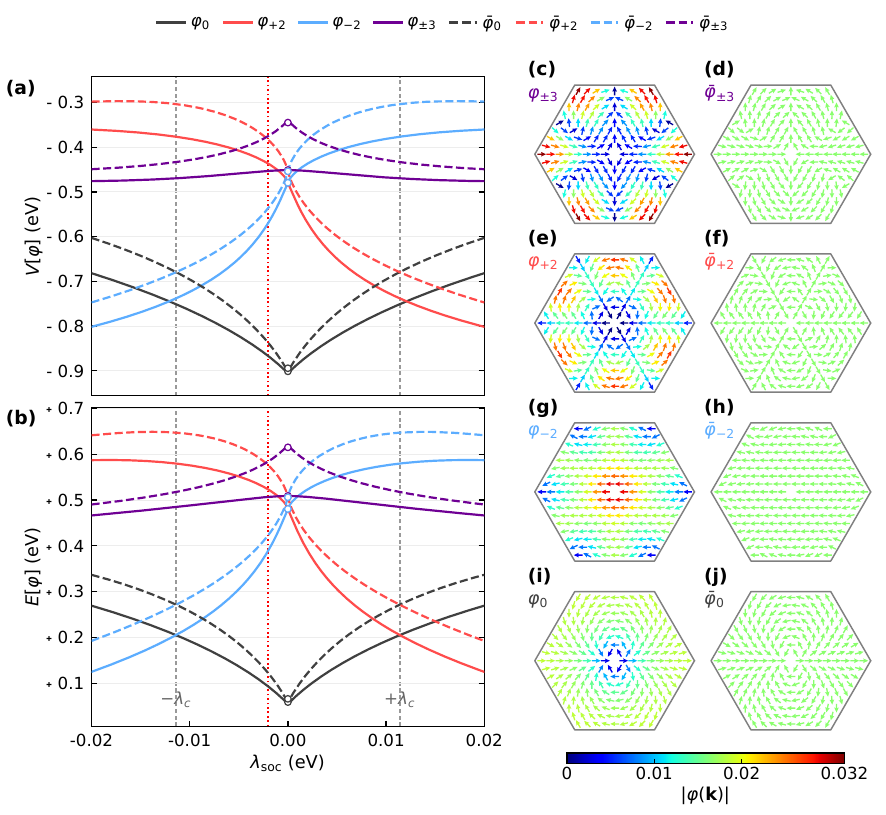}}
\put(10,255.5){\color{white}\rule{11bp}{43bp}}
\put(15.5,277){\makebox(0,0){\rotatebox{90}{\fontsize{9}{11}\selectfont $K[\varphi]$ (eV)}}}
\put(10,97.5){\color{white}\rule{11bp}{43bp}}
\put(15.5,119){\makebox(0,0){\rotatebox{90}{\fontsize{9}{11}\selectfont $E[\varphi]$ (eV)}}}
\end{picture}}
\caption{Relative-class-resolved energies and phase-preserving
amplitude controls. Panels (a,b) show the interaction and total
energies $K[\varphi]$ and $E[\varphi]$, respectively.
Solid and dashed curves represent
native states $\varphi_{\mathcal I}$ and constant-amplitude
controls $\bar\varphi_{\mathcal I}$. grey, red, light blue and
purple identify the classes $0,+2,-2,\pm3$, respectively.
The red dotted vertical line in each energy panel marks
$\lambda_{\mathrm{soc}}=-0.002\,\mathrm{eV}$, used for all
wavefunction maps. grey dashed vertical lines mark
$\pm\lambda_c$, with $\lambda_c\simeq0.0114\,\mathrm{eV}$,
where the native Chern-locked and zero-winding branches cross.
Open circles at zero SOC are averages of the two nearest nonzero
samples, $\lambda_{\mathrm{soc}}=\pm0.0002\,\mathrm{eV}$;
they are not separately computed zero-SOC states.
Panels (c,d), (e,f), (g,h) and (i,j) show the
native/constant-amplitude pairs for
$\mathcal I=\pm3,+2,-2,0$, respectively.
Fixed-length arrows encode phase, colors encode magnitude,
and grey hexagons mark the FBZ. The wavefunction maps use a
common 216-point subset of the full energy-summation mesh.
The labels $+3$ and $-3$ denote the same relative class.}
\label{fig:si-four-classes}
\end{figure}

The state label can also select a branch in each relative class
$\mathcal I$. Applying the same amplitude flattening to these
branches gives the native/flattened pairs compared in
figure~\ref{fig:si-four-classes}(a,b). The figure shows their
interaction and total energies
across SOC. The red dotted line selects
$\lambda_{\mathrm{soc}}=-0.002\,\mathrm{eV}$ for the wavefunction
maps in (c)--(j), where $C_{\mathrm{exc}}=2$.
From top to bottom, the native/constant-amplitude pairs belong to
$\mathcal I=\pm3,+2,-2,0$.
Their sampled core windings are $-1,4,0,2$, respectively, and
their energy ranks at this SOC are $3,4,2,1$.
The unchanged phases within each pair and the contrasting energy
curves extend the amplitude-controlled comparison to the additional
relative classes. The grey crossing guides refer specifically to
the native Chern-locked and zero-winding branches.

\subsection{SOC reversal and phase evolution within relative sectors}
\label{sec:si-soc-reversal}

\begin{figure}[p]
\centering
\includegraphics[width=\textwidth]{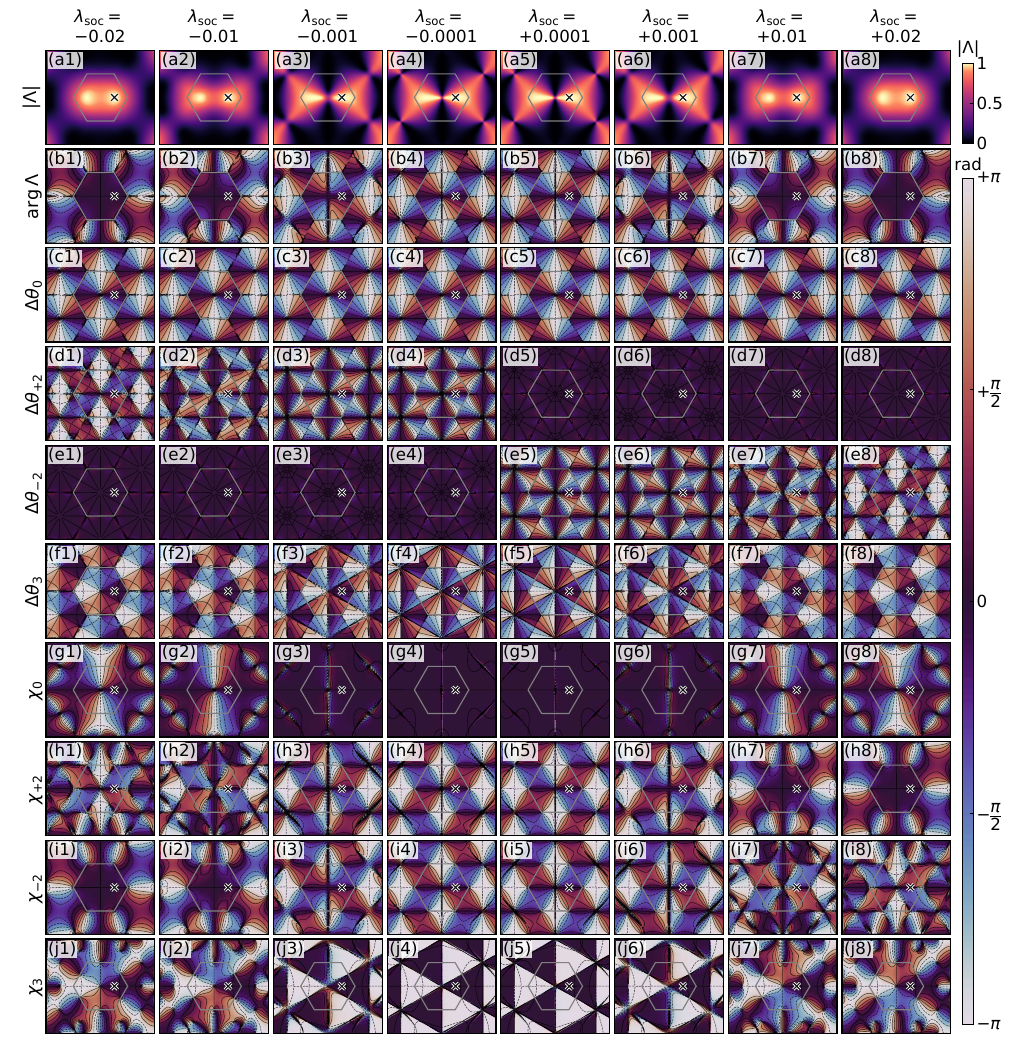}
\caption{Phase decomposition across relative sectors and SOC signs.
Columns 1--8 show
$\lambda_{\mathrm{soc}}=-0.02,-0.01,-0.001,-0.0001,
+0.0001,+0.001,+0.01,+0.02\,\mathrm{eV}$.
Each panel label combines its row letter and column number.
Rows (a,b) show $|\Lambda|$ and $\operatorname{Arg}\Lambda$;
rows (c)--(f) show $\Delta\theta_{\mathcal I}$ for
$\mathcal I=0,+2,-2,3$, respectively, and rows (g)--(j) show
the corresponding $\chi_{\mathcal I}$.
Maps vary the target momentum $\mathbf k'$ for the fixed
source $a_0\mathbf k=(1.0771,0)$ in the displayed frame
(cross). grey hexagons delimit the first Brillouin zone.
Phase contours are spaced by $\pi/6$ in $[-\pi,\pi)$;
$+\pi$ and $-\pi$ represent the same phase.
The separate phases depend on the Bloch frame, whereas
$\chi_{\mathcal I}$ is gauge invariant on nonzero interaction
links. Each sector row follows the selected branch, whose
energy rank can change with SOC. The labels $+3$ and $-3$
denote the same class.}
\label{fig:si-all-sector-phases}
\end{figure}

\begin{figure}[p]
\centering
\includegraphics[width=\textwidth]{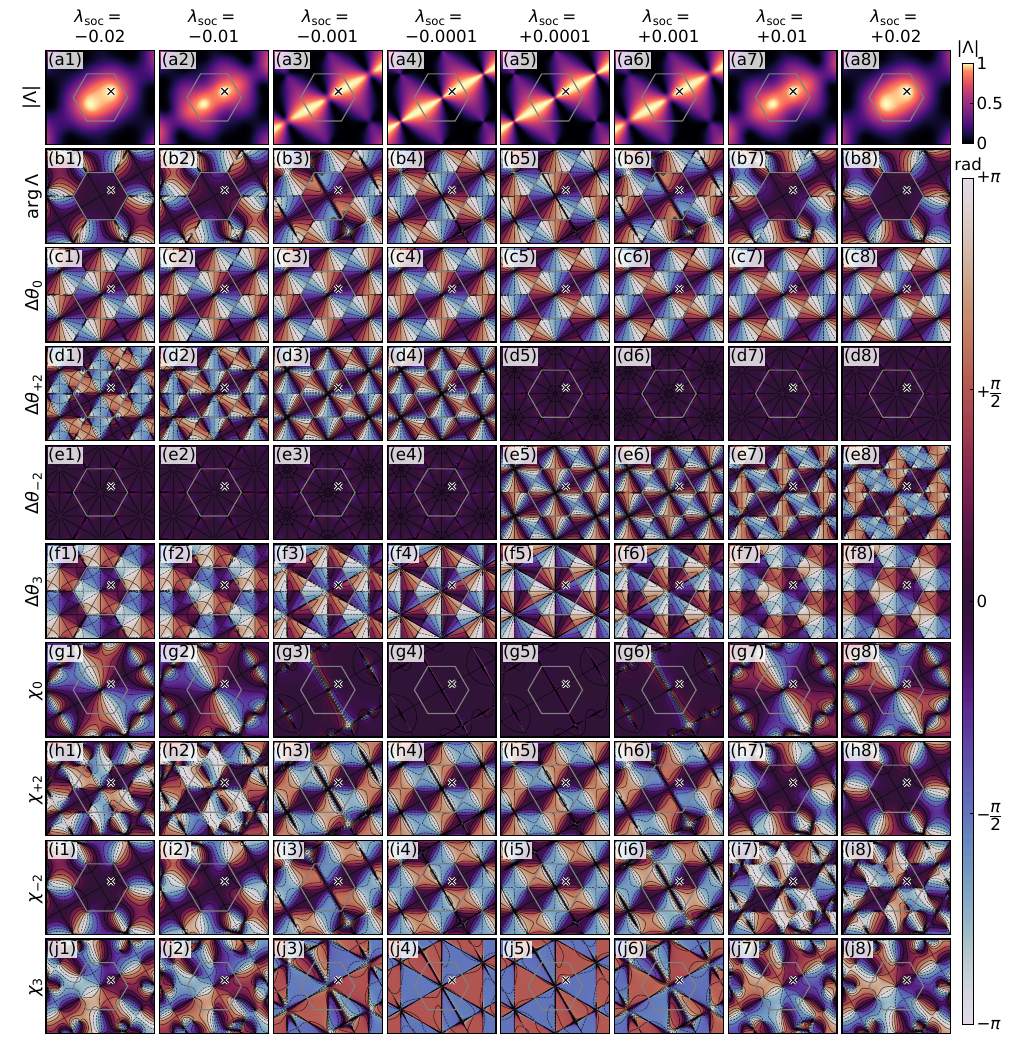}
\caption{Relative-sector phase maps at a second fixed source momentum.
The eight SOC values, row definitions, phase contours and panel-label
convention are the same as in figure~\ref{fig:si-all-sector-phases}.
Here the cross marks $a_0\mathbf k=(0.80784,0.46641)$ in the
displayed frame. Each map varies the target momentum $\mathbf k'$;
grey hexagons delimit the FBZ. The same selected branches in
$\mathcal I=0,+2,-2,3$ are used. Comparing these maps with
figure~\ref{fig:si-all-sector-phases} shows the source dependence
of the link-phase textures within the same relative classes.}
\label{fig:si-all-sector-phases-source2}
\end{figure}

Figures~\ref{fig:si-all-sector-phases} and
\ref{fig:si-all-sector-phases-source2} extend the phase decomposition
to the selected low-energy branches in the classes
$\mathcal I=0,+2,-2,3$ and to both signs of SOC, using two fixed
source momenta. The definitions of $\Delta\theta_{\mathcal I}$ and
$\chi_{\mathcal I}$ are those of the preceding sections; the
subscripts identify the relative class, not an energy rank or the
coefficient winding. In figure~\ref{fig:si-all-sector-phases},
panels (g4,g5) show the nearly phase-aligned Chern-locked branch
at $\lambda_{\mathrm{soc}}=\mp0.0001\,\mathrm{eV}$.
Panels (h4,h5) and (i4,i5) retain the respective residual chirality
of the $+2$ and $-2$ sectors across SOC reversal, whereas (h4,i4)
and (h5,i5) compare the two opposite chiralities at fixed SOC.
The second source changes the detailed textures, allowing the
sector correspondence to be distinguished from the appearance
of a particular momentum-resolved map.

The change of coefficient winding across zero SOC must be
distinguished from a degeneracy of exciton energies. Denote
one-sided small-SOC quantities by superscripts $+$ and $-$.
In our convention $C_{\mathrm{exc}}^+=-2$ and
$C_{\mathrm{exc}}^-=+2$. Following a fixed relative class gives
\begin{equation}
m_{\mathcal I}^{+}-m_{\mathcal I}^{-}
\equiv-4\pmod 6.
\label{eq:si-soc-winding-shift}
\end{equation}
Table~\ref{tab:si-soc-windings} lists the computed windings.
The first three rows have an integer difference of $-4$;
the class $3$ has $+2$, the same residue modulo six.
This relation describes the change of Bloch background for all
four classes and does not itself require two levels to meet.
All Chern and relative-class labels in this comparison refer
to the nonzero-SOC sides; no band Chern number is assigned at
the band-touching point itself.

To connect these labels to symmetry, we first define the physical
rotation character of the excitation. Let
$\widehat{\mathcal R}_6$ be the many-body crystal rotation by
$\pi/3$, and define the rotation phase of the filled-band
reference by
$\widehat{\mathcal R}_6|\Omega\rangle=\xi_\Omega|\Omega\rangle$.
For an exciton $|\Phi_s\rangle$ of definite rotation character,
formed with $\varphi_s$ as in equation~\eqref{eq:si-pair-state},
define $\xi_s$ through
\begin{equation}
\widehat{\mathcal R}_6|\Phi_s\rangle
=\xi_\Omega\xi_s|\Phi_s\rangle.
\label{eq:si-excitation-rotation-definition}
\end{equation}
Thus $\xi_s$ is the unit-modulus phase acquired by the excitation
under a $60^\circ$ rotation, relative to the filled-band reference.
It includes both the envelope and the electron--hole Bloch frame.

The band sewing in equations~\eqref{eq:si-valence-sector} and
\eqref{eq:si-conduction-sector} specifies the latter contribution.
For the interband frame
$P_{\mathbf k}=|u_{c\mathbf k}\rangle\langle u_{v\mathbf k}|$,
the orbital rotation obeys
\begin{equation}
\widehat R_6 P_{\mathbf k}\widehat R_6^{-1}
=\exp[\mathrm i(r_v+r_c)\pi/3]P_{R_6\mathbf k}.
\label{eq:si-pair-frame-sewing}
\end{equation}
For a coefficient of definite sixfold character, the core winding
$m_s\equiv m_{\Gamma,s}$ determines
$\varphi_s(R_6\mathbf k)=\exp(\mathrm i m_s\pi/3)\varphi_s(\mathbf k)$
in the stated frame. Relabelling the momentum sum after rotation
evaluates the envelope at $R_6^{-1}\mathbf k$, supplying the phase
$\exp(-\mathrm i m_s\pi/3)$. Combining the frame and envelope gives
\begin{equation}
\xi_s
=\exp\!\left[\frac{\mathrm i\pi}{3}(r_v+r_c-m_s)\right].
\label{eq:si-pair-rotation-character}
\end{equation}
In the present YYK sewing convention, the band offsets satisfy
\begin{equation}
r_v+r_c-C_{\mathrm{exc}}
=\begin{cases}
-3,&\lambda_{\mathrm{soc}}<0,\\
+3,&\lambda_{\mathrm{soc}}>0.
\end{cases}
\label{eq:si-rotation-offset}
\end{equation}
These offsets are equivalent modulo six. Substituting
$m_s\equiv C_{\mathrm{exc}}+\mathcal I_s\pmod6$ therefore
makes the rotation character a function of the relative class:
\begin{equation}
\xi_{\mathcal I}
=\exp\!\left[\frac{\mathrm i\pi}{3}(3-\mathcal I)\right].
\label{eq:si-relative-rotation-character}
\end{equation}
In particular, the Chern-locked class has $\xi_0=-1$ on both SOC
sides. For the displayed negative-SOC state, $r_v+r_c=-1$ and
$m_0=2$ give the phase $\exp(-\mathrm i\pi)$; for its
positive-SOC counterpart, $r_v+r_c=1$ and $m_0=-2$ give
$\exp(\mathrm i\pi)$. Chern locking cancels the relative winding
but leaves this common Bloch-frame rotation offset. Consequently,
$\mathcal I=0$ does not denote the trivial rotation representation:
the excitation changes sign under one $60^\circ$ rotation, while
its probability density is unchanged. The offset in
equation~\eqref{eq:si-rotation-offset} is specific to the band
sewing used here, rather than a consequence of the relative-index
definition alone.

\begin{table}[htbp]
\centering
\caption{One-sided winding labels of the selected low-energy branches.
The integers are computed at
$\lambda_{\mathrm{soc}}=\mp0.0001\,\mathrm{eV}$ in the stated
calculation frame, as in columns 4 and 5 of the phase maps.
The last column gives the excitation's sixfold rotation eigenvalue,
relative to the filled-band reference, from
equation~\eqref{eq:si-relative-rotation-character}.}
\label{tab:si-soc-windings}
\begin{tabular}{c c c c}
\hline
$\mathcal I$ & $m_{\mathcal I}^{-}$ & $m_{\mathcal I}^{+}$
& $\xi_{\mathcal I}$ \\
\hline
$0$ & $2$ & $-2$ & $-1$ \\
$+2$ & $4$ & $0$ & $\mathrm e^{\mathrm i\pi/3}$ \\
$-2$ & $0$ & $-4$ & $\mathrm e^{-\mathrm i\pi/3}$ \\
$3\equiv-3$ & $-1$ & $1$ & $1$ \\
\hline
\end{tabular}
\end{table}

These rotation characters determine the action of SOC reversal.
With
$S=\tau_x\otimes\mathbf 1_3$, which exchanges the orbital
blocks, the Hamiltonian above satisfies
\begin{equation}
H(\widetilde{\mathbf k},-\lambda_{\mathrm{soc}})
=S H(\widetilde{\mathbf k},\lambda_{\mathrm{soc}})^*S^\dagger.
\label{eq:si-soc-reversal-bloch}
\end{equation}
The antiunitary operation $\mathcal A=S\mathcal K$, with
$\mathcal K$ complex conjugation, has $\mathcal A^2=1$ and
commutes with the physical sixfold rotation. For isolated
projected bands, consistent transformation of the Bloch frame,
reciprocal continuation and interaction preserves this relation
in the electron--hole problem.

Because $\mathcal A$ is antiunitary and commutes with rotation,
it conjugates $\xi_{\mathcal I}$ and maps the class
$\mathcal I$ to $-\mathcal I$. Corresponding partner branches
therefore satisfy
\begin{equation}
E_{\mathcal I}(\lambda_{\mathrm{soc}})
=E_{-\mathcal I}(-\lambda_{\mathrm{soc}}),
\label{eq:si-soc-partner-energy}
\end{equation}
where the branch correspondence is fixed by this symmetry,
rather than by an assumed constant energy rank.

At a fixed small nonzero SOC, the selected $+2$ and $-2$
branches have coefficient windings differing by four, as
seen in table~\ref{tab:si-soc-windings}.
If these branches admit a common symmetry-preserving projected
limit at zero SOC, the distinct conjugate rotation characters
of $\mathcal I=+2$ and $-2$ require distinct equal-energy
partners in that limit. Their approach is supported by the
discrete spectrum: both relevant levels at zero SOC lie near
$482.245\,\mathrm{meV}$.

For the class labelled $\mathcal I=\pm3$ in table I of the
main text, $[-\mathcal I]_6=[\mathcal I]_6$: sign reversal
produces no distinct relative sector. With the self-conjugate
character $\xi_3=1$, SOC reversal therefore does not require
an additional degenerate partner or a level crossing at zero
SOC for this class; an accidental degeneracy is not excluded.
The Chern-locked class is also self-conjugate, since
$\xi_0=-1$ is real; it likewise has no distinct sector partner
enforced by this operation with $\mathcal A^2=1$.
For the source in figure~\ref{fig:si-all-sector-phases}, panels
(j4,j5) illustrate the class-$3$ continuation through nearly
identical $0$-or-$\pi$ patterns of $\chi_3$, with $+\pi$ and $-\pi$
representing the same phase. This particular phase pattern is
source dependent: the self-conjugate class does not require
identical or $0$-or-$\pi$ link phases at every source momentum,
as the corresponding panels of
figure~\ref{fig:si-all-sector-phases-source2} demonstrate.

\begin{figure}[p]
\centering
\includegraphics[width=\textwidth]{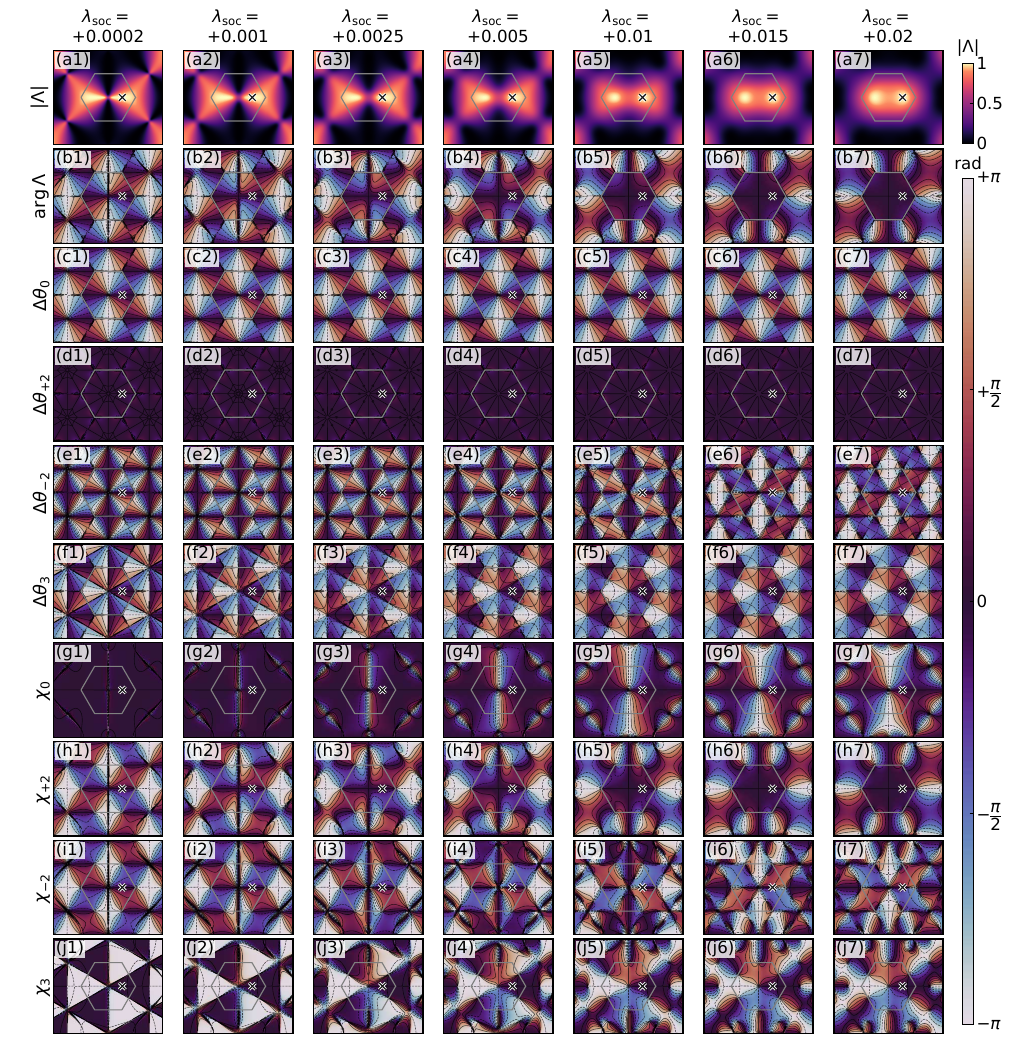}
\caption{Intermediate-SOC phase maps for a source on $\Gamma$--$K_1$.
Columns 1--7 correspond to
$\lambda_{\mathrm{soc}}=0.0002,0.001,0.0025,0.005,0.01,0.015,
0.02\,\mathrm{eV}$. The source is
$a_0\mathbf k=(1.0771,0)$, as in
figure~\ref{fig:si-all-sector-phases}.
Rows, contour spacing and phase conventions follow that figure.
The seven columns are computed SOC samples, not interpolations
between SOC values.
Panels (i5,i6) bracket the rapid source-relative phase rotation
of the $\mathcal I=-2$ branch between $0.01$ and
$0.015\,\mathrm{eV}$. The contrast with
figure~\ref{fig:si-positive-source2} separates this source-factor
effect from the evolving target texture and form factor.
The selected relative classes and calculation frame are unchanged.}
\label{fig:si-positive-source1}
\end{figure}

\begin{figure}[p]
\centering
\includegraphics[width=\textwidth]{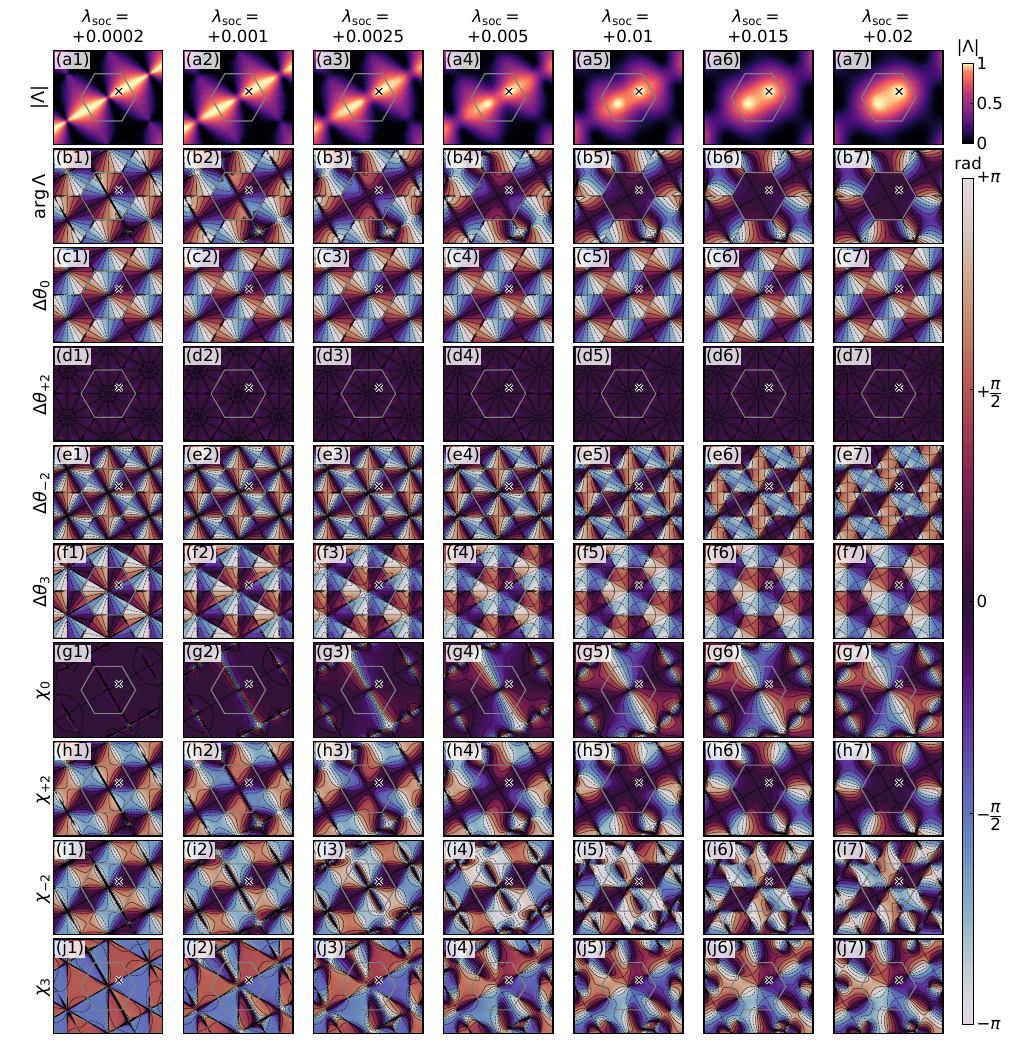}
\caption{Intermediate-SOC phase maps at the second source.
The seven SOC values, row definitions, contour spacing and phase
conventions are the same as in figure~\ref{fig:si-positive-source1}.
The source is
$a_0\mathbf k=(0.80784,0.46641)$, as in
figure~\ref{fig:si-all-sector-phases-source2}.
In particular, (h1)--(h7) and (i1)--(i7) resolve the progressive
evolution of $\chi_{+2}$ and $\chi_{-2}$ while the selected
relative classes remain unchanged.}
\label{fig:si-positive-source2}
\end{figure}

At a fixed sign of SOC, substantial changes in the link-phase
texture can occur within the same relative class.
Figures~\ref{fig:si-positive-source1} and
\ref{fig:si-positive-source2} resolve this evolution at seven
positive values of $\lambda_{\mathrm{soc}}$ between $0.0002$ and
$0.02\,\mathrm{eV}$.

For the source on $\Gamma$--$K_1$, a rapid phase rotation
accompanies this evolution. It is particularly visible
between (i5,i6) of figure~\ref{fig:si-positive-source1}, at
$\lambda_{\mathrm{soc}}=0.01$ and $0.015\,\mathrm{eV}$, and
accounts for much of the contrast between (i7,i8) of
figure~\ref{fig:si-all-sector-phases}. The negative-SOC counterpart
is the change of $\chi_{+2}$ between (h1,h2) of the latter figure.
Tracking the stored eigenstates near
$|\lambda_{\mathrm{soc}}|\simeq0.013\,\mathrm{eV}$ confirms
smooth branch evolution within the same relative class.
At this source the amplitude becomes very small, while its phase
relative to other momenta rotates by approximately $\pi$.
The conjugate factor $\varphi_{\mathcal I}(\mathbf k)^*$ enters
every outgoing link, so the source rotation produces an
approximately common shift of the link phases, superposed on
changes in the target wavefunction and form factor. The small
source amplitude simultaneously weakens the links; a large
phase-map change therefore need not imply a large energy change.
Figure~\ref{fig:si-geometry}(d2,e7) shows opposite signs of
the nearly real coefficients on either side of a low-amplitude
region along $\Gamma$--$K_1$. Comparing (d2) with (d1), and
(e7) with (e8), shows that this region moves outwards between
$|\lambda_{\mathrm{soc}}|=0.01$ and $0.02\,\mathrm{eV}$.
Its passage through the selected source produces the phase rotation
in this calculation frame.

For the source used in
figure~\ref{fig:si-all-sector-phases-source2}, the intermediate
maps in figure~\ref{fig:si-positive-source2}(h1)--(h7) and
(i1)--(i7) show the progressive deformation of $\chi_{+2}$ and
$\chi_{-2}$, respectively. The nearly constant coherence phase
of the $+2$ branch in row (d) leaves its link phase primarily
following the form-factor phase in row (b). For the $-2$ branch,
the evolving wavefunction coherence in row (e) also contributes.
The intermediate contours connect the markedly different endpoint
textures without a change of the selected relative class.

For the off-axis source, the phase evolution instead follows the
changing target coherence and form factor. Together, the two source
choices distinguish an approximately common source-phase shift
from a spatially varying target contribution, both within the same
relative sectors.

\clearpage
\subsection{Normalized amplitude variation and the vortex core}
\label{sec:si-near-core}

The phase maps describe individual interaction links, whereas the
amplitude profile is determined by a normalized, nonlocal eigenvalue problem.
To connect these descriptions, we first derive the fixed-phase
amplitude variation and then evaluate it for the constant-amplitude
states defined in section~\ref{sec:si-amplitude-controls}.
For the amplitude variation and gradient maps below, $i,j$ index the $N_k$ native FBZ momenta,
$\mathbf k_i,\mathbf k_j$. The Hermitian matrix
$W_{ij}=\Delta_i\delta_{ij}+K_{ij}$ contains the band separation
and the interaction summed over all seven retained reciprocal images.
Thus the sum over $j$ below includes all image contributions through
$K$; the images are not additional independent amplitudes.
In particular, the omitted zero-transfer term does not set the full
$K_{ii}$ to zero, because nonzero reciprocal images also contribute.

For one selected state, temporarily suppress its class label and
write $\varphi_i=r_i e^{\mathrm i\theta_i}$, where $r_i\geq0$ and
$\sum_i r_i^2=1$. At fixed phases the energy is
$E[r]=r^{\mathrm T}H_\theta r$, with the real symmetric matrix
\begin{equation}
(H_\theta)_{ij}
=\Delta_i\delta_{ij}
+\operatorname{Re}\!\left[
e^{-\mathrm i\theta_i}K_{ij}e^{\mathrm i\theta_j}\right].
\label{eq:si-core-fixed-phase-matrix}
\end{equation}
For a real amplitude direction $v$, consider
$r(\epsilon)=(r+\epsilon v)/\|r+\epsilon v\|$, with $\epsilon$
small enough that the amplitudes remain nonnegative. Its normalized
energy is exactly
\[
E[r(\epsilon)]
=\frac{E[r]+2\epsilon v^{\mathrm T}H_\theta r
+\epsilon^2v^{\mathrm T}H_\theta v}
{1+2\epsilon v^{\mathrm T}r+\epsilon^2v^{\mathrm T}v}.
\]
Differentiating the numerator and the normalization denominator gives
\begin{equation}
\left.\frac{\mathrm dE[r(\epsilon)]}{\mathrm d\epsilon}
\right|_{\epsilon=0}
=2v^{\mathrm T}(H_\theta-E[r]\mathbf 1_{N_k})r.
\label{eq:si-core-normalized-variation}
\end{equation}
Here $\mathbf 1_{N_k}$ is the identity matrix. The second term is
essential: changing one amplitude also rescales every component
through normalization.

We now restore the relative-class label and use the native phase
$\theta_{\mathcal I,i}=\operatorname{Arg}\varphi_{\mathcal I,i}$
to form $\bar\varphi_{\mathcal I,i}=\bar r
 e^{\mathrm i\theta_{\mathcal I,i}}$, where
$\bar r=N_k^{-1/2}$.
Its own expectation value is
$\bar E_{\mathcal I}=\bar\varphi_{\mathcal I}^{\dagger}
W\bar\varphi_{\mathcal I}$; this is not the native eigenenergy.
For a change at one momentum, the normalized trial path is
\[
\bar\varphi_{\mathcal I}^{(i)}(\epsilon)
=\frac{\bar\varphi_{\mathcal I}
+\epsilon e^{\mathrm i\theta_{\mathcal I,i}}e_i}
{\sqrt{1+2\epsilon\bar r+\epsilon^2}},
\]
where $e_i$ is the $i$th unit vector. Choosing $v=e_i$ in
\eqref{eq:si-core-normalized-variation} defines the pointwise
amplitude gradient
\begin{equation}
\begin{split}
g_{\mathcal I,\mathbf k_i}\equiv g_{\mathcal I,i}
&=\left.\frac{\mathrm dE[\bar\varphi_{\mathcal I}^{(i)}(\epsilon)]}
{\mathrm d\epsilon}\right|_{\epsilon=0}\\
&=2\left\{\operatorname{Re}\!\left[
e^{-\mathrm i\theta_{\mathcal I,i}}
\sum_j W_{ij}\bar\varphi_{\mathcal I,j}\right]
-\bar E_{\mathcal I}\bar r\right\}.
\end{split}
\label{eq:si-core-amplitude-gradient}
\end{equation}
A positive $g_{\mathcal I,i}$ means that a small decrease
of the corresponding amplitude lowers the trial energy; a negative
value favors an increase. The gradient has energy units because the
discrete amplitudes and $\epsilon$ are dimensionless. It obeys
$\sum_i\bar r g_{\mathcal I,i}=0$, and is unchanged by a common
shift of the energy origin. For a common amplitude variation on a
$C_6$ orbit, the derivative is the sum of $g_{\mathcal I,i}$ over
that orbit. The inference therefore also applies to a variation
that preserves the state's rotational character.

The gradient describes the initial fixed-phase relaxation of the
constant-amplitude trial. At a native eigenstate it vanishes, and
the amplitudes satisfy
\begin{equation}
(\Delta_i+K_{ii}-E_{\mathcal I})\varphi_{\mathcal I,i}
=-\sum_{j\ne i}K_{ij}\varphi_{\mathcal I,j}.
\label{eq:si-core-coherent-field}
\end{equation}

Figures~\ref{fig:si-gradient-reversal} and
\ref{fig:si-gradient-positive} show
$g_{\mathcal I,\mathbf k}$ for $\mathcal I=0,+2,-2,3$, together
with the corresponding native $|\varphi_{\mathcal I}|$ insets.
Every native momentum is evaluated separately, without radial
averaging. Blue denotes $g>0$, red denotes $g<0$, and the black
curves mark $g=0$. For $\mathcal I=0$, positive gradients near the
centre favor amplitude depletion, consistent with the depleted
native envelopes. At fixed SOC, $g_0$ increases towards the
innermost resolved region; it is the derivative for an amplitude
decrease, $-g_0$, that becomes more negative.
Among the sampled positive SOCs, the innermost gradient is largest
near $\lambda_{\mathrm{soc}}=10^{-3}\,\mathrm{eV}$.

\begin{figure}[p]
\centering
\includegraphics[width=\textwidth]{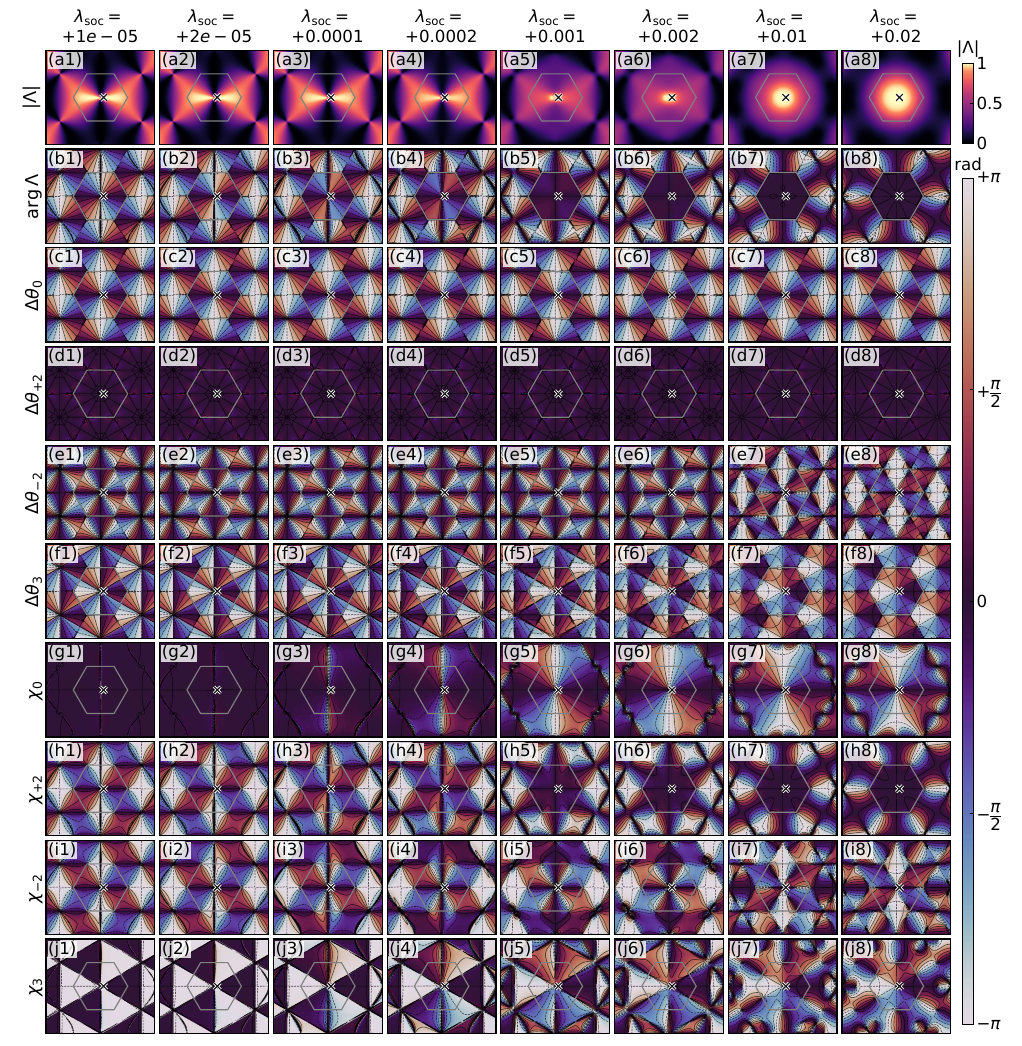}
\caption{Near-core phase matching across four relative classes
for a fixed source on the positive horizontal axis, at
$\widetilde{\mathbf k}=(0.23936,0)$ in the plotted coordinates.
Columns 1--8 correspond to
$\lambda_{\mathrm{soc}}=10^{-5},2\times10^{-5},10^{-4},
2\times10^{-4},10^{-3},2\times10^{-3},0.01,0.02\,\mathrm{eV}$.
Rows (a,b) show $|\Lambda_{\mathbf k,\mathbf k'}|$ and
$\operatorname{Arg}\Lambda_{\mathbf k,\mathbf k'}$.
Rows (c)--(f) show $\Delta\theta_{\mathcal I}$ for
$\mathcal I=0,+2,-2,3$, respectively; rows (g)--(j) show the
corresponding $\chi_{\mathcal I}$. Each panel label combines
its row letter and SOC-column number.
The cross marks the fixed source, the grey hexagon marks the
FBZ, and $\mathbf k'$ spans the displayed target plane.
Phase contours are spaced by $\pi/6$, with $-\pi$ and $+\pi$
representing the same phase. The phases are defined in
section~\ref{sec:si-phase}; the subscripts denote relative
classes, with $3$ and $-3$ equivalent modulo six.
For the positive SOC shown here, the Chern-locked and
zero-winding branches are $\varphi_0$ and $\varphi_{+2}$,
respectively. Residual mismatch in (g3,g4) is most visible
near transverse low-overlap directions, while the central
high-overlap region becomes nearly phase aligned in (h5)--(h8).}
\label{fig:si-core-source-zero}
\end{figure}

\begin{figure}[p]
\centering
\includegraphics[width=\textwidth]{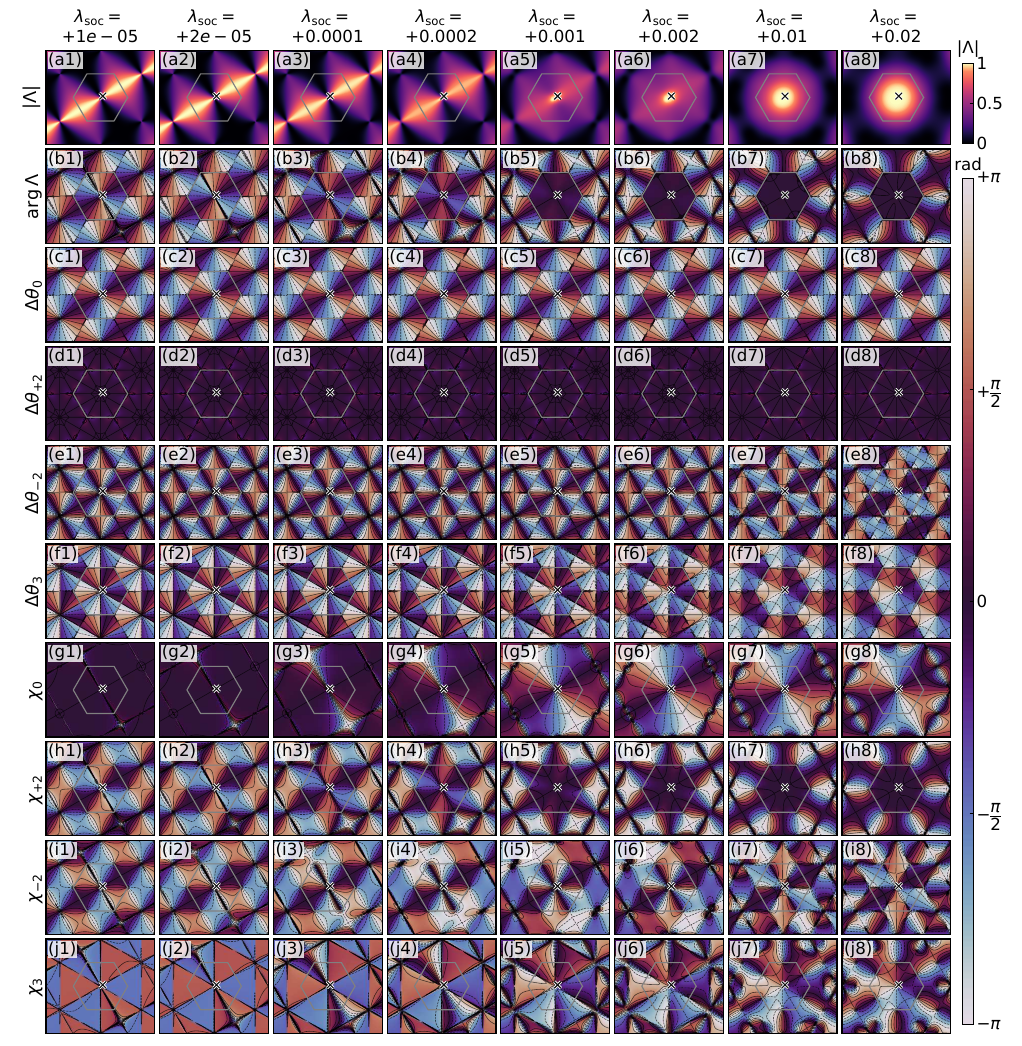}
\caption{Near-core phase matching for a source with plotted polar
angle $\phi_{\mathbf k}=\pi/6$, at
$\widetilde{\mathbf k}=(0.17952,0.10365)$.
The eight positive SOC values, rows (a)--(j), relative-class
labels, contour intervals and symbols follow
figure~\ref{fig:si-core-source-zero}.
Both sources lie at $4/35$ of the directional distance from
$\Gamma$ to the FBZ boundary; their absolute radii differ because
the boundary distances differ along the two directions.
Comparing (g1)--(g4) with (g5)--(g8) shows the evolution of the
Chern-locked interaction phase at this second source direction.
The zero-winding branch in row (h) becomes nearly phase aligned
over the central high-overlap region as SOC increases.}
\label{fig:si-core-source-pisix}
\end{figure}

\begin{figure}[p]
\centering
\includegraphics[width=\textwidth]{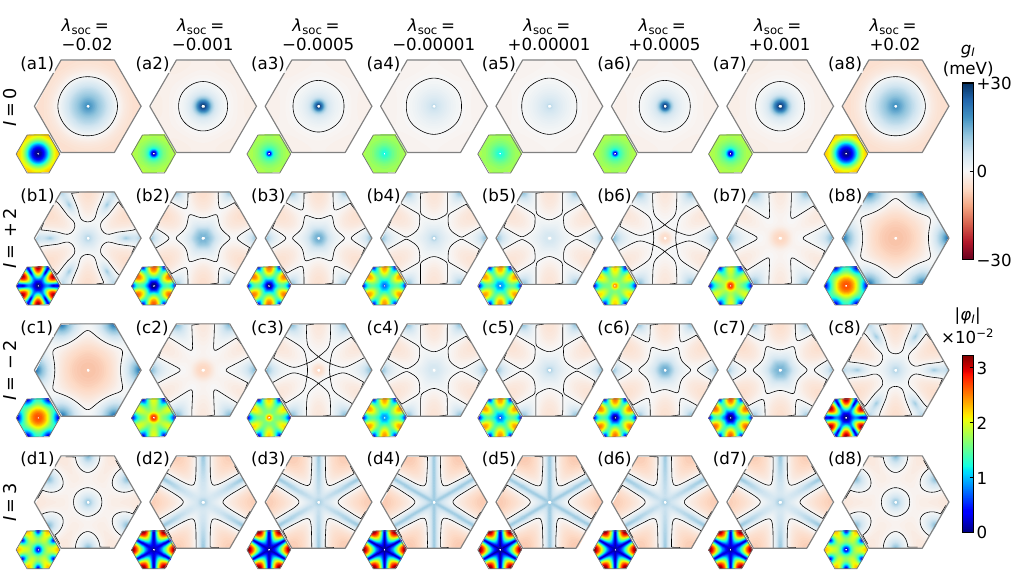}
\caption{Normalized amplitude gradients across SOC reversal.
Rows (a)--(d) show $g_{\mathcal I,\mathbf k}$ from
\eqref{eq:si-core-amplitude-gradient} for
$\mathcal I=0,+2,-2,3$, respectively. Columns 1--8 have $\lambda_{\mathrm{soc}}$ (in eV) equal to
$-0.02$, $-0.001$, $-0.0005$, $-0.00001$,
$+0.00001$, $+0.0005$, $+0.001$, and $+0.02$.
Each gradient is evaluated at the constant-amplitude trial
$\bar\varphi_{\mathcal I}$ with the native state's phases and
the full seven-image Hermitian kernel.
Blue ($g>0$) favors an initial amplitude reduction, and red
($g<0$) favors an increase, with normalization included.
Black contours mark $g=0$; grey hexagons mark the FBZ.
Insets show the native eigenstate magnitude $|\varphi_{\mathcal I}|$
at the same SOC and relative class, with the common lower color
scale. The main panels share the signed scale in meV.
Maps interpolate individual native-momentum values; white centres
mark the unsampled neighbourhood of $\Gamma$. The Chern-locked branch remains in row (a),
whereas the zero-winding branch switches between rows (c) and (b)
when SOC changes from negative to positive.}
\label{fig:si-gradient-reversal}
\end{figure}

\begin{figure}[p]
\centering
\includegraphics[width=\textwidth]{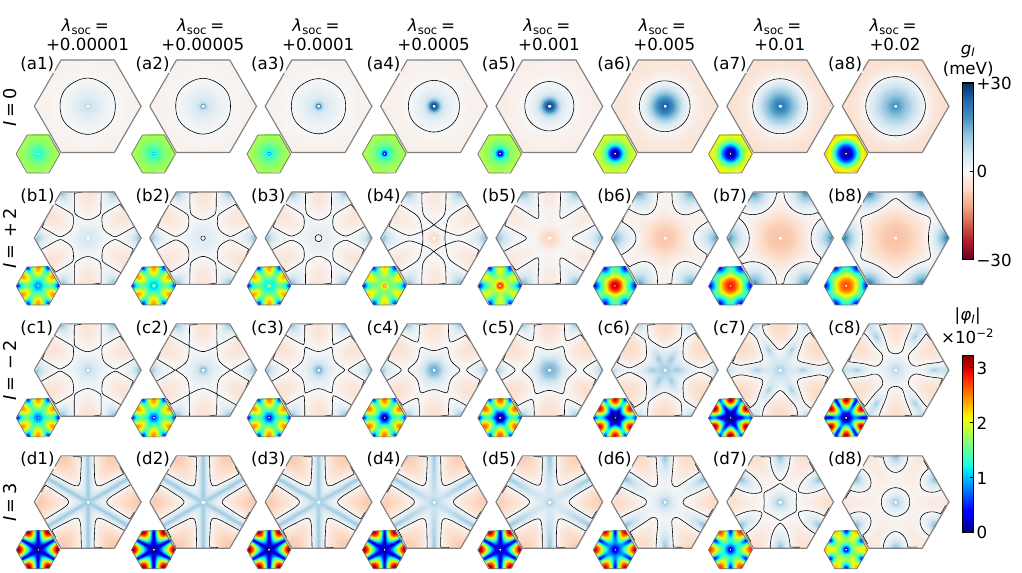}
\caption{SOC dependence of the amplitude gradient and native
magnitude at positive SOC. Rows, scales and symbols follow
figure~\ref{fig:si-gradient-reversal}; columns 1--8 now correspond
to $\lambda_{\mathrm{soc}}=10^{-5},5\times10^{-5},10^{-4},
5\times10^{-4},10^{-3},0.005,0.01,0.02\,\mathrm{eV}$.
The Chern-locked branch in row (a) favors depletion at the
innermost sampled momenta throughout this range. Its weaker
small-SOC gradient accompanies a more nearly uniform native
magnitude outside the unresolved core.
For the zero-winding branch in row (b), the innermost gradient
is positive in (b1) but negative in (b2)--(b8), indicating a
change in the preferred initial amplitude redistribution.
Rows (c,d) give the other selected classes for comparison.
The main panels evaluate flattened trials; the insets show native
eigenstate magnitudes.}
\label{fig:si-gradient-positive}
\end{figure}

The relation to phase matching is illustrated by the near-core
maps in figures~\ref{fig:si-core-source-zero} and
\ref{fig:si-core-source-pisix}. Their fixed source positions are
$\widetilde{\mathbf k}=(0.23936,0)$ and $(0.17952,0.10365)$,
respectively, each at $4/35$ of the directional FBZ radius.
At positive SOC, $C_{\mathrm{exc}}=-2$: the Chern-locked branch
has $\mathcal I=0$ and $m_0=-2$, while the zero-winding branch
has $\mathcal I=+2$ and $m_{+2}=0$.
At the smallest SOC, the high-overlap region largely has
$\chi_0\simeq0$; the larger residual phases occur mainly where
$|\Lambda|$ is small. Their energy weight is consequently reduced.
At larger SOC, the mismatch extends into the high-overlap region,
and suppresses the coherent attractive contribution near the core.
Comparison with the outer-source maps in
figures~\ref{fig:si-positive-source1} and
\ref{fig:si-positive-source2} shows that the phase texture also
depends on the distance of the source from $\Gamma$.
Large phase variations alone, however, do not determine the
weighted sum or the sign of $g$.

To make this distinction explicit, define the off-diagonal
attractive field of the flattened trial by
$\bar F_{\mathcal I,i}=-\sum_{j\ne i}K_{ij}
\bar\varphi_{\mathcal I,j}$ and let
$\bar D_{\mathcal I,i}=\Delta_i+K_{ii}-\bar E_{\mathcal I}$.
Then
\[
g_{\mathcal I,i}
=2\left[\bar D_{\mathcal I,i}\bar r
-\operatorname{Re}\!\left(e^{-\mathrm i\theta_{\mathcal I,i}}
\bar F_{\mathcal I,i}\right)\right].
\]
Both the link magnitudes and their phases enter $\bar F$, and the
diagonal and normalization terms remain in $\bar D$.
At the smallest SOCs, weak cancellation at the resolved inner
momenta makes the link magnitudes important to the core penalty.
Cancellation becomes substantial near
$\lambda_{\mathrm{soc}}=10^{-3}\,\mathrm{eV}$, where its balance
with $\bar D$ produces the larger gradient.
The phase maps display the central image, whereas $g$ includes
all seven images.

For the positive-SOC zero-winding sector, the innermost gradient
changes from positive in figure~\ref{fig:si-gradient-positive}(b1)
to negative in (b2)--(b8), favouring the central enhancement seen
in the corresponding insets. The nearly aligned high-overlap
phases in panels (h5)--(h8) of figures~\ref{fig:si-core-source-zero}
and \ref{fig:si-core-source-pisix} support this trend.
Under SOC reversal the zero-winding sector changes from
$\mathcal I=+2$ to $-2$.

For the Chern-locked sector, weaker depletion at resolved momenta
as SOC decreases suggests a narrowing core. We quantify its
recovery using the refined mesh below.

\begin{figure}[p]
\centering
\includegraphics[width=\textwidth]{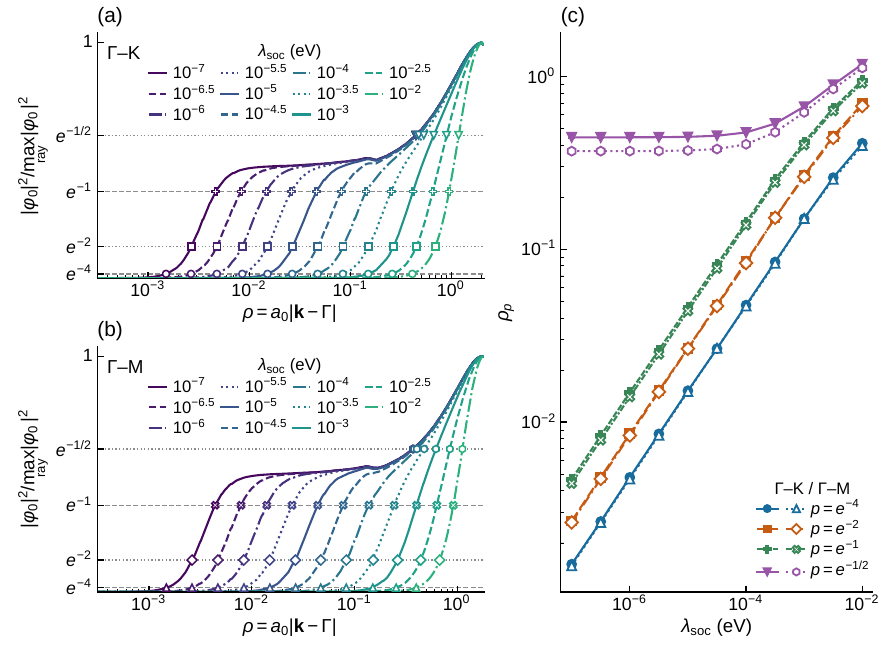}
\caption{Inner-core and outer-envelope density recovery of the
$\mathcal I=0$ branch on a refined momentum mesh.
(a,b), left: Density normalized to its own ray maximum along
$\Gamma$--$K$ and $\Gamma$--$M$, respectively, for the eleven
positive SOC values in the legend.
The dimensionless radius is $\rho=a_0|\mathbf k-\Gamma|$.
Horizontal lines and symbols identify the density fractions
$p=e^{-4},e^{-2},e^{-1},e^{-1/2}$.
(c), right: First upward-crossing radii $\rho_p$ for the two directions.
At the smallest SOCs shown, the three low-threshold radii
follow approximately square-root trends, providing numerical
evidence for a vanishing momentum-space core as
$\lambda_{\mathrm{soc}}\to0^+$. The higher-threshold radius
follows the broad outer rise and remains approximately constant.
Curves interpolate the scalar density of the weighted eigenstates
using the mesh and fixed transfer cutoff specified in the text.}
\label{fig:si-density-recovery}
\end{figure}

\paragraph{Refined-grid density and recovery radii.}
For figure~\ref{fig:si-density-recovery}, we replace the periodic
Voronoi cells of $\Gamma$ and the 18 native points within twice
the nearest-neighbour mesh spacing by a $C_6$-symmetric quadrature.
Each triangular fan from $\Gamma$ to a patch boundary edge is
parametrized by $\widetilde{\mathbf k}=s[(1-t)\mathbf a+t\mathbf b]$,
where $\mathbf a,\mathbf b$ are its outer vertices. Four-point
Gauss--Legendre quadrature in $s$ is applied on
$[2^{-\ell-1},2^{-\ell}]$ for $\ell=0,\ldots,7$ and
$[0,2^{-8}]$, with two-point quadrature in $t\in[0,1]$.
The weights include the Jacobian $s|\det(\mathbf a,\mathbf b)|$;
retained outer points keep their periodic Voronoi-cell areas.
The resulting 5814-point mesh reaches
$\rho=a_0|\mathbf k-\Gamma|\simeq3.44\times10^{-5}$ and
excludes exact $\Gamma$.

The physical area weights $A_i$ are the dimensionless quadrature
areas divided by $a_0^2$. Set $w_i=A_i/(2\pi)^2$ and
$c_i=\sqrt{w_i}\,\varphi_0(\mathbf k_i)$. The Hermitian matrix
acting on $c$ is
\[
H^{(c)}_{ij}=\Delta_i\delta_{ij}
+\sqrt{w_iw_j}\,\frac{L_{ij}+L_{ji}^{*}}{2},
\qquad
L_{ij}=\sum_{\mathbf G}
U(|\mathbf k_i-\mathbf k_j-\mathbf G|)\Lambda_{i,j+\mathbf G},
\]
where $U$ is the unweighted interaction in
Eq.~\eqref{eq:si-rk}. The seven-image sum omits transfers
$a_0|\mathbf k_i-\mathbf k_j-\mathbf G|<10^{-3}$ image by image.
Both inner and outer amplitudes are updated in double precision
by Davidson iteration in the rotation sector selected by the
native $\mathcal I=0$ state. Convergence requires
$\|H^{(c)}c-Ec\|/(\|H^{(c)}c\|+|E|\|c\|)\leq10^{-8}$
and an energy change below $10^{-8}\,\mathrm{eV}$.
The normalization is
$\sum_i|c_i|^2=\sum_iw_i|\varphi_0(\mathbf k_i)|^2=1$,
so the continuous-normalization density is $|c_i|^2/w_i$.

We interpolate the scalar density $|\varphi_0|^2$ directly,
using a local linear radial-basis function with 12 neighbours
and no smoothing, along $d=\Gamma$--$K$ and $\Gamma$--$M$,
without angular averaging. Define
\[
D_{\lambda_{\mathrm{soc}}}^{(d)}(\rho)=
\frac{|\varphi_0(\Gamma+(\rho/a_0)\mathbf e_d;\lambda_{\mathrm{soc}})|^2}
{\max_{\rho'\,\mathrm{on\ ray}\ d}
 |\varphi_0(\Gamma+(\rho'/a_0)\mathbf e_d;\lambda_{\mathrm{soc}})|^2},
\qquad \lambda_{\mathrm{soc}}>0,
\]
where $\mathbf e_d$ is the unit vector along the selected ray. The density recovery radius $\rho_p^{(d)}$ is the
first upward crossing of $D_{\lambda_{\mathrm{soc}}}^{(d)}=p$.
Its associated momentum scale is $\rho_p^{(d)}/a_0$.
These radii are operational measures of a momentum-space profile,
with the reference maximum chosen separately for each SOC and ray.
The short continuation to the exact $K/M$ endpoints uses extrapolation.

Figures~\ref{fig:si-density-recovery}(a,b) show an inner rise,
an intermediate plateau and a broader increase towards the outer
FBZ at small SOC. The plateau lies above $e^{-1}$ and below
$e^{-1/2}$, separating the low-threshold core radii from the
higher-threshold envelope scale. Thus a nearly constant
$\rho_{e^{-1/2}}$ is compatible with a contracting inner core.

The radii for $p=e^{-4},e^{-2},e^{-1}$ decrease as SOC is reduced
on both rays. Log--log fits over the five smallest SOCs,
$10^{-7}$--$10^{-5}\,\mathrm{eV}$, give slopes close to $1/2$;
extending the fit range to $10^{-4}\,\mathrm{eV}$ preserves
this trend for the two lowest thresholds.
Within the sampled finite-cutoff model, these results provide
numerical evidence for a vanishing inner-core scale as
$\lambda_{\mathrm{soc}}\to0^+$, while the higher-threshold
radius remains nearly constant.
The broad outer rise reflects a nonuniform envelope shaped by
band geometry and the nonlocal interaction, beyond the narrow
vortex structure near $\Gamma$.

This separation of scales complements
figures~\ref{fig:si-core-source-zero}--\ref{fig:si-gradient-positive}.
The phase and form-factor maps resolve the coherence and overlap
of individual links,
and the trial gradients describe the initial amplitude response.
The refined eigenstates include the joint response of link
magnitudes, phases, band separation and normalization, resolving
the narrow core beneath the broader native-mesh profile.

\paragraph{Small-SOC limit and numerical stability.}
A shrinking core with a nonzero limiting inner plateau suggests
nonuniform density recovery. If the finite-SOC envelope is
continuous and vanishes at $\Gamma$, this interpretation gives
\[
\lim_{\rho\to0^+}\lim_{\lambda_{\mathrm{soc}}\to0^+}
D_{\lambda_{\mathrm{soc}}}^{(d)}(\rho)=D_*^{(d)}>0,
\qquad
\lim_{\lambda_{\mathrm{soc}}\to0^+}\lim_{\rho\to0^+}
D_{\lambda_{\mathrm{soc}}}^{(d)}(\rho)=0.
\]
These conditional ordered limits describe the extrapolated trend;
they do not assign a path-independent joint limit or a complex
wavefunction at the band touching, where the isolated-band class
is undefined.

Changing the refined-patch area between $1/16$ and $16$ times
the baseline, at fixed minimum inner resolution, preserves the
plateau and approximately square-root contraction. The tested
recovery radii for $p=e^{-2},e^{-1},e^{-1/2}$ change by at most
about $6\%$. This patch-sensitivity check supports the separation
of inner and outer scales at the stated transfer cutoff.


\clearpage

\section{Limits of a local momentum-space energy description}
\label{sec:si-local-coulomb}

The phase-matching analysis in section~S3 retains the interaction between
finitely separated momenta. Here we examine whether the corresponding
energy can instead be reconstructed from a local expression proportional
to $|D\varphi|^2$. The distinction concerns the energetic approximation
in momentum space, rather than whether the Wannier equation contains
the screened Coulomb interaction. Berry connection, covariant derivative
and Berry curvature describe local geometric information. Reconstructing
the Coulomb energy from this information requires control of the
finite-separation contributions omitted by the expansion.

\subsection{From the exact deviation cost to a local gradient approximation}
\label{sec:si-local-expansion}

We retain the uniform-mesh normalization
$\sum_{\mathbf k}|\varphi(\mathbf k)|^2=1$ and the reciprocal continuation
of section~S1. The source $\mathbf k$ lies in the FBZ, while
$\mathbf k'$ runs over the sampled points in the central FBZ and six
adjacent BZs. Write $\ell=(\mathbf k,\mathbf k')$ for a retained
ordered pair, with $u_\ell=\Delta A\,U(|\mathbf k'-\mathbf k|)/(2\pi)^2<0$.
The quadrature measure is thus already included in $u_\ell$.
For $\Lambda_\ell\ne0$, define
$J_\ell=-u_\ell|\Lambda_\ell|\geq0$ and
$P_\ell=\Lambda_\ell/|\Lambda_\ell|$.
Writing $\varphi_{\mathbf k}=r_{\mathbf k}e^{\mathrm i\theta_{\mathbf k}}$,
the combined phase is
$\chi_\ell=\theta_{\mathbf k'}-\theta_{\mathbf k}
+\operatorname{Arg}\Lambda_\ell$ modulo $2\pi$.
The deviation cost of section~S3.1 is
\begin{equation}
d_\ell=J_\ell r_{\mathbf k}r_{\mathbf k'}(1-\cos\chi_\ell),
\qquad
K[\varphi]=-\sum_\ell J_\ell r_{\mathbf k}r_{\mathbf k'}
+\sum_\ell d_\ell.
\label{eq:si-local-phase-cost}
\end{equation}
A zero-weight pair contributes zero without requiring its phase.
All sums include the retained reciprocal images and ordered pairs.

An exact quadratic decomposition makes the relation to a covariant
gradient explicit:
\begin{equation}
\begin{split}
K[\varphi]&=B[\varphi]+\mathcal E_{\mathrm{NL}}[\varphi],\\
B[\varphi]&=-\frac12\sum_\ell J_\ell
\bigl(r_{\mathbf k}^2+r_{\mathbf k'}^2\bigr),\\
\mathcal E_{\mathrm{NL}}[\varphi]
&=\frac12\sum_\ell J_\ell
\left|P_\ell\varphi_{\mathbf k'}-\varphi_{\mathbf k}\right|^2\\
&=\frac12\sum_\ell J_\ell
\bigl(r_{\mathbf k'}-r_{\mathbf k}\bigr)^2+\sum_\ell d_\ell.
\end{split}
\label{eq:si-local-exact-decomposition}
\end{equation}
The factors of one half belong to this quadratic decomposition.
They do not introduce an extra factor into the interaction expectation.
The positive quantity $\mathcal E_{\mathrm{NL}}$ contains both amplitude
variation and phase mismatch. Its generally state-dependent negative
baseline $B$ must be retained when reconstructing $K$.

Use dimensionless momenta $\widetilde{\mathbf k}=a_0\mathbf k$ and
$\vec{\delta}_\ell=a_0(\mathbf k'-\mathbf k)$.
In a smooth local frame,
$P_\ell\varphi_{\mathbf k'}-\varphi_{\mathbf k}
=\delta_\ell^a D_a\varphi_{\mathbf k}+O(|\vec{\delta}_\ell|^2)$,
where $D_a=\partial/\partial\widetilde k_a-\mathrm i\widetilde A_{{\rm exc},a}$
and $\widetilde A_{{\rm exc},a}=A_{{\rm exc},a}/a_0$.
Restricting transfers to $|\vec{\delta}_\ell|\leq Q$ therefore gives
the candidate local energy
\begin{equation}
E_{\mathrm{loc},Q}[\varphi]
=\sum_{\mathbf k}(D_a\varphi_{\mathbf k})^*
\kappa_Q^{ab}(\mathbf k)D_b\varphi_{\mathbf k},
\qquad
\kappa_Q^{ab}(\mathbf k)
=\frac12\sum_{\mathbf k'}^{|\vec{\delta}_\ell|\leq Q}
J_\ell\delta_\ell^a\delta_\ell^b.
\label{eq:si-local-tensor}
\end{equation}
Repeated Cartesian indices are summed. A scalar expression
$\kappa\sum_{\mathbf k}|D\varphi_{\mathbf k}|^2$ additionally replaces
this tensor by a momentum-independent isotropic coefficient.
This approximation requires controlled variation of the full product
$\varphi_{\mathbf k}^*\Lambda_{\mathbf k\mathbf k'}
\varphi_{\mathbf k'}$ over the retained separations.
A well-defined derivative at $\mathbf k$ does not establish that
condition throughout the momentum range coupled by the interaction.

\subsection{Cutoff dependence of the scalar stiffness}
\label{sec:si-local-kappa}

The scalar-stiffness test uses the second moment of the bare screened
interaction, before inserting the form-factor magnitude:
\begin{equation}
\kappa_{\mathrm b}(Q)
=-\frac{1}{4N_k}
\sum_\ell^{10^{-3}\leq|\vec{\delta}_\ell|\leq Q}
u_\ell|\vec{\delta}_\ell|^2.
\label{eq:si-local-bare-kappa}
\end{equation}
The factor $1/N_k$ averages over source momenta.
This is a bare-kernel diagnostic, distinct from the
form-factor-weighted tensor in equation~\eqref{eq:si-local-tensor}.
It is independent of the exciton state and, at fixed interaction and
mesh, of SOC. With dimensionless momentum coordinates its units are
eV. The corresponding coefficient for derivatives with respect to
physical momentum is $\kappa_{\mathrm b}/a_0^2$.

\begin{figure}[tbp]
\centering
\includegraphics[width=110mm]{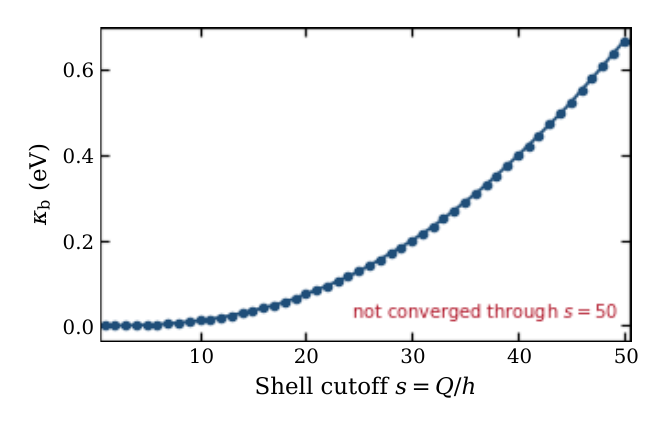}
\caption{Cutoff dependence of the bare scalar stiffness
$\kappa_{\mathrm b}(sh)$ in equation~\eqref{eq:si-local-bare-kappa}.
The calculation includes the central FBZ and six reciprocal images
on the $N=35$ mesh. The coefficient has not reached a plateau by
$s=50$. The stiffness is expressed in eV using the dimensionless momentum
coordinates of the calculation.}
\label{fig:si-local-kappa}
\end{figure}

For the $N=35$ mesh, the nearest-neighbour spacing is
$h=2\pi/(3N)\simeq0.05984$. Figure~\ref{fig:si-local-kappa}
shows the cumulative coefficient for $Q=sh$, with all seven reciprocal
images retained. It continues to increase over the tested shells
$s=1,\ldots,50$, reaching approximately $0.665$~eV without a plateau.
The scan therefore does not select a cutoff-independent stiffness
within this range. The factor $|\vec{\delta}|^2$ enhances the
sensitivity of this moment to larger transfers.
The scan therefore supplies no controlled scalar stiffness for the
full energy. Including larger transfers to increase $\kappa$ is
justified by a local expansion only if the corresponding remainder
remains small. Saturation after exhausting the finite set of retained
transfers would not establish that condition.

\subsection{A rescaled nearest-neighbour deformation energy}
\label{sec:si-local-rescaled}

To examine the local-gradient description on the calculated excitons,
the nearest-neighbour test evaluates
\begin{equation}
E_{\mathrm{def}}^{0}[\varphi]
=-\frac12\sum_{\ell\in\mathcal L_h}
u_\ell\left|\Lambda_\ell\varphi_{\mathbf k'}
-\varphi_{\mathbf k}\right|^2,
\qquad
E_{\mathrm{def}}^{(\kappa)}[\varphi]
=\frac{\kappa}{\kappa_1}E_{\mathrm{def}}^{0}[\varphi],
\label{eq:si-local-stencil}
\end{equation}
where $\mathcal L_h$ contains directed first-neighbour pairs and
$\kappa_1=\kappa_{\mathrm b}(h)\simeq4.17\times10^{-5}$~eV.
The superscript $0$ denotes the unrescaled first-shell energy,
not zero SOC. This finite-step diagnostic uses the full $\Lambda$, including its
magnitude. It approaches a covariant-gradient expression only when
the nearest-neighbour expansion is accurate, and is not identical
to the $J$-weighted quantity in equation~\eqref{eq:si-local-exact-decomposition}.

\begin{figure}[!htbp]
\centering
\includegraphics[width=\linewidth]{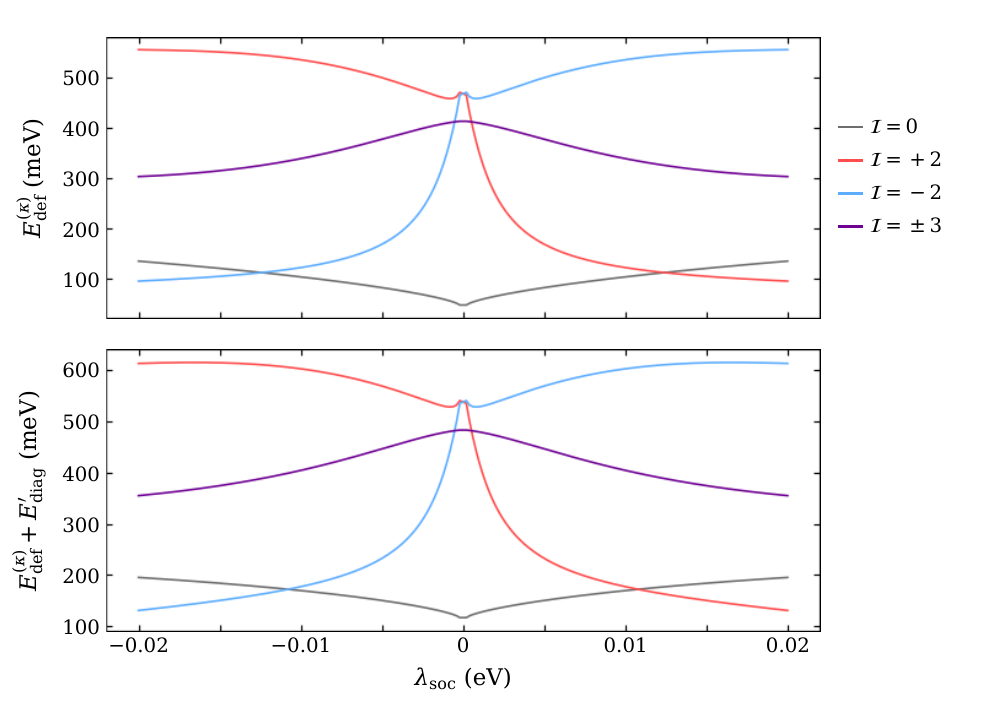}
\caption{Nearest-neighbour deformation diagnostic with prescribed
$\kappa=0.10$~eV. The upper panel shows
$E_{\mathrm{def}}^{(\kappa)}$ and the lower panel adds
$E'_{\mathrm{diag}}$. The reference is the $\Gamma$ band gap at
$\lambda_{\mathrm{soc}}=0.02$~eV, approximately $0.891$~eV.
The horizontal axis is $\lambda_{\mathrm{soc}}$ in eV.
The displayed deformation values at zero SOC are interpolated from
$\lambda_{\mathrm{soc}}=\pm0.0002$~eV, and do not assign a relative
index at the band touching. The labels $+3$ and $-3$ denote the same
modulo-six class.}
\label{fig:si-local-energy}
\end{figure}

Figure~\ref{fig:si-local-energy} uses the prescribed value
$\kappa=0.10$~eV, rather than a converged result of the shell scan.
It increases the first-shell energy by a factor of approximately
$2.40\times10^3$. Before rescaling, the saved first-shell energies
are only $0.008$--$0.56$~meV over the evaluated states and SOC values.
The resulting rescaled curves retain a qualitative
competition and crossing between the vortex and zero-winding
excitons, resembling the broad trend in Main Figure~2(c).
Their small-SOC structure differs, however. Within the displayed range,
the $\mathcal I=\pm2$ and $\mathcal I=3$ curves remain separated,
whereas the corresponding branches in Main Figure~2(c) approach one
another. The deformation values at zero SOC are interpolated between
finite-SOC samples, while the lower panel retains the stored diagonal
contribution there. These connectors do not define isolated-band
indices at the band touching.

The lower panel adds
$E'_{\mathrm{diag}}[\varphi]=\sum_{\mathbf k}
\Delta_{\mathbf k}|\varphi_{\mathbf k}|^2-\Delta_{\mathrm{ref}}$,
with the same reference gap $\Delta_{\mathrm{ref}}$ for every state
and SOC. This constant subtraction changes the energy origin.
It does not supply the state-dependent attractive baseline or the
omitted finite-transfer contributions. The plotted quantity is
therefore a test of a chosen local reconstruction, rather than the
full exciton energy. Its qualitative crossing after a large empirical
rescaling does not show that the microscopic local approximation
explains the observed level crossing.

\subsection{Finite-separation distribution of the deviation cost}
\label{sec:si-local-maps}

The momentum-resolved maps expose information that a scalar stiffness
does not retain. For a fixed source momentum, define
$z_\ell=u_\ell\varphi_{\mathbf k}^*\Lambda_\ell\varphi_{\mathbf k'}$.
The plotted quantity is
\begin{equation}
\mathcal C_\ell=2\bigl(\operatorname{Re}z_\ell+|z_\ell|\bigr)
=2J_\ell r_{\mathbf k}r_{\mathbf k'}(1-\cos\chi_\ell)
=2d_\ell.
\label{eq:si-local-map-cost}
\end{equation}
It includes the interaction weight, overlap magnitude and both
coefficient amplitudes. Its factor of two follows the diagnostic
pair-plus-conjugate convention. The full ordered-pair sum of
deviation costs in equation~\eqref{eq:si-local-phase-cost} uses
$\mathcal C_\ell/2$.

\begin{figure}[tbp]
\centering
\includegraphics[width=\linewidth]{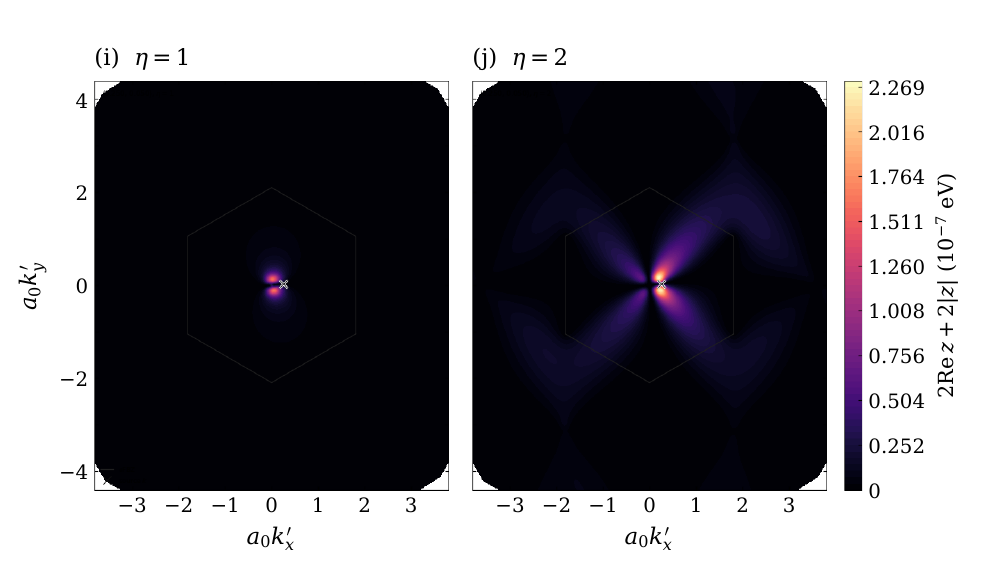}
\caption{Fixed-source deviation-cost maps
$\mathcal C_\ell=2d_\ell$ for the two lowest excitons
at $\lambda_{\mathrm{soc}}=0.0002$~eV.
The original diagnostic panel labels are retained: (i) $\eta=1$
and (j) $\eta=2$. The cross marks the source and the central
hexagon marks the FBZ. The $\mathbf k'$ axes are in dimensionless
coordinates and cover the retained BZs. Contributions at different
$\mathbf k'$ points are displayed separately. The maps interpolate the complex $z_\ell$ before
evaluating equation~\eqref{eq:si-local-map-cost} and use a common
linear colour scale in eV. Native pairs with
$|\vec{\delta}|<10^{-3}$ are excluded. Interpolated pixels near
the source do not represent additional short-transfer samples.}
\label{fig:si-local-cost-maps}
\end{figure}

Figure~\ref{fig:si-local-cost-maps} compares the two lowest
energy-ranked states at $\lambda_{\mathrm{soc}}=0.0002$~eV.
The source is fixed at
$\widetilde{\mathbf k}\simeq(0.2591,0.02992)$ while the extended
target momentum varies. The second state displays a pronounced,
anisotropic cost over finite separations, rather than a contribution
confined to an infinitesimal neighbourhood of the source.
The first state has a different and substantially more concentrated
distribution. These differences arise from the combined phase and
its weights, not from the transfer magnitude alone.

The relevant question is whether the region that contributes to the
energy lies within the validity range of the local expansion.
The anisotropic finite-separation cost cannot be assumed to follow
from the small-separation expansion merely by increasing its scalar
coefficient. The following decomposition tests this issue directly
by separating the locally approximated contribution from the omitted
finite-transfer energy.

\subsection{A fixed-state test of the omitted contribution}
\label{sec:si-local-splitting}

A separate stored calculation supplies this comparison using the
exact decomposition in equation~\eqref{eq:si-local-exact-decomposition}.
Let $\mathcal E_{<,Q}$ contain its transfers with
$|\vec{\delta}_\ell|\leq Q$,
$\mathcal E_{>,Q}=\mathcal E_{\mathrm{NL}}-\mathcal E_{<,Q}$,
and $R_Q=\mathcal E_{<,Q}-E_{\mathrm{loc},Q}$.
Then
\begin{equation}
E[\varphi]=E_{\mathrm{diag}}[\varphi]+B[\varphi]
+E_{\mathrm{loc},Q}[\varphi]+\mathcal E_{>,Q}[\varphi]+R_Q[\varphi].
\label{eq:si-local-complete-energy}
\end{equation}
This test retains the full, state-dependent baseline $B$ and replaces
only the positive nonlocal term by its local approximation.
It therefore separates a missing-baseline error from the effect
of omitted finite transfers.

For the same $N=35$, seven-image YYK model at
$\lambda_{\mathrm{soc}}=0.010$~eV, take a hard cutoff $Q=5h$.
The local derivatives are obtained by a gauge-covariant
least-squares fit over the first-neighbour stencil, with weights
$1/|\vec{\delta}|^2$. The hard window $Q=5h$ was the largest tested
window passing a common 5\% accuracy criterion for the short-transfer
energy, including a check on longer pairs excluded from the derivative
fit, for the two competing states at six tested SOC values.
It is a diagnostic cutoff, not a universal interaction range.
For this separate calculation, the stored vectors were promoted to
complex128 precision and individually divided by their norms before
evaluating the decomposition.
Table~\ref{tab:si-local-splitting} compares the resulting energy
differences between the two excitons.
\clearpage
\begin{table}[!htbp]
\centering
\caption{Fixed-state decomposition of
$\Delta X=X_{m=0}-X_{m=-2}$ at
$\lambda_{\mathrm{soc}}=0.010$~eV for the $N=35$ mesh and $Q=5h$.
All entries are in meV and use the same two normalized eigenstates
of the full Coulomb problem, retaining the central and six adjacent BZs.
The local reconstruction retains the exact interaction baseline $B$.}
\label{tab:si-local-splitting}
\begin{tabular}{lr}
\hline
Contribution & $\Delta X$ (meV)\\
\hline
Diagonal energy $E_{\mathrm{diag}}$ & $-10.90$\\
Exact interaction baseline $B$ & $-25.67$\\
Local tensor energy $E_{\mathrm{loc},Q}$ & $+0.33$\\
Finite-transfer term $\mathcal E_{>,Q}$ & $+66.48$\\
Short-transfer remainder $R_Q$ & $-0.02$\\
\hline
$E_{\mathrm{diag}}+B+E_{\mathrm{loc},Q}$ & $-36.24$\\
Full energy $E$ & $+30.23$\\
\hline
\end{tabular}
\end{table}
The local reconstruction reverses the energy ordering of these fixed
states. The small short-transfer remainder does not rescue it,
because the omitted finite-transfer term sets the sign of the
splitting. These are expectation values of the same selected states,
not eigenvalues of a newly diagonalized local Hamiltonian.
At this SOC the local energies reproduce the two exact short-transfer
energies to better than 1\%, but this accurate short-transfer
approximation still misses the contribution that determines the ordering.

The limitation is therefore one of energetic self-consistency.
Restricting the expansion to a range where its short-transfer energy
is accurate yields a contribution too small to determine the full
ordering. Increasing a scalar stiffness by hand does not recover
the omitted momentum-pair dependence and provides no controlled
explanation of the observed level crossing. The cutoff dependence,
the prescribed-stiffness test and the finite-transfer comparison
motivate retaining $a_{\mathcal I}$ and $\chi_{\mathcal I}$ over the
full retained momentum-pair set when analysing the YYK energy ordering.

The partial qualitative agreement with local geometric diagnostics
can be understood from the angular phase structure discussed in
Main section~III. In the displayed frame at small, nonzero SOC,
$\chi_{\mathcal I}\simeq(m_{\mathcal I}-C_{\mathrm{exc}})\delta\phi$,
where $\delta\phi$ is the azimuthal angle difference and the integer
difference is not reduced modulo six.
At fixed pair weights, this relation favours the Chern-locked state
$m_0=C_{\mathrm{exc}}$ by suppressing its deviation cost.
For neighbouring points on a smooth core-surrounding loop with
approximately constant amplitude, the same compensation gives
$D_\parallel\varphi_0\simeq0$.
A geometric interpretation based on Berry phases or Wilson loops,
together with the coefficient winding, and a covariant-derivative
diagnostic can therefore reflect the same phase-matching tendency.
This correspondence helps explain qualitative agreement without
establishing the interaction-energy scale.

\looseness=-1
Such agreement relies on the approximate angular form of both the
Bloch-overlap phase and the exciton phase over the relevant momentum
region. It requires additional structure of the Bloch Hamiltonian
and a suitable parameter regime, rather than following from the
band Chern numbers alone. As this angular form changes, loop phases
and local gradients do not determine the weighted contributions
between finitely separated momenta, or the resulting crossing.
Thus, qualitative agreement after rescaling does not validate a
local reconstruction of the Coulomb energy.
In the general case, the energetic analysis must retain the full
nonlocal momentum-pair contributions, as in Main section~III,
unless the error of a local reduction is independently controlled.

\clearpage

\section{Optical matrix elements and polarization-resolved spectra}
\label{sec:si-optical-methods}

\subsection{Bond-based optical operator}

We evaluate normalized absorption using a nearest-neighbour
approximation to the interband momentum matrix element
\cite{berghauser2014analytical}. The orbital momentum overlaps are
represented by a common prefactor multiplying the bond direction;
this prefactor, including its conventional phase, is absorbed into
the normalization. The resulting bond sum retains the momentum
dependence from the orbital geometry and Bloch coefficients.
Let $i,j$ label the orbitals in the tight-binding basis,
and let $\mathbf r_i$ denote their intracell positions. For each translation
$\mathbf R$ in the set
$\mathcal R_{\mathrm{opt}}=\{\mathbf 0,\pm2\mathbf a_1,\pm2\mathbf a_2\}$,
we define the displacement
\begin{equation}
\mathbf b_{ij}^{\mathbf R}
=\mathbf R+\mathbf r_i-\mathbf r_j.
\label{eq:si-opt-bond}
\end{equation}
Here $\mathbf a_1$ and $\mathbf a_2$ are the geometric vectors used to
construct the kagome unit cell, and the displacements are expressed in the
same dimensionless length units as the tight-binding geometry. The bond
indicator $s_{ij}^{\mathbf R}$ is one when
$\bigl|\lvert\mathbf b_{ij}^{\mathbf R}\rvert-1\bigr|<0.1$ and zero
otherwise. The Cartesian optical matrices used in the calculation are
\begin{equation}
[\hat O_\mu(\mathbf k)]_{ij}
=\sum_{\mathbf R\in\mathcal R_{\mathrm{opt}}}
s_{ij}^{\mathbf R}\,b_{ij,\mu}^{\mathbf R}
\exp\!\left(\mathrm{i}\widetilde{\mathbf k}\cdot\mathbf b_{ij}^{\mathbf R}\right),
\qquad \mu\in\{x,y\}.
\label{eq:si-opt-operator}
\end{equation}
The corresponding interband matrix element is
\begin{equation}
M_\mu(\mathbf k)
=\langle u_{c\mathbf k}|\hat O_\mu(\mathbf k)|u_{v\mathbf k}\rangle.
\label{eq:si-opt-matrix}
\end{equation}
We use $M_\pm=M_x\pm\mathrm{i}M_y$ for the two circular components,
labelled by $\sigma=\pm$. The common factor $1/\sqrt{2}$ for unit-normalized circular vectors
also cancels from the normalized spectra. The figures therefore
report polarization-resolved absorption within this common-overlap
approximation, with an undetermined overall scale.

\subsection{Coherent transition amplitudes and line shape}
\label{sec:si-opt-coherence}

Under the Bloch-frame transformation of
section~\ref{sec:si-gauge-index}, the interband optical matrix
element acquires the same phase as the envelope:
\begin{equation}
M_\sigma'(\mathbf k)
=e^{\mathrm i\zeta(\mathbf k)}M_\sigma(\mathbf k).
\label{eq:si-optical-matrix-gauge}
\end{equation}
The transition function
$T_{\sigma,\eta}=M_\sigma\varphi^{(\eta)*}$ is therefore
gauge invariant. On a common contour where both factors are
continuous and nonzero, its winding is
\begin{equation}
w_c(T_{\sigma,\eta})
=w_c(M_\sigma)-w_c(\varphi^{(\eta)}).
\label{eq:si-transition-winding-identity}
\end{equation}
Gauge covariance does not require rotational symmetry. Its connection to the polarization selection rule for
states of definite rotation character is derived in
section~\ref{sec:si-opt-selection}.

The exciton eigenvectors on the uniform production mesh satisfy
$\sum_{\mathbf k}|\varphi^{(\eta)}(\mathbf k)|^2=1$.
Their optical transition amplitudes are coherent sums over this same mesh,
\begin{equation}
\mathcal T_{\sigma,\eta}
=\sum_{\mathbf k}M_\sigma(\mathbf k)
\varphi^{(\eta)*}(\mathbf k),
\label{eq:si-opt-amplitude}
\end{equation}
with oscillator weights
$\Theta_\sigma^{(\eta)}=|\mathcal T_{\sigma,\eta}|^2$.
The uniform quadrature factor is absorbed into the common response
normalization.

The Elliott response follows from the coherent excitonic polarization
\cite{elliott1957intensity,berghauser2014analytical}. Writing the photon
energy as $E_\gamma=\hbar\omega>0$, we evaluate
\begin{equation}
\mathcal S_\sigma(E_\gamma;\lambda_{\mathrm{soc}})
=\frac{1}{E_\gamma}
\sum_\eta
\frac{\gamma\,\Theta_\sigma^{(\eta)}}
{\bigl(E^{(\eta)}-E_\gamma\bigr)^2+\gamma^2}.
\label{eq:si-opt-response}
\end{equation}
The Lorentzian is the imaginary part of
$[E^{(\eta)}-E_\gamma-\mathrm{i}\gamma]^{-1}$, with half width
$\gamma=10\,\mathrm{meV}$. The photon-energy factor $1/E_\gamma$
is retained, and all 50 computed exciton states enter the sum at
each SOC.

For the spectrum at $\lambda_{\mathrm{soc}}=0.010\,\mathrm{eV}$,
both circular-polarization curves share one normalization,
\begin{equation}
\widetilde{\alpha}_\sigma(E_\gamma)
=\frac{\mathcal S_\sigma(E_\gamma;0.010\,\mathrm{eV})}
{\displaystyle\max_{\tau=\pm,\,E\in\mathcal E_{\mathrm{line}}}
\mathcal S_\tau(E;0.010\,\mathrm{eV})},
\label{eq:si-opt-normalized-line}
\end{equation}
where $\mathcal E_{\mathrm{line}}$ spans 10--1000\,meV.
This preserves the relative intensity of the two polarizations.
For the SOC maps, each polarization is normalized over its full
sampled energy range (10--760\,meV) and SOC range,
\begin{equation}
\widetilde{\alpha}_\sigma(E_\gamma;\lambda_{\mathrm{soc}})
=\frac{\mathcal S_\sigma(E_\gamma;\lambda_{\mathrm{soc}})}
{\displaystyle\max_{E\in\mathcal E_{\mathrm{map}},\,\lambda_{\mathrm{soc}}'
\in\mathcal L_{\mathrm{map}}}
\mathcal S_\sigma(E;\lambda_{\mathrm{soc}}')},
\label{eq:si-opt-normalized-map}
\end{equation}
where $\mathcal E_{\mathrm{map}}$ and $\mathcal L_{\mathrm{map}}$ denote the
sampled photon energies and spin--orbit couplings. Thus the intensity
variation with $\lambda_{\mathrm{soc}}$ is retained within each panel.
The two panels have independent overall intensity scales.

\subsection{Spectral reflection and the projected optical character}
\label{sec:si-zero-t2-character}

We first derive the projected optical character needed for the
selection rule in section~\ref{sec:si-opt-selection}.
At $t_2=0$, subtract the common on-site energy by defining
$H_0(\widetilde{\mathbf k})=H(\widetilde{\mathbf k})-\epsilon_{\rm orb}\mathbf 1_6$.
Here $\widetilde{\mathbf k}=a_0\mathbf k$, as in the Hamiltonian above.
Using Pauli matrices $\tau_\mu$ acting on the two three-orbital blocks gives
\begin{equation}
 H_0(\widetilde{\mathbf k})=\tau_x\otimes B(\widetilde{\mathbf k})
 +\tau_z\otimes A(\widetilde{\mathbf k})+\mathbf 1_2\otimes C(\widetilde{\mathbf k}),
\label{eq:si-zero-t2-blocks}
\end{equation}
where $A=\operatorname{Re}D_+$ and
$C=\mathrm i\operatorname{Im}D_+$ are evaluated at
$\widetilde{\mathbf k}$ and $t_2=0$.
The matrices $A$ and $B$ are real symmetric, while $C$ is purely
imaginary and Hermitian. Let $Q=\tau_y\otimes\mathbf 1_3$ and let
$\mathcal K$ denote complex conjugation. The antiunitary operation
$\mathcal C=Q\mathcal K$ produces a fixed-$\mathbf k$ Bloch-matrix
spectral reflection,
\begin{equation}
 QH_0^*(\widetilde{\mathbf k})Q^\dagger=-H_0(\widetilde{\mathbf k}),
\label{eq:si-spectral-reflection}
\end{equation}
and $\mathcal C^2=-\mathbf 1_6$.
For nondegenerate conduction and valence bands forming the central
positive--negative energy pair, with a nonzero gap, their vectors can
therefore be related by
\begin{equation}
 |u_{c\mathbf k}\rangle
 =\mathrm e^{\mathrm i\alpha(\mathbf k)}Q|u_{v\mathbf k}^*\rangle.
\label{eq:si-central-conjugate-pair}
\end{equation}
Here $\alpha(\mathbf k)\in\mathbb R$ accounts for the relative phases
of the two normalized Bloch eigenvectors.

To establish the projected rotation character, we compare the
implemented bond sum with the complete sixfold bond set. Denote
the implemented five-translation operator by
$O_\mu^{(5)}(\mathbf k)\equiv\hat O_\mu(\mathbf k)$,
as defined in equation~\eqref{eq:si-opt-operator}.
Introduce $O_\mu^{(7)}$ solely as an algebraic comparison, obtained by
including the two real-space optical translations $\pm2\mathbf a_3$
in the same bond sum. These are distinct from the reciprocal images
retained in the Coulomb kernel. At the stated embedding, squared bond
lengths are integers, so the prescribed tolerance around unit length
selects precisely the unit bonds. Direct enumeration gives
$O_x^{(5)}-O_x^{(7)}=0$ and
\begin{equation}
 O_y^{(5)}-O_y^{(7)}=
 \begin{pmatrix}L&0\\0&L^{\mathrm T}\end{pmatrix},
\label{eq:si-opt-missing-bonds}
\end{equation}
where
\begin{equation}
 L=\begin{pmatrix}
 0&0&0\\
 0&0&\mathrm e^{-\mathrm i\kappa_3}\\
 0&-\mathrm e^{\mathrm i\kappa_3}&0
 \end{pmatrix}.
\label{eq:si-opt-defect-block}
\end{equation}
Writing this difference as $\Delta O_y$, the matrix
$Q^\dagger\Delta O_y$ is antisymmetric under transpose.
Consequently, for the conjugate band pair above,
\begin{equation}
 \langle u_{c\mathbf k}|\Delta O_y|u_{v\mathbf k}\rangle=0.
\label{eq:si-opt-projected-cancellation}
\end{equation}
The same equality holds for the $x$ component, so both operators have
identical projected matrix elements in this channel at $t_2=0$.

The complete bond sum transforms as a Cartesian vector under the
sixfold orbital rotation. Combining this transformation with the
Bloch-frame sewing phases in equations~\eqref{eq:si-valence-sector}
and~\eqref{eq:si-conduction-sector} gives
\begin{equation}
 M_\pm(R_6\mathbf k)
 =\exp[\mathrm i(r_v+r_c\pm1)\pi/3]M_\pm(\mathbf k).
\label{eq:si-opt-projected-character}
\end{equation}
This establishes the optical character of the implemented projected
channel in the zero-$t_2$ model, without replacing its operator.
The exact cancellation in the selection rule additionally requires
an exciton of definite rotation character and a symmetry-complete
momentum sum.

Finite $t_2$ adds a real block-diagonal term that removes the spectral
reflection used in this argument, while the Hamiltonian retains its
$C_6$ spatial symmetry. Departures from the projected cancellation
therefore need not indicate broken lattice rotation symmetry.

\subsection{Rotational selection rule and the sixfold realization}
\label{sec:si-opt-selection}
The coherent sum in equation~\eqref{eq:si-opt-amplitude} makes the
selection rule explicit for general $C_n$ symmetry
\cite{cao2018unifying,zhang2018optical}.
Use the gauge-invariant transition function
$T_{\sigma,\eta}(\mathbf k)$ defined in
section~\ref{sec:si-opt-coherence}, and let
$R_n$ denote a rotation by $2\pi/n$. Suppose that
the selected exciton and optical channel have a definite character,
\begin{equation}
T_{\sigma,\eta}(R_n\mathbf k)
=\exp\!\left(\frac{2\pi\mathrm{i}\ell_{\sigma,\eta}}{n}\right)
T_{\sigma,\eta}(\mathbf k),
\label{eq:si-opt-character}
\end{equation}
where $\ell_{\sigma,\eta}$ is an integer defined modulo $n$.
For a symmetry-complete mesh with equal integration weights on each
rotation orbit,
\begin{equation}
\sum_{j=0}^{n-1}T_{\sigma,\eta}(R_n^j\mathbf k)
=T_{\sigma,\eta}(\mathbf k)
\sum_{j=0}^{n-1}\exp\!\left(\frac{2\pi\mathrm{i}j\ell_{\sigma,\eta}}{n}\right).
\label{eq:si-opt-orbit-sum}
\end{equation}
The geometric sum vanishes unless $\ell_{\sigma,\eta}\equiv0\pmod n$.
Equivalently, rotation leaves the complete weighted sum unchanged
while multiplying its summand by the character; a nontrivial
character therefore forces a zero transition amplitude. This latter
argument also includes momenta with shorter rotation orbits.

On a rotation-invariant contour on which $T_{\sigma,\eta}$ is continuous
and nonzero, its phase change over one $n$th of the contour is
$2\pi\ell_{\sigma,\eta}/n$ up to an integer multiple of $2\pi$.
Thus
\begin{equation}
w(T_{\sigma,\eta})\equiv\ell_{\sigma,\eta}\pmod n.
\label{eq:si-opt-winding-character}
\end{equation}
On a common contour where the factors are nonzero, the product
$T_{\sigma,\eta}=M_\sigma\varphi^{(\eta)*}$ gives the necessary condition
$w(M_\sigma)-m^{(\eta)}\equiv0\pmod n$.
Equivalently, the relative class must satisfy
$[m^{(\eta)}-C_{\rm exc}]_n
\equiv[w(M_\sigma)-C_{\rm exc}]_n$.
Setting $n=6$ gives the selection rule of the main text. A permitted
class need not have a nonzero oscillator weight because the coherent
sum can undergo further cancellation. Within a degenerate exciton
subspace, states must be chosen with definite rotation character
before this state-by-state rule is applied.

For the projected YYK channel at $t_2=0$,
equation~\eqref{eq:si-opt-projected-character} gives the allowed
class $\mathcal I\equiv r_v+r_c\pm1-C_{\mathrm{exc}}\pmod6$
for $\sigma_\pm$. On the negative-SOC side,
$r_v+r_c=-1$ and $C_{\mathrm{exc}}=+2$; on the positive-SOC
side these are $+1$ and $-2$. Both substitutions select
$\mathcal I=-2$ for $\sigma_+$ and $\mathcal I=+2$ for
$\sigma_-$. The Chern-locked class $\mathcal I=0$ is
therefore excluded from both circular channels under these
assumptions.

\clearpage
\makeatletter
\immediate\write\@auxout{\string\citation{apsrevControl}}
\makeatother
\bibliographystyle{apsrev4-2}
\bibliography{bib}